\documentclass[11pt]{article}

\usepackage[left=1in,top=1in,right=1in,bottom=1in,head=.1in,nofoot]{geometry}

\usepackage{setspace,url,bm,amsmath} 

\usepackage{titlesec} 
\titlelabel{\thetitle.\quad} 

\usepackage[margin=20pt]{subcaption}

\usepackage{amsmath,amsfonts,amsthm,amssymb}
\usepackage{graphicx}

\usepackage[colorlinks=true,linkcolor=black,citecolor=blue,urlcolor=blue]{hyperref}
\usepackage{threeparttable}
\usepackage{manyfoot}%
\usepackage{booktabs}%
\usepackage{algorithm}%
\usepackage{algorithmicx}%
\usepackage{algpseudocode}%
\usepackage{listings}%
\usepackage{dsfont}%
\usepackage{bm}
\usepackage{etoolbox}
\usepackage{natbib}
\usepackage{adjustbox}
\usepackage{threeparttable}
\usepackage{dcolumn}

\theoremstyle{definition}

\newtheorem*{theorem*}{Theorem}
\newtheorem{theorem}{Theorem}[section]

\newtheorem{lemma}{Lemma}[section]
\newtheorem{remark}{Remark}[section]
\newtheorem{example}{Example}[section]

\newtheorem*{corollary*}{Corollary}

\newtheorem{condition}{Condition}

\newtheorem{proposition}{Proposition}[section]

\DeclareFontFamily{U}{mathx}{\hyphenchar\font45}
\DeclareFontShape{U}{mathx}{m}{n}{
      <5> <6> <7> <8> <9> <10>
      <10.95> <12> <14.4> <17.28> <20.74> <24.88>
      mathx10
      }{}
\DeclareSymbolFont{mathx}{U}{mathx}{m}{n}
\DeclareFontSubstitution{U}{mathx}{m}{n}
\DeclareMathAccent{\widecheck}{0}{mathx}{"71}
\DeclareMathAccent{\wideparen}{0}{mathx}{"75}

\allowdisplaybreaks[4]

\def\T{\text{T}}

\def\v{{\varepsilon}}
\def\pr{P}
\def\Perp{\perp\!\!\!\!\perp}

\title{\bf  Inference under covariate-adaptive
randomization with many strata}
\date{}

\author{
\small
{
Jiahui Xin$^{1}$, \ \ Hanzhong Liu$^{2}$, \ \ Wei Ma$^{1}$\thanks{\small{Correspondence: \texttt{mawei@ruc.edu.cn}}}
}
\\ \\
{\small $^{1}$ Institute of Statistics and Big Data, Renmin University of China, Beijing, China}\\
{\small $^{2}$ Center for Statistical Science, Department of Industrial Engineering, Tsinghua University, Beijing, China}
}

\begin{document}
\doublespacing

\maketitle


\begin{abstract}
Covariate-adaptive randomization is widely employed to balance baseline covariates in interventional studies such as clinical trials and experiments in development economics. Recent years have witnessed substantial progress in inference under covariate-adaptive randomization with a fixed number of strata. However, concerns have been raised about the impact of a large number of strata on its design and analysis, which is a common scenario in practice, such as in multicenter randomized clinical trials. In this paper, we propose a general framework for inference under covariate-adaptive randomization, which extends the seminal works of \cite{bugni2018inference, Bugni2019} by allowing for a diverging number of strata. Furthermore, we introduce a novel weighted regression adjustment that ensures efficiency improvement. On top of establishing the asymptotic theory, practical algorithms for handling situations involving an extremely large number of strata are also developed. Moreover, by linking design balance and inference robustness, we highlight the advantages of stratified block randomization, which enforces better covariate balance within strata compared to simple randomization. This paper offers a comprehensive landscape of inference under covariate-adaptive randomization, spanning from fixed to diverging to extremely large numbers of strata.

\vspace{12pt}
\end{abstract}

\section{Introduction}
Randomization is widely regarded as the gold standard for interventional studies such as clinical trials \citep{Rosenberger2015} and experiments in development economics \citep{duflo2007using,bruhn2009pursuit}. However, simple randomization may result in imbalances in important baseline covariates \citep{Cornfield1959}. Covariate-adaptive randomization is proposed to address this issue by ensuring balance among treatment groups with respect to covariates. Most notably, stratified block randomization \citep{Zelen1974} first stratifies experimental units into strata based on stratification variables and then assigns treatment using block randomization within each stratum separately. Recent surveys indicate that approximately 70\% of randomized trials have utilized stratified block randomization \citep{Lin2015, Ciolino2019}, in addition to its application in COVID-19 treatment or vaccine trials \citep{wang2020remdesivir, Baden2021}. The minimization method \citep{Pocock1975} aims to achieve balance across covariates' margins and has been extended to control various types of imbalance measures \citep{Hu2012}. This approach is commonly used in practice when dealing with a large number of strata. Nonetheless, minimization and other covariate-adaptive randomization methods have primarily been developed based on a fixed number of strata \citep{Hu2012,Hu2020,Zhao2024}; consequently, studies on inference also assume a fixed number of strata, typically requiring a correctly specified regression model for the outcome \citep{Shao2010, Ma2015}.

In recent years, robust inference under covariate-adaptive randomization has received much attention;  here, ``robust'' signifies that valid inference can be drawn without imposing a true data generating model. The seminal studies by \cite{bugni2018inference,Bugni2019} focused on regressions based on stratification indicators, although there remains a possibility of imbalances in other baseline variables \citep{liu2022balancing}. Recommendations from regulatory agencies such as the European Medicines Agency \citep{EMA2015} and the U.S. Food and Drug Administration \citep{FDA2023} endorse covariate adjustment to address the imbalances. It has been theoretically demonstrated that regression adjustment with additional baseline covariates improves efficiency for covariate-adaptive randomization with the same assignment probability across strata \citep{ma2022regression, ye2022inference,gu2023regression,liu2023lasso}.  Furthermore, nonparametric adjustment to further enhance efficiency or achieve semiparametric efficiency bounds are also being considered \citep{bannick2023general,Rafi2023,tu2023unified,wang2023model}. Still, all the above-mentioned works are concerned with the fixed number of strata.

However, a large number of strata is commonly seen in practice. This is particularly the case when the randomization procedure involves sites. In multicenter clinical trials, multiple sites are typically enrolled to meet sample size requirements and to include participants from diverse demographics, thus enhancing the generalizability of findings \citep{guideline1999statistical}. The number of strata can increase quickly when combining sites with other stratification variables. For example, even in the simple case of 10 sites and two prognostic factors, with 2 and 5 levels respectively, the total number of strata would be 100.
Clearly, there is a gap between the common scenario with a large and possibly diverging number of strata and the restricted theory built on a fixed number of strata. As noted by \cite{FDA2023}, ``The statistical properties of such methods [inference under covariate-adaptive randomization] are best understood when the number of strata is small relative to the sample size." The consequence of this lack of theory for many strata is that practitioners may intentionally or unintentionally remove some stratification variables from the analysis, losing efficiency and ignoring heterogeneity.

To fill this gap, this paper lays a solid theoretical foundation for inference under covariate-adaptive randomization with a diverging number of strata: (i) establishing the asymptotic properties of the stratified difference-in-means estimator under mild assumptions about the randomization procedure; (ii) proposing two forms of regression adjustment (unweighted and weighted) with delicate adjustments for degrees of freedom; (iii) presenting important randomization examples to fulfill the assumptions required by the asymptotic theory. These are accomplished by integrating several key results and techniques, including stratified sampling results from \cite{Bickel1984}, the coupling technique in \cite{bugni2018inference, Bugni2019}, the use of triangular arrays to manage the diverging number of strata, and tail bounds in non-asymptotic statistics. On top of developing theory for cases with a diverging number of strata, practical algorithms are also developed and investigated for handling situations involving an extremely large number of strata, each possibly containing very few units.

Our contributions lie in multiple aspects. First, we extend the results of \cite{bugni2018inference,Bugni2019} allowing for a diverging number of strata. Second, we provide a novel weighted regression adjustment that guarantees efficiency improvement even under various assignment probabilities across strata. Third, we point out the benefit of stratified block randomization in the case of an extremely large number of strata by linking design balance and inference robustness. To conclude, by establishing both rigorous theory and practical methods, we offer a comprehensive landscape of inference under covariate-adaptive randomization, spanning from fixed to diverging to extremely large numbers of strata.

\section{Setup and Assumptions}

\subsection{Setup}
We adopt the Neyman-Rubin potential outcome model. Each unit is represented by $W_{ni}^\T = (Y_i(0), Y_i(1), B_i^\T, X_i^\T)$, $i = 1, \ldots, n$ which are independently and identically distributed (i.i.d.). Let $W^{(n)} = (W_{n1}, \ldots, W_{nn})^\T$. For treatment $a \in \{0, 1\}$, $Y_i(a)$ denotes the potential outcome. $B_i \in \mathbb{R}^{r_n}$ serves as the covariate for randomization, and if $B_i$ is continuous, we discretize it before randomization. $X_i \in \mathbb{R}^p$ is the additional covariate for regression, discussed further in Section~\ref{section:regress}. We allow $r_n$, the dimension of $B_i$, to increase as $n$ grows. 

$S_n(\cdot): \mathrm{supp}(B_i) \to \mathcal{S}_n$ is a function of the covariate  for randomization $B_i$, with the stratification indicator $S_{ni} = S_n(B_i)$. We use the triangular-array-like notation $\{S_{ni}\}_{i=1}^n$ to accommodate the diverging number of strata. Let $S^{(n)} = (S_{n1}, \ldots, S_{nn})^\T$. The experimenter then selects a vector of treatment assignments $A^{(n)}$ based on $S^{(n)}$, where $A^{(n)} = (A_{n1}, \ldots, A_{nn})^\T$ and $A_{ni}$ takes values from $\{0, 1\}$. After treatment assignments, we only observe \[Y_{ni} = A_{ni}Y_i(1) + (1 - A_{ni})Y_i(0).\] Our interest lies in inferring the average treatment effect (ATE): \[\tau = E\{Y_i(1) - Y_i(0)\}\] from the observed data $\{Y_{ni}, A_{ni}, S_{ni}, X_i\}_{i=1}^n$.

The following notations are introduced. Define $[a, s]$ as the index $\{i: S_{ni} = s, A_{ni} = a\}$, $[\cdot, s]$ as the index $\{i: S_{ni} = s\}$, and $[a, \cdot]$ as the index $\{i: A_{ni} = a\}$. Let the stratum probability $p_{n}(s) = \pr(S_{ni} = s)$. Define the stratum-specific treatment group size, stratum size, and treatment group size as $n_a(s)$, $n(s)$, and $n_a$, respectively. The target stratum-specific and overall assignment probabilities, as well as the actual stratum-specific and overall assignment proportions, are denoted as $\pi_{na}(s)$, $\pi_{na} = \sum_s p_n(s)\pi_{na}(s)$, and $\hat{\pi}_{na}(s) = n_a(s)/n(s)$, $\hat{\pi}_{na} = n_a/n$, respectively.

For any i.i.d. (multivariate) potential outcome $f_{ni}(a)$ from the population $f_n(a)$, $a \in \{0, 1\}$, we introduce high-level notations for convenience. If the subscript does not depend on $n$, such as $f = Y$ or $X$, we allow the degenerate form $f_{ni}(a) = f_i(a)$ from the population $f_n(a) = f(a)$. In each treatment arm within each stratum, the mean, second-order moment, and variance of $f$ are denoted as $\mu(f, a, s) = E\{f_{ni}(a)|S_{ni} = s, A_{ni} = a\}$, $\mu_2(f, a, s) = E\{f_{ni}(a) {f_{ni}}^\T(a)|S_{ni} = s, A_{ni} = a\}$, and $\sigma^2(f, a, s) = \mathrm{var}\{f_{ni}(a)|S_{ni} = s, A_{ni} = a\}$, respectively. Correspondingly, the sample mean, sample second-order moment, and sample variance of $f$ are defined as $\hat{\mu}(f, a, s) = \sum_{i \in [a, s]}{f_{ni}(a)}/{n_a(s)}$, $\hat{\mu}_2(f, a, s) = \sum_{i \in [a, s]}{f_{ni} (a){f_{ni}}^\T(a)}/{n_a(s)}$, and $\hat{\sigma}^2(f, a, s) = \sum_{i \in [a, s]}{\{f_{ni}(a) - \hat{\mu}(f, a, s)\}\{f_{ni}(a) - \hat{\mu}(f, a, s)\}^\T}\allowbreak/{\{n_a(s) - 1\}}$, respectively. In the bracket, replacing $a$ or $s$ with $\cdot$ signifies averaging the corresponding indices, e.g., 
$[\cdot,s]$ and $\sigma^2(f,\cdot,s)$. Additionally, define $\tilde{f}_{ni}(a) = f_{ni}(a) - E\{f_{ni}(a)|S_{ni}\}$ and $\check{f}_{ni}(a) = \tilde{f}_{ni}(a)/\sqrt{\pi_{na}(S_{ni})}$.

\subsection{Assumptions}
Subsequently, we introduce a series of assumptions.
We commence with a set of assumptions regarding potential outcomes.
\newline

(A1) For $a\in\{0,1\}$, for some $\v>0$,
    \[
    \sup_{n\ge1,s\in\mathcal S_n}E\{\mid \tilde Y_i(a)\mid^{2+\v}\mid S_{ni}=s\}<\infty.
    \]

(A2) For $a\in\{0,1\}$,
\[
\liminf_{n\to\infty}E[\mathrm{var}\{Y_i(a)\mid S_{ni}\}]>0.
\]

\begin{remark}
Assumptions (A1) and (A2) largely correspond to Assumption 2.1 in \cite{bugni2018inference,Bugni2019}, although there are some slight differences, particularly in Assumption (A1). Our generalization accommodates scenarios where the number of strata diverges. Specifically, Assumption (A1) originated from \cite{Bickel1984}, where we weaken the original third-order moment to a $(2+\v)$-order moment. Assumption (A2) is introduced to ensure non-degeneracy and comparable with Assumption 3.1 (a) with $m_{d,n}=0$ in \cite{bai2024covariate}.
\end{remark}

The subsequent set of assumptions pertains to the experimental design.
\newline

(B1) $A^{(n)}\Perp W^{(n)}\mid S^{(n)}.$

(B2) For $a\in\{0,1\}$,
\[
0<\inf_{n\ge1,s\in\mathcal S_n}\pi_{na}(s)\le\sup_{n\ge1,s\in\mathcal S_n}\pi_{na}(s)<1.
\]

(B3) For $a\in\{0,1\}$,
    \[
    \sup_{s\in\mathcal S_n}\mid \hat\pi_{na}(s)-\pi_{na}(s)\mid =o_P(1)
    \]
    and
    \[
\lim_{n\to\infty}\pr\left(\inf_{s\in\mathcal S_n}n_a(s)\ge1\right)=1.
    \]
\begin{remark}
Assumptions (B1) and (B3) correspond closely to Assumption 2.2 (a) and (b) in \cite{Bugni2019}, respectively.
Assumptions (B1) and (B2) are  standard requirements commonly known as ``unconfoundedness" and ``overlap" respectively in i.i.d. observational studies \citep{Imbens2015}. Covariate-adaptive randomization naturally ensures unconfoundedness and overlap but it could be challenging to verify unconfoundedness and overlap in observational studies and even impossible.
Assumption (B2) allows for different assignment probabilities across different strata, similar to \cite{Bugni2019}.
Assumption (B3) stipulates the uniform convergence of the actual assigned proportions and prohibits the presence of any empty treatment arm within any stratum in an asymptotic sense.
If the number of strata remains fixed within the asymptotic framework, Assumptions (B1)-(B3) naturally reduce to Assumption 2.2 (a) and (b) in \cite{Bugni2019}.

\end{remark}

\begin{remark} 
    Matched-pair experiments \citep{Bai2022,bai2024covariate} adhere to Assumptions~(B1)--(B3), yet their framework differs from ours. More specifically, matching algorithm relies on a single realization of all covariates for randomization $\{B_{i}\}_{i=1}^n$ while our pre-specification of some stratification scheme could sequentially assign units. Despite this difference in theoretical framework, our findings in Section 4 and 5 reveal some similarity in results of \cite{Bai2022,bai2024covariate}.
\end{remark}
\section{Examples}
Experiments often involve many strata. Thus, it is important to approximate the sampling distribution with the number of strata $|\mathcal{S}_n|$ growing with the total sample size $n$ at a certain rate. We argue that this type of diverging asymptotic regime is for building theory, and in practice, experimenters may only observe one pair of $(n, |\mathcal{S}_n|)$. In this section, we briefly outline some examples that satisfy Assumptions (B1)--(B3). Let $A^{(n,k)} = (A_{n1}, \ldots, A_{nk})^\T$ and $S^{(n,k)} = (S_{n1}, \ldots, S_{nk})^\T$. Without loss of generality, we assume that $|\mathcal{S}_n|$ is non-decreasing and $\inf_s p_n(s)$ is non-increasing as $n$ grows.

\begin{example}[Simple randomization]
Simple randomization involves $n$ independent assignments. For $1\leq k\leq n$ and $a\in\{0,1\}$, the assignment probability is given by:
\[
\pr\{A_{nk}=a\mid (S^{(n,k)},A^{(n,k-1)})\}=\pr\{A_{nk}=a\mid S_{nk}\}\
=\pi_{na}(S_{nk}).
\]
Simple randomization is a special case where all strata have the same assignment probability $\pi_{na}$.
\end{example}

\begin{proposition}
\label{prop:sr}
Under Assumption (B2), if 
    \[\log(|\mathcal S_n|)=o\left(n \inf_s p^2_n(s)\log\{|\inf_s p_n(s)|\} \right)\] 
and 
\[
n \inf_s p^2_n(s)\log\{|\inf_s p_n(s)|\}\to\infty,
\]then
simple randomization satisfies Assumptions (B1)--(B3). If additionally assuming every stratum shares the same stratum probability $p_n(s)=1/|\mathcal S_n|$, then simple randomization with $|\mathcal S_n|=O(n^{\kappa})$ for any $0\leq\kappa<1/2$ satisfies Assumptions (B1)--(B3).
\end{proposition}

\begin{example}[Stratified block randomization]
Within every stratum $s$ in $\mathcal S_n$, stratified block randomization \citep{Zelen1974} allocates
$
n_1(s)=\lfloor \pi_{n1}(s)n(s) \rfloor
$ units to treatment. The remaining units are assigned to control with equal probability, calculated as:
\begin{align*}
\frac{1}{\binom{n(s)}{n_1(s)}}.
\end{align*}
\end{example}
\begin{proposition}
\label{prop:sbr}
Under Assumption (B2), if
    \[
2{|\log\{\inf_s p_n(s)\}|}{n^2\inf_s p^2_n(s)} -\log|\mathcal S_n|\to\infty
\]
and
\[
n\inf_s p_n(s)\to\infty,
\]
then stratified block randomization satisfies Assumptions (B1)--(B3). If additionally assume that every stratum shares the same stratum probability $p_n(s)=1/|\mathcal S_n|$, then stratified block randomization with $\mid \mathcal S_n\mid=O(n^{\kappa})$ for any $0\leq\kappa<1$ satisfies Assumptions (B1)--(B3).
\end{proposition}

\begin{remark}[Minimization]
    The minimization method \citep{Pocock1975} aims to balance the marginal distribution of covariates rather than achieving balance within strata. The assignment probability is defined as follows:
\begin{align*}
    P\left\{A_{nk}=1|S^{(n,k)}, A^{(n,k-1)}\right\}= \begin{cases}\frac{1}{2} & \text { if } \mathrm{Imb}_{nk}=0 \\ \lambda & \text { if } \mathrm{Imb}_{nk}<0 \\ 1-\lambda & \text { if } \mathrm{Imb}_{nk}>0\end{cases},
\end{align*}
where $1/2\leq\lambda\leq1$, $\mathrm{Imb}_{nk}=\mathrm{Imb}(S^{(n,k)},A^{(n,k-1)})$, and $\mathrm{Imb}(\cdot)$ represents some weighted average of marginal imbalance.
In Hu and Hu's method \citep{Hu2012} which generalizes minimization, $\mathrm{Imb}(\cdot)$ can account for imbalance both marginally and within strata. This helps achieve a more balanced allocation, especially when the number of strata is relatively large.
However, both minimization and Hu and Hu's method introduce complex dependencies across samples. This dependence poses challenges in verifying whether the design condition is sufficient to deduce Assumption (B3). Hence, we consider the specific design conditions under minimization as a conjecture.
\end{remark}

\begin{remark}
    Literature on inference under many strata also expands within other design frameworks. In the finite population framework \citep{splawa1990application, Imbens2015}, \cite{Liu2020} considered regression adjustment accommodating a growing number of strata in stratified randomized experiments. Our design framework pre-specifies a stratification scheme within a super-population. This results in stratum sizes following a multinomial distribution, mirroring actual data collection processes like clinic recruitment. Superpopulation assumption is commonly made in experimental design literature \citep{benkeser2021improving,armstrong2022asymptotic}.
\end{remark}

\section{Stratified Difference-in-means}

\cite{ma2022regression} pointed that the stratified difference-in-means estimator is equivalent to the fully saturated linear regression \citep{Bugni2019} and recommended the stratified difference-in-means because of its optimality among regression-adjusted estimators only with stratum indicators. But their setting is that the number of strata is fixed. In the following we establish asymptotic normality of the stratified difference-in-means that allows a diverging number of strata.

We give the third group of assumptions concerning the limit of variances and regression coefficients. Assumption~(C1) guarantees existence of the limit of variance within strata $V_{nW}(Y)$ and variance between strata $V_{nB}(Y)$.
\newline

(C1) $V_{W}(Y,a)=\lim_{n\to\infty} V_{nW}(Y,a)$ and $V_{B}(Y)=\lim_{n\to\infty} V_{nB}(Y)$ both exist for $a\in\{0,1\}$, where
\begin{align*}
V_{nW}(Y,a)&=E\left[\frac{\mathrm{var}\{Y_i(a)\mid S_{ni}\}}{\pi_{na}(S_{ni})}\right],\\
V_{nB}(Y)&=\mathrm{var}[E\{Y_i(1)-Y_i(0)\mid S_{ni}\}].
\end{align*}

\begin{theorem}
\label{thm:var}
Let $\hat\tau$ be the stratified difference-in-means estimator
\[
\hat\tau=\sum_{s\in\mathcal S_n} \frac{n(s)}{n} \{\hat\mu(Y,1,s)-\hat\mu(Y,0,s)\}.
\] 
If Assumptions~(A1)--(A2), (B1)--(B3) and (C1) hold, 
we have
$$
\sqrt{n}\left(\hat{\tau}-\tau\right) \xrightarrow{d} \mathcal N\left(0,V_{\hat\tau}\right),
$$
where
\[V_{\hat\tau}=V_W(Y,0)+V_W(Y,1)+V_{B}(Y).
\]
\end{theorem}

\begin{remark}
    Theorem~\ref{thm:var} extends findgings of Theorem 3.1 in \cite{bugni2018inference} from a scenario with a fixed number of strata to one with a diverging number of strata. \cite{Rafi2023} showed the semiparametric efficiency bound under covariate-adaptive randomization with a fixed number of strata. When no additional covariates are employed for adjustment—meaning all covariates are discrete and utilized for randomization—the stratified difference-in-means estimator \citep{Bugni2019, ma2022regression} achieves this bound. Theorem~\ref{thm:var} establishes that the asymptotic variance of the stratified difference-in-means estimator retains the same form even in the case with a diverging number of strata.
\end{remark}

\begin{remark}
    \cite{Rafi2023} established the semiparametric bound under covariate-adaptive randomization with a fixed number of strata. To the best of our knowledge, there exists no literature that has established the efficiency bound under covariate-adaptive randomization in scenarios allowing for a diverging number of strata. It remains an open question whether, subject to certain regularity conditions using triangular arrays, the semiparametric efficiency bound with a diverging number of strata maintains the same form. If affirmed, this would imply that the stratified difference-in-means estimator also achieves the efficiency bound in cases with no additional covariates for adjustment and an increasing number of strata, as demonstrated in Theorem~\ref{thm:var}. 
\end{remark}

\begin{remark}
    $V_{\hat\tau}$ shares the same form as that of matched-pair experiments where matching requires all baseline covariates. In particular, we allow the dimension of the stratification variable to grow as the total sample size increases and the assignment of unit $i$ only need stratification variables of the first $i$ units.
     In a similar but different setting, \cite{bai2023efficiency} studied the efficiency bound of finely stratified experiments and showed that the naive moment estimator could achieve the efficiency bound if matching becomes finer and finer in their sense.
\end{remark}

\section{\label{section:regress}Regression Adjustment}
In this section, we focus on further improving the efficiency of the stratified difference-in-means estimator discussed in Section 4.  We present the asymptotic properties of both unweighted and weighted regression adjustments. Unweighted regression adjustment achieves efficiency improvement even under homogeneous propensity. Moreover, by employing propensity to weight covariance matrix, weighted regression adjustment enhances efficiency under heterogeneous propensity.

We define several population-level regression coefficients for following theorems. Let $\Sigma_{R Q}=E\left[\{R-E(R)\}\{Q-E(Q)\}^{\mathrm{T}}\right]$ be the covariance between two random vectors $R$ and $Q$. We assume that $\Sigma_{\tilde{X} \tilde{{X}}}$ is strictly positive-definite. 

\subsection{Unweighted regression adjustment}

Define the unweighted covariance matrices: 
\[
\Sigma_{\tilde X \tilde X}=E\{\mathrm{cov}(X_i,X_i\mid S_{ni})\},\quad \Sigma_{\tilde X \tilde Y(a)}=E\{\mathrm{cov}(X_i,Y_i(a)\mid S_{ni})\}.
\]For $a\in\{0,1\}$, denote  $\beta_n(a)=\Sigma_{\tilde{{X}} \tilde{{X}}}^{-1} \Sigma_{\tilde{{X}}\tilde{Y}(a)}$ as the population regression coefficient for regressing $\tilde{Y}_i(a)=Y_i(a)-E\left\{Y_i(a)|S_{ni}\right\}$ on $\tilde{{X}}_i={X}_i-E\left({X}_i|S_{ni}\right)$.

Define $r_{ni}(a)=Y_i(a)-X_i^\T\beta_{n}$, where 
\begin{align*}
    \beta_{n}&=\left\{\frac{1}{\pi_{n0}}\Sigma_{\tilde X\tilde X}+\frac{1}{\pi_{n1}}\Sigma_{\tilde X\tilde X}\right\}^{-1}\left\{\frac{1}{\pi_{n0}}\Sigma_{\tilde X\tilde Y(0)}+\frac{1}{\pi_{n1}}\Sigma_{\tilde X\tilde Y(1)}\right\}\\
    &=\pi_{n1}\Sigma_{\tilde X\tilde X}^{-1}\Sigma_{\tilde X\tilde Y(0)}+\pi_{n0}\Sigma_{\tilde X\tilde X}^{-1}\Sigma_{\tilde X\tilde Y(1)}\\
    &=\pi_{n1}\beta_n(0)+\pi_{n0}\beta_n(1).
\end{align*}

Define $\hat r_{ni}(a)=Y_i(a)-X_i^\T\hat\beta_{n}$,
where
\begin{align*}
    \hat\beta_{n}&=\pi_{n1}\hat\beta_n(0)+\pi_{n0}\hat\beta_n(1)\\
    &=\pi_{n1}\hat\Sigma_{\tilde X(0)\tilde X(0)}^{-1}\hat\Sigma_{\tilde X(0)\tilde Y(0)}+\pi_{n0}\hat\Sigma_{\tilde X(1)\tilde X(1)}^{-1}\hat\Sigma_{\tilde X(1)\tilde Y(1)},\\
    \hat\Sigma_{\tilde X(a)\tilde X(a)}&=\sum_{s\in\mathcal S_n} \frac{n(s)}{n}\frac{1}{n_a(s)-1}\sum_{i\in[a,s]}\{X_i-\hat\mu(X,a,s)\}\{X_i-\hat\mu(X,a,s)\}^\T,\\
    \hat\Sigma_{\tilde X(a)\tilde Y(a)}&=\sum_{s\in\mathcal S_n} \frac{n(s)}{n}\frac{1}{n_a(s)-1}\sum_{i\in[a,s]}\{X_i-\hat\mu(X,a,s)\}\{Y_i-\hat\mu(Y,a,s)\}.
\end{align*}

\begin{remark}
    In comparison with \cite{ma2022regression}, we have slightly modified the (co-)variance estimator $\hat\Sigma_{\tilde X(a)\tilde X(a)}$ and $\hat\Sigma_{\tilde X(a)\tilde Y(a)}$ by adjusting the degrees of freedom and employing unweighted regression. Consequently, the estimator differs from the one proposed in \citet{ma2022regression} but asymptotically equivalent with some conditions (See Section~\ref{section:comparison} in Appendix).
\end{remark}
\begin{remark}
    Our regression coefficient $\hat\beta_n(a),a\in\{0,1\}$ is same as ``the weighted least squares estimator based on the treated units" for many small strata in \cite{Liu2020}. To clarify, the term ``weighted" in \cite{Liu2020} pertains to units, while our ``unweighted" refers to the covariance matrix. 
\end{remark}

We make a stronger Assumption~(A1') and (A2') than Assumption~(A1) and (A2), respectively. We make the similar Assumption~(A3) about $X$ as Assumption~(A1') about $Y(a)$. 
\newline

(A1') For $a\in\{0,1\}$,
\[\sup_{n\ge1,s\in\mathcal S_n} E\{\tilde Y_{i}^4(a)\mid S_{ni}=s\}<\infty.\]

(A2') For $a\in\{0,1\}$,
\[
\liminf_{n\to\infty}E[\mathrm{var}\{r_{ni}(a)\mid S_{ni}\}]>0.
\]

(A3) For $a\in\{0,1\}$,
for all coordinate $k$,
    \[\sup_{n\ge1,s\in\mathcal S_n} E(\tilde X_{ik}^4\mid S_{ni}=s)<\infty.\]
\begin{remark}
    Assumption~(A1') requires fourth-order moments, stronger than ($2+\varepsilon$)-order moments due to covariance estimation in the triangular arrays. Assumption~(A3) is formulated for the same reason. In the context of paired stratified experiments within the finite population framework, \cite{fogarty2018regression} also assumed bounded fourth-order moments. Additionally, under the finite population framework but for ``coarsely stratified experiments" (in sense of \cite{fogarty2018mitigating}), \cite{Liu2020} proposed a weaker assumption concerning the maximum stratum-specific squared range. Assumption~(A2') maintains non-degeneracy similar to Assumption~(A2).
\end{remark}

Similar to \cite{Liu2020}, covariance estimation necessitates at least two samples per treatment arm within every stratum. We introduce the stronger Assumption~(B3') compared to Assumption~(B3).
\newline

(B3') For $a\in\{0,1\}$,
    \[
    \sup_{s\in\mathcal S_n}\mid \hat\pi_{na}(s)-\pi_{na}(s)\mid =o_P(1)
    \]
    and
    \[
    \lim_{n\to\infty}\pr\left(\inf_{s\in\mathcal S_n}n_a(s)\ge2\right)=1.
    \]

We make the similar Assumption~(C2) as Assumption~(C1) to keep the existence of asymptotic variance.
\newline

(C2) $\beta=\lim_{n\to\infty}\beta_{n}$, $V_{W}(r,a)=\lim_{n\to\infty} V_{nW}(r,a)$ and  $V_{B}(r)=\lim_{n\to\infty} V_{nB}(r)$ all exist for $a\in\{0,1\}$,  where
\begin{align*}
r_i(a)&=Y_i(a)-X_i^\T\beta,\\
V_{nW}(r,a)&=E\left[\frac{\mathrm{var}\{r_i(a)\mid S_{ni}\}}{\pi_{na}(S_{ni})}\right],\\
V_{nB}(r)&=\mathrm{var}[E\{r_{ni}(1)-r_{ni}(0)\mid S_{ni}\}].
\end{align*}
\begin{theorem}
\label{thm:var2}
Let $\hat\tau_{adj}$ be the stratified difference-in-means estimator with unweighted regression adjustment
\[
\hat\tau_{adj}=\sum_{s\in\mathcal S_n}\frac{n(s)}{n}\{\hat\mu(\hat r,1,s)-\hat\mu(\hat r,0,s)\}.
\]
If  Assumptions~(A1'), (A2'), (A3), (B1), (B2), (B3') and (C2) hold,  we have
$$
\sqrt{n}\left(\hat\tau_{adj}-\tau\right) \xrightarrow{d} \mathcal N\left(0, V_{\hat\tau_{adj}}\right),
$$
where
\[
V_{\hat\tau_{adj}}=V_W(r,0)+V_W(r,1)+V_B(r).
\]
\end{theorem}

\begin{remark}
    Given the adjustment function $X_i^\T\beta_n$, its estimate $X_i^\T\hat\beta_n$ is the working model to estimate the weighted conditional mean $\pi_{n1}E\{Y_i(0)\mid X_i\}+\pi_{n0}E\{Y_i(1)\mid X_i\}$. $V_{\hat\tau_\mathrm{adj}}$ shares the same form as that of covariate adjustment in matched-pair experiments. It can be checked similarly as Step 4 of Proof of Theorem 3.1 in \cite{bai2024covariate} while matching needs all baseline variables. Similar as Remark 3.2 in \cite{bai2024covariate}, if assignment probability is the same across strata and
    \begin{align*} & E\left\{\pi_{n1}Y_i(0)+\pi_{n0}Y_i(1)|S_{ni},  X_i\right\}-E\left\{\pi_{n1}Y_i(0)+\pi_{n0}Y_i(1)|S_{ni}\right\}\\
    =&X_i^\T\beta_n-E\left\{X_i^\T\beta_n\mid S_{ni}\right\}\end{align*}
with probability one, $V_{\hat\tau_\mathrm{adj}}$ attains the efficiency bound in \cite{Rafi2023},  although efficiency bound is not clear if the number of strata increases.
\end{remark}
\subsection{Weighted regression adjustment}

Define the weighted covariance matrices: 
\[
\Sigma_{\check X(a) \check X(a)}=E\left\{\frac{1}{\pi_{na}(S_{ni})}\mathrm{cov}(X_i,X_i\mid S_{ni})\right\},\quad \Sigma_{\check X(a) \check Y(a)}=E\left\{\frac{1}{\pi_{na}(S_{ni})}\mathrm{cov}(X_i,Y_i(a)\mid S_{ni})\right\}.
\]

Define $r^*_{ni}(a)=Y_i(a)-X_i^\T\beta^*_{n}$, where 
\begin{align*}
    \beta^*_{n}=\{\Sigma_{\check X(0)\check X(0)}+\Sigma_{\check X(1)\check X(1)}\}^{-1}\{\Sigma_{\check X(0)\check Y(0)}+\Sigma_{\check X(1)\check Y(1)}\}.
\end{align*}

Define $\hat r^*_{ni}(a)=Y_i(a)-X_i^\T\hat\beta^*_{n}$,
where
\begin{align*}
    \hat\beta^*_{n}&=\{\hat\Sigma_{\check X(0)\check X(0)}+\hat\Sigma_{\check X(1)\check X(1)}\}^{-1}\{\hat\Sigma_{\check X(0)\check Y(0)}+\hat\Sigma_{\check X(1)\check Y(1)}\},\\
    \hat\Sigma_{\check X(a)\check X(a)}&=\sum_{s\in\mathcal S_n} \frac{n(s)}{n}\frac{n(s)}{n_a(s)}\frac{1}{n_a(s)-1}\sum_{i\in[a,s]}\{X_i-\hat\mu(X,a,s)\}\{X_i-\hat\mu(X,a,s)\}^\T,\\
    \hat\Sigma_{\check X(a)\check Y(a)}&=\sum_{s\in\mathcal S_n} \frac{n(s)}{n}\frac{n(s)}{n_a(s)}\frac{1}{n_a(s)-1}\sum_{i\in[a,s]}\{X_i-\hat\mu(X,a,s)\}\{Y_i-\hat\mu(Y,a,s)\}.
\end{align*}

\begin{remark}
    Both \cite{ma2022regression} and \cite{gu2023regression} employed weighted regression adjustment. They refrain from adjusting degrees of freedom (using $1/n_a(s)$ instead of $1/\{n_a(s)-1\}$) and thus utilize the weight $\sqrt{n_a(s)/n(s)}$, which approximately equals $\sqrt{\pi_{na}(s)}$ if the number of strata is fixed. In this context, we adopt the reciprocal, $1/\sqrt{\pi_{na}(s)}$, as the optimal weight to guarantee efficiency improvement.
\end{remark}

We make similar Assumption~(A2'') and (C3) as Assumption~(A2') and (C2), respectively.
\newline

(A2'') For $a\in\{0,1\}$,
\[
\liminf_{n\to\infty}E[\mathrm{var}\{r^*_{ni}(a)\mid S_{ni}\}]>0.
\]

(C3) $\beta^*=\lim_{n\to\infty}\beta^*_{n}$, $V_{W}(r^*,a)=\lim_{n\to\infty} V_{nW}(r^*,a)$ and  $V_{B}(r^*)=\lim_{n\to\infty} V_{nB}(r^*)$ all exist for $a\in\{0,1\}$,  where
\begin{align*}
r^*_i(a)&=Y_i(a)-X_i^\T\beta^*,\\
V_{nW}(r^*,a)&=E\left[\frac{\mathrm{var}\{r^*_i(a)\mid S_{ni}\}}{\pi_{na}(S_{ni})}\right],\\
V_{nB}(r^*)&=\mathrm{var}[E\{r^*_i(1)-r^*_i(0)\mid S_{ni}\}].
\end{align*}

The following theorem shows efficiency gain of weighted regression adjustment even under varying assignment probability.
\begin{theorem}
\label{thm:var3}
Let $\hat\tau^*_{adj}$ be the stratified difference-in-means estimator with weighted regression adjustment
\[
\hat\tau^*_{adj}=\sum_{s\in\mathcal S_n}\frac{n(s)}{n}\{\hat\mu(\hat r^*,1,s)-\hat\mu(\hat r^*,0,s)\}.
\]
If Assumptions~(A1'), (A2''), (A3), (B1), (B2), (B3') and (C3) hold,  we have
$$
\sqrt{n}\left(\hat\tau^*_{adj}-\tau\right) \xrightarrow{d} \mathcal N\left(0, V_{\hat\tau^*_{adj}}\right),
$$
where
\[
V_{\hat\tau_{adj}^*}=V_W(r^*,0)+V_W(r^*,1)+V_B(r^*).
\]
Furthermore, $V_{\hat\tau_{adj}^*}\leq V_{\hat\tau}$.
\end{theorem}
\begin{remark}
\label{rmk:wt reg}
The weighted regression adjustment ensures efficiency improvement even when assignment probabilities differ across strata. In fact, if the assignment probability is the same across strata, the weighted regression adjustment reduces to the unweighted version, as discussed in Remark 13. However, achieving the efficiency bound (presumably the same as the case of fixed strata) requires linearity of the conditional mean, which is almost never satisfied.
In the context of covariate-adaptive randomization with a fixed number of strata, \cite{Rafi2023} demonstrated that nonparametric adjustment using a cross-fitted kernel estimator can achieve the efficiency bound. \cite{tu2023unified} gave similar results using local linear kernel and generic machine learning. Notably, both kernel methods and sample splitting technique require more samples than regression we considered. In our scenario with relatively few samples within many strata, nonparametric estimation and machine learning methods might lack theoretical guarantees and exhibit poor empirical performance. In comparison, our weighted regression adjustment offers a feasible and acceptable method to further improve efficiency.
\end{remark}

\section{Variance Estimator}
For variance estimation, we introduce Assumption (A1'') and (A3') with raw moments which are similar but stronger than  Assumption~(A1') and (A3) with central moments, respectively. 
\newline

(A1'') For $a\in\{0,1\}$,
    \[\sup_{n\ge1,s\in\mathcal S_n} E\{Y_i^4(a)\mid S_{ni}=s\}<\infty.\]

(A3') For all coordinate $k$,
    \[\sup_{n\ge1,s\in\mathcal S_n} E(X_{ik}^4\mid S_{ni}=s)<\infty.\]

We present three non-parametric variance estimators and demonstrate their consistency.
\begin{theorem}
\label{thm:est}
If Assumptions~ (A1''), (B1), (B2), (B3') and (C1) hold, we have
$$
\hat V_{nB}(Y) \xrightarrow{P}  V_{B}(Y)
$$
and
$$\hat V_{nW}(Y,a)\xrightarrow{P} V_{W}(Y,a)
$$
for $a\in\{0,1\}$, where 
\begin{align*}
\hat V_{nB}(Y)&=\sum_{s\in\mathcal S_n} \frac{n(s)}{n}\sum_{a\in\{0,1\}}\{\hat \mu_2(Y,a,s) - \hat\sigma^2(Y,a,s)\}-2\sum_{s\in\mathcal S_n}\frac{n(s)}{n}\left\{\hat\mu(Y,0,s)\hat\mu(Y,1,s)\right\} - \hat\tau^2
\end{align*}
and
$$
\hat V_{nW}(Y,a)=\sum_{s\in\mathcal S_n}\frac{n(s)}{n}\left\{\frac{n(s)}{n_a(s)}\hat\sigma^2(Y,a,s)\right\}.
$$
\end{theorem}

\begin{theorem}
\label{thm:est2}
If Assumptions~(A1''), (A3'), (B1), (B2), (B3') and (C2) hold,  we have
$$
\hat V_{nB}(\hat r)\xrightarrow{P} V_{B}(r)
$$
and
$$\hat V_{nW}(\hat{r},a)\xrightarrow{P} V_W(r,a)
$$
for $a\in\{0,1\}$, where 

\begin{align*}
\hat V_{nB}(\hat r)&=\sum_{s\in\mathcal S_n} \frac{n(s)}{n}\sum_{a\in\{0,1\}}\{\hat \mu_2(\hat r,a,s) - \hat\sigma^2(\hat r,a,s)\}-2\sum_{s\in\mathcal S_n}\frac{n(s)}{n}\left\{\hat\mu(\hat r,0,s)\hat\mu(\hat r,1,s)\right\} - \hat\tau_{adj}^2,
\end{align*}
and
\begin{align*}
\hat V_{nW}(\hat r,a)=\sum_{s\in\mathcal S_n}\frac{n(s)}{n}\left\{\frac{n(s)}{n_a(s)}\hat\sigma^2(\hat r,a,s)\right\}.
\end{align*}

\end{theorem}

\begin{theorem}
\label{thm:est3}
If Assumptions~(A1''), (A3'), (B1), (B2), (B3') and (C3) hold,  we have
$$
\hat V_{nB}(\hat r^*)\xrightarrow{P} V_{B}(r^*)
$$
and
$$\hat V_{nW}(\hat r^*,a)\xrightarrow{P} V_W(r^*,a)
$$
for $a\in\{0,1\}$, where 

\begin{align*}
\hat V_{nB}(\hat r^*)&=\sum_{s\in\mathcal S_n} \frac{n(s)}{n}\sum_{a\in\{0,1\}}\{\hat \mu_2(\hat r^*,a,s) - \hat\sigma^2(\hat r^*,a,s)\}-2\sum_{s\in\mathcal S_n}\frac{n(s)}{n}\left\{\hat\mu(\hat r^*,0,s)\hat\mu(\hat r^*,1,s)\right\} - \hat\tau_{adj}^{*2},
\end{align*}
and
\begin{align*}
\hat V_{nW}(\hat r^*,a)=\sum_{s\in\mathcal S_n}\frac{n(s)}{n}\left\{\frac{n(s)}{n_a(s)}\hat\sigma^2(\hat r^*,a,s)\right\}.
\end{align*}

\end{theorem}
\begin{remark}
    The new variance estimators perform a delicate adjustment for degrees of freedom unlike those presented in \cite{ma2022regression}. In the Appendix, we compute the exact difference between our new variance estimator $\hat V_{\hat\tau}=\hat V_{nW}(Y,0)+\hat V_{nW}(Y,1)+\hat V_{nB}(Y)$ and that proposed in \cite{ma2022regression}, revealing that our new variance estimator always yields larger values. Simulation results also show that the old variance estimators often underestimate the true variance, while our proposed ones have coverage probabilities closer to the pre-specified values. However, given the stronger assumption that
$\inf_{s\in\mathcal S_n}n(s)\xrightarrow{P}\infty,$
the difference tends to zero in probability. In practice, we recommend new estimators to avoid underestimating true variances.
The asymptotic assumption that $\inf_{s\in\mathcal S_n}n(s)\xrightarrow{P}\infty$ is motivated by theoretical considerations rather than practical necessity. As demonstrated in Section 8, the simulation results show good numerical performance regardless of the number of units within each stratum, whether large or small. 
\end{remark}

\section{Extension to an Extremely Large Number Strata}
To estimate the mean or variance within strata, we require at least one or two sample(s) in each treatment arm within each stratum, as Assumption (B3) or (B3'), respectively. In most clinical trials, the number of strata is relatively small compared to the total number of units, thus satisfying both assumptions. 

We have extended the fixed number of strata setting to the diverging number of strata setting and established the asymptotic results in the previous sections. In this section, we will further consider the extreme scenario where there are so many strata that the proportion of strata with fewer than four samples is non-negligible. We provide two practical algorithms to handle it. One way is to discard strata without sufficient samples to estimate and only retain complete strata as described in Algorithm~\ref{alg:complete}. Another more elegant approach is to impute with estimates of other strata, as detailed in Algorithm~\ref{alg:impute}. More specifically, Algorithm~\ref{alg:impute} imputes with linear combination of other estimates in the same cluster. (Similar samples are more likely to be within the same cluster.)

\begin{remark}
    Algorithms~\ref{alg:complete} and~\ref{alg:impute} are extensions of the methods outlined in Sections 4 and 5, respectively. If both treatment arms contain at least 2 samples in every stratum, Algorithms 1 and 2 will reduce to the results previously established.
    Notably, Algorithm 2 offers flexibility in selecting $C$ and $w_a$. In particular, setting $w_a(s)=n_a(s)/n_a$ results in the imputed estimate being equivalent to the combined estimate of the entire cluster.
\end{remark}

\begin{algorithm}
\caption{Complete case}
\label{alg:complete}
\begin{algorithmic}[1]
\State  Input assignment vector $A^{(n)}\in\mathbb R^n$, stratum vector $S^{(n)}\in\mathbb R^n$, observed outcome vector $v^{(n)}\in\mathbb R^n$.
\Comment{Assume $S^{(n)}$ takes values from $1$ to $|\mathcal{S}_n|$.}
\State Initialize observed stratum indices $\mathrm{IND}_\mathrm{est}$ and $\mathrm{IND}_\mathrm{se}$ as vectors in $\mathbb R^{|\mathcal S_n|}$. 
\For{$s=1$ to $|\mathcal S_n|$}
    \State $n_0(s)=\sum_{i=1}^n I\{A_{ni}=0,S_{ni}=s\}$.\Comment{$I\{\cdot\}$ represents the indicator function.}
    \State $n_1(s)=\sum_{i=1}^n I\{A_{ni}=1,S_{ni}=s\}$.
    \If{$n_0(s)<2$ || $n_1(s)<2$}
        \State $\mathrm{IND}_\mathrm{se}[s]=0$
    \Else
        \State 
        $\mathrm{IND}_\mathrm{se}[s]=1$
    \EndIf

    \If{$n_0(s)<1$ || $n_1(s)<1$}
        \State $\mathrm{IND}_\mathrm{est}[s]=0$
    \Else
        \State 
        $\mathrm{IND}_\mathrm{est}[s]=1$
    \EndIf
\EndFor
\State Estimate effect only with strata with $\mathrm{IND}_\mathrm{est}[s]=1$.
\State Estimate variance only with strata with $\mathrm{IND}_\mathrm{se}[s]=1$.
\end{algorithmic}
\end{algorithm}

\begin{algorithm}
\caption{Imputation case}
\label{alg:impute}
\begin{algorithmic}[1]
\State  Input assignment vector $A^{(n)}\in\mathbb R^n$, stratum vector $S^{(n)}\in\mathbb R^n$, observed outcome vector $v^{(n)}\in\mathbb R^n$, cluster-stratum vector $C\in\mathbb R^{|\mathcal S_n|}$,stratum weight vector $w_a\in\mathbb R^{|\mathcal S_n|}$.\Comment{Assume $S^{(n)}$ takes values from $1$ to $|\mathcal{S}_n|$. $C[s]$ is the cluster of $s$th stratum. Set $w_a[s]=n(s)/n$ as default.}
\State For $a\in\{0,1\}$, initialize observed stratum indices $\mathrm{IND}_\mathrm{est,a}$ and $\mathrm{IND}_\mathrm{se,a}$ as vectors in $\mathbb R^{|\mathcal S_n|}$.
\For{$s=1$ to $|\mathcal S_n|$}
\For{$a=0$ to $1$}
    \State $n_a(s)=\sum_{i=1}^n I\{A_{ni}=a,S_{ni}=s\}$.\Comment{$I\{\cdot\}$ represents the indicator function.}
    \If{$n_a(s)<2$}
        \State
        $\mathrm{IND}_\mathrm{se,a}[s]=0$
        \State $\hat \sigma^2(v,a,s)=0$
    \Else
        \State 
        $\mathrm{IND}_\mathrm{se,a}[s]=1$
        \State Calculate $\hat \sigma^2(v,a,s)$
    \EndIf
    \If{$n_a(s)<1$}
        \State $\mathrm{IND}_\mathrm{est,a}[s]=0$
        \State $\hat \mu(v,a,s)=\hat \mu_2(v,a,s)=0$
    \Else
        \State 
        $\mathrm{IND}_\mathrm{est,a}[s]=1$
        \State Calculate $\hat \mu(v,a,s)$ and $\hat \mu_2(v,a,s)$
    \EndIf
    \EndFor
\EndFor

\For{$s=1$ to $|\mathcal S_n|$}
\For{$a=0$ to $1$}
\If{$\mathrm{IND}_\mathrm{se,a}[s]=0$}
\State $\hat \sigma^2(v,a,s)=\frac{\sum_{s'}I\{C[s]==C[s']\}w_a[s']\hat \sigma^2(v,a,s')}{\sum_{s'}I\{C[s]==C[s']\}w_a[s']}$
\EndIf
\If{$\mathrm{IND}_\mathrm{est,a}[s]=0$}
\State $\hat \mu(v,a,s)=\frac{\sum_{s'}I\{C[s]==C[s']\}w_a[s']\hat \mu(v,a,s')}{\sum_{s'}I\{C[s]==C[s']\}w_a[s']}$
\State $\hat \mu_2(v,a,s)=\frac{\sum_{s'}I\{C[s]==C[s']\}w_a[s']\hat \mu_2(v,a,s')}{\sum_{s'}I\{C[s]==C[s']\}w_a[s']}$
\EndIf
\EndFor
\EndFor
\State Estimate effect with all strata.

\State Estimate variance with all strata.
\end{algorithmic}
\end{algorithm}

\section{Numerical Study}

\subsection{Simulation with a diverging number of strata}

In this section, we will evaluate the empirical performance of our proposed three estimators, $\hat{\tau}$, $\hat{\tau}_{adj}$, and $\hat{\tau}^*_{adj}$, as well as two estimators from \cite{ma2022regression}—the versions without adjustment for degrees of freedom of $\hat{\tau}$ and $\hat{\tau}_{adj}$—as the number of strata diverges. We consider both linear and non-linear models.

For $a\in\{0,1\}$ and $1\leq i\leq n$, the potential outcomes are generated using the equation:
\[
Y_i(a)=\mu_a+g_a(B_i,X_i)+\sigma_a(B_i,X_i)\v_i(a),
\]
where $\mu_a$, $g_a(B_i,X_i)$, $\sigma_a(B_i,X_i)$, and $\v_i(a)$ are specified as follows. In each model, $\{B_i,X_i,\v_i(0),\v_i(1)\}_{i=1}^n$ are independent and identically distributed (i.i.d.), and $\v_i(a)$ follows a standard normal distribution.

Without loss of generality, we set $B_i=X_i$ in the following models. For each model, we consider three settings: $\lvert\mathcal{S}_n\rvert=25$, $n=500$; $\lvert\mathcal{S}_n\rvert=50$, $n=1500$; and $\lvert\mathcal{S}_n\rvert=1000$, $n=4000$. For each setting under Model 1 and 3 below, we consider two cases: a uniform assignment probability of $0.5$, and varying assignment probabilities in a uniform grid $[0.2,0.8]$. For each setting under Model 2 below, we consider two cases: a uniform assignment probability of $0.5$, and half of strata with assignment probability of $0.2$ and half of strata with assignment probability of $0.8$.

Model 1: $\boldsymbol{X}_i$ is a five-dimensional vector,
$$
g_0\left(\boldsymbol{X}_i\right)=g_1\left(\boldsymbol{X}_i\right)=\sum_{j=1}^5 \beta_j X_{i j},
$$
where $X_{i 1} \sim \operatorname{Beta}(2,2), X_{i 2}$ takes values in $\{1,2\}$ with equal probability, $X_{i 3} \sim$ Unif $[-2,3], X_{i 4}$ takes values in $\{1,2,3,4,5\}$ with equal probability, $X_{i 5} \sim \mathcal{N}(0,1)$, and they are all independent of each other. $\sigma_0\left(\boldsymbol{X}_i\right)=1, \sigma_1\left(\boldsymbol{X}_i\right)=2, \beta=(2,8,10,3,6)^{\mathrm{T}}.$ 
$X_{i 2},X_{i 3}$ and $X_{i 4}$ are used for randomization, and $X_{i 1}$ and $X_{i 3}$ are used as the additional covariates.

Setting 1: $|\mathcal S_n|=25,n=500$. $S_{ni}=(\lceil X_{i3}+2\rceil,X_{i4})$.

Setting 2: $|\mathcal S_n|=50,n=1500$. $S_{ni}=(X_{i2},\lceil X_{i3}+2\rceil,X_{i4})$.

Setting 3: $|\mathcal S_n|=100,n=4000$. $S_{ni}=(X_{i2},2\lceil X_{i3}+2\rceil,X_{i4})$.
\newline

Model 2: $\boldsymbol{X}_i$ is a five-dimensional vector,
$$
\begin{aligned}
& g_0\left(X_i\right)=\sum_{j=1}^5(\beta_{0j}I\{S_{ni}\text{ is odd}\}+\beta_{1j}I\{S_{ni}\text{ is even}\})X_{ij}, \\
& g_1\left(X_i\right)=\sum_{j=1}^5\beta_{1j}X_{ij},
\end{aligned}
$$
where $X_{i 1} \sim 5\cdot\mathrm{Beta}(2,2), X_{i 2}$ takes values in $\{1,2,3,4,5\}$ with equal probability, $X_{i 3} \sim \mathrm{Unif}[-2,3]$, $X_{i4}$ takes values in $\{1,2,3,4,5\}$ with equal probability, $X_{i 5} \sim \mathcal N(0,5)$, and they are all independent of each other. $\sigma_0\left(\boldsymbol{X}_i\right)=1, \sigma_1\left(\boldsymbol{X}_i\right)=1, \beta_0=(5,4,3,2,1)^{\mathrm{T}}, \beta_1=(1,2,3,4,5)^{\mathrm{T}}$.  $X_{i 1},X_{i 2}$ and $X_{i 4}$ are used for randomization, and $X_{i 1}$, $X_{i 3}$ and $X_{i 5}$ are used as the additional covariates. 
\begin{align*}
    \pi_{n1}(s)= \begin{cases}0.2, & \text { if $s$ is odd,} \\ 0.8, & \text { if $s$ is even.}\end{cases}
\end{align*}

Setting 1: $|\mathcal S_n|=25,n=500$. $S_{ni}=X_{i2}+5\times(X_{i4}-1)$.

Setting 2: $|\mathcal S_n|=50,n=1500$.  $X_{i 1 S}$ is the stratified variable of $X_{i 1}$, taking values $\{1,2\}$ determined by its relative value of $2.5$.  $S_{ni}=X_{i1S}+2\times(X_{i2}-1)+2\times5\times(X_{i4}-1)$.

Setting 3: $|\mathcal S_n|=100,n=4000$.  $X_{i 1 S}$ is the stratified variable of $X_{i 1}$, taking values $\{1,2,3,4\}$ determined by its relative value of first, second and third quartiles of $5\cdot\mathrm{Beta}(2,2)$. $S_{ni}=X_{i1S}+4\times(X_{i2}-1)+4\times5\times(X_{i4}-1)$.
\newline

Model 3: $\boldsymbol{X}_i$ is a four-dimensional vector,
$$
g_0\left(\boldsymbol{X}_i\right)=\sum_{j=1}^4 \beta_j X_{i j}, \quad g_1\left(\boldsymbol{X}_i\right)=\beta_1 \log \left(X_{i 1}\right) X_{i 4},
$$
where $X_{i 1} \sim \operatorname{Beta}(3,4), X_{i 2} \sim \operatorname{Unif}[-2,2], X_{i 3}=X_{i 1} X_{i 2}, X_{i 4}$ takes values in $\{1,2,3,4,5\}$ with equal probability. $X_{i 1}, X_{i 2}, X_{i 4}$ are independent of each other. $\sigma_0\left(\boldsymbol{X}_i\right)=X_{i 3 S}, \sigma_1\left(\boldsymbol{X}_i\right)=2 X_{i 2 S}$, where $X_{i 2 S}=\lceil1.25(X_{i2}+2)\rceil$ and $X_{i 3 S}=I\{X_3>0\}+1$, and $\beta=(20,7,5,6)^{\mathrm{T}}$.  $X_{i1},X_{i 2}$ and $X_{i 4}$ are used for randomization, and $X_{i 1}$ and $X_{i 3}$ are used as the additional covariates.

Setting 1: $|\mathcal S_n|=25,n=500$. 
$S_{ni}=(X_{i2S},X_{i3S})$.

Setting 2: $|\mathcal S_n|=50,n=1500$.  $X_{i 1 S}$ is the stratified variable of $X_{i 1}$, determined by its relative value of $0.5$. $S_{ni}=(X_{i1S},X_{i2S},X_{i3S})$.

Setting 3: $|\mathcal S_n|=100,n=4000$. $X_{i 1 S}=\lceil 4X_{i1}\rceil$. $S_{ni}=(X_{i1S},X_{i2S},X_{i3S})$.
\newline

We present simulation results for five estimators under three different randomization schemes: simple randomization, minimization, and stratified block randomization as shown in Tables 1, 3, and 5 (with the same assignment probability) and Tables 2, 4, and 6 (with varying assignment probabilities). Minimization employed a biased-coin probability of $0.75$ and equal weight. The bias, standard deviation (SD), and root mean squared error (RMSE) of the treatment effect estimators, standard error (SE) estimators, and empirical coverage probabilities (CP) of 95\% confidence intervals were computed over $2000$ replications.

Tables 1, 3, and 5 present simulation results under the same assignment probability. Firstly, all estimators exhibit small finite-sample bias, with the standard deviation (SD) decreasing as the total sample size $n$ increases. Secondly, among the three randomization schemes, simple randomization yields the largest SD, while stratified block randomization, known for achieving strong balance, displays the least SD across most settings. Thirdly, adjusting for additional covariates enhances efficiency. Fourthly, the old variance estimators (SE) of old estimates in \cite{ma2022regression} typically underestimate SD, leading to lower SE than SD and hence lower coverage probabilities than $0.95$. The new variance estimators (SE) are slightly larger, a scenario more common under simple randomization compared to stratified block randomization and minimization. Fifthly, under same assignment probability, there is no significant difference between $\hat\tau_{adj}$ and $\hat\tau_{adj}^*$, which corresponds with our theory (See Remark \ref{rmk:wt reg}).

Tables 2, 4, and 6 present simulation results under varying assignment probabilities, exhibiting behaviors similar to those under the same assignment probability. Only if with varying assignment probability, $\hat\tau_{adj}^*$ could have both lower SD and SE than $\hat\tau_{adj}$. Particularly, see Table 4. We always recommend $\hat\tau_{adj}^*$ for regression adjustment.
\begin{table}
    \centering
    \caption{Simulation results for Model 1 under same assignment probability ($\pi=0.5$).}
    \begin{adjustbox}{width=1\textwidth}
    \begin{threeparttable}
    \begin{tabular}{lllcccccccccccccccc}
    \toprule
    & & & \multicolumn{5}{c}{SR} & 
    \multicolumn{5}{c}{MIN} & \multicolumn{5}{c}{SBR} &
    \\ \cmidrule(lr){4-8}\cmidrule(lr){9-13}\cmidrule(lr){14-18}
    Setting & Est & D.F. & \multicolumn{1}{c}{Bias} & \multicolumn{1}{c}{SD} & \multicolumn{1}{c}{RMSE} & \multicolumn{1}{c}{SE} & \multicolumn{1}{c}{CP} & \multicolumn{1}{c}{Bias} & \multicolumn{1}{c}{SD} & \multicolumn{1}{c}{RMSE} & \multicolumn{1}{c}{SE} & \multicolumn{1}{c}{CP} & \multicolumn{1}{c}{Bias} & \multicolumn{1}{c}{SD} & \multicolumn{1}{c}{RMSE} & \multicolumn{1}{c}{SE} & \multicolumn{1}{c}{CP} \\ \midrule
    $|\mathcal S_n|=25,n=500$ & $\hat\tau$ & w/o & 0.01 & 0.73 & 0.73 & 0.71 & 0.94 & 0.01 & 0.72 & 0.72 & 0.70 & 0.94 & 0.00 & 0.70 & 0.70 & 0.69 & 0.94 \\
    &  & w & 0.01 & 0.73 & 0.73 & 0.73 & 0.94 & 0.01 & 0.72 & 0.72 & 0.72 & 0.95 & 0.00 & 0.70 & 0.70 & 0.71 & 0.95 \\
    & $\hat\tau_{adj}$ & w/o & 0.01 & 0.68 & 0.68 & 0.66 & 0.94 & 0.02 & 0.67 & 0.67 & 0.65 & 0.94 & 0.00 & 0.65 & 0.65 & 0.64 & 0.95 \\
    &  & w & 0.01 & 0.68 & 0.68 & 0.68 & 0.95 & 0.02 & 0.67 & 0.67 & 0.67 & 0.95 & 0.00 & 0.65 & 0.65 & 0.66 & 0.96 \\
    & $\hat\tau_{adj}^*$ & w & 0.01 & 0.68 & 0.68 & 0.68 & 0.94 & 0.02 & 0.67 & 0.67 & 0.67 & 0.95 & 0.00 & 0.65 & 0.65 & 0.66 & 0.96 \\
    \midrule
    $|\mathcal S_n|=50,n=1500$ & $\hat\tau$ & w/o & -0.01 & 0.36 & 0.36 & 0.35 & 0.94 & -0.01 & 0.36 & 0.36 & 0.35 & 0.95 & 0.00 & 0.35 & 0.35 & 0.35 & 0.95 \\
    &  & w & -0.01 & 0.36 & 0.36 & 0.36 & 0.95 & -0.01 & 0.36 & 0.36 & 0.36 & 0.95 & 0.00 & 0.35 & 0.35 & 0.35 & 0.95 \\
    & $\hat\tau_{adj}$ & w/o & -0.01 & 0.33 & 0.33 & 0.32 & 0.95 & -0.01 & 0.32 & 0.32 & 0.32 & 0.94 & 0.00 & 0.32 & 0.32 & 0.31 & 0.94 \\
    &  & w & -0.01 & 0.33 & 0.33 & 0.33 & 0.95 & -0.01 & 0.32 & 0.32 & 0.32 & 0.95 & 0.00 & 0.32 & 0.32 & 0.32 & 0.95 \\
    & $\hat\tau_{adj}^*$ & w & -0.01 & 0.33 & 0.33 & 0.33 & 0.95 & -0.01 & 0.32 & 0.32 & 0.32 & 0.95 & 0.00 & 0.32 & 0.32 & 0.32 & 0.95 \\ \midrule
    $|\mathcal S_n|=100,n=4000$ & $\hat\tau$ & w/o & 0.00 & 0.21 & 0.21 & 0.2 & 0.94 & 0.00 & 0.21 & 0.21 & 0.20 & 0.94 & 0.00 & 0.21 & 0.21 & 0.20 & 0.93 \\
    &  & w & 0.00 & 0.21 & 0.21 & 0.20 & 0.95 & 0.00 & 0.21 & 0.21 & 0.20 & 0.95 & 0.00 & 0.21 & 0.21 & 0.20 & 0.93 \\
    & $\hat\tau_{adj}$ & w/o & -0.01 & 0.20 & 0.20 & 0.20 & 0.95 & 0.00 & 0.20 & 0.20 & 0.20 & 0.94 & 0.00 & 0.20 & 0.20 & 0.19 & 0.94 \\
    &  & w & -0.01 & 0.20 & 0.20 & 0.20 & 0.95 & 0.00 & 0.20 & 0.20 & 0.2 & 0.95 & 0.00 & 0.20 & 0.20 & 0.20 & 0.94 \\
    & $\hat\tau_{adj}^*$ & w & -0.01 & 0.20 & 0.20 & 0.20 & 0.95 & 0.00 & 0.20 & 0.20 & 0.20 & 0.95 & 0.00 & 0.20 & 0.20 & 0.20 & 0.94 \\ \bottomrule
    \end{tabular}
    \begin{tablenotes}
    \item SR, (stratified) simple randomization; SBR, stratified block randomization; MIN, minimization; CP, coverage probability; RMSE, root mean squared error; SD, standard deviation; SE, standard error; Est, estimator. D.F. w/o: without adjustment for degrees of freedom; D.F. w: with adjustment for degrees of freedom.
    \end{tablenotes}
    \end{threeparttable}
    \end{adjustbox}
\end{table}

\begin{table}
    \centering
    \caption{Simulation results for Model 1 under varying assignment probability ($\pi\in[0.2,0.8]$).}
    \begin{adjustbox}{width=1\textwidth}
    \begin{threeparttable}
    \begin{tabular}{lllccccccccccccccc}
    \toprule
    & & & \multicolumn{5}{c}{SR} & \multicolumn{5}{c}{SBR} &
    \\ \cmidrule(lr){4-8}\cmidrule(lr){9-13}
    Setting & Est & D.F. & \multicolumn{1}{c}{Bias} & \multicolumn{1}{c}{SD} & \multicolumn{1}{c}{RMSE} & \multicolumn{1}{c}{SE} & \multicolumn{1}{c}{CP} & \multicolumn{1}{c}{Bias} & \multicolumn{1}{c}{SD} & \multicolumn{1}{c}{RMSE} & \multicolumn{1}{c}{SE} & \multicolumn{1}{c}{CP} \\ \midrule
    $|\mathcal S_n|=25,n=500$ & $\hat\tau$ & w/o & 0.06 & 0.84 & 0.84 & 0.76 & 0.92 & 0.01 & 0.77 & 0.77 & 0.73 & 0.93 \\
 &  & w & 0.06 & 0.84 & 0.84 & 0.81 & 0.94 & 0.01 & 0.77 & 0.77 & 0.77 & 0.95 \\
 & $\hat\tau_{adj}$ & w/o & 0.03 & 0.76 & 0.76 & 0.71 & 0.93 & 0.01 & 0.71 & 0.71 & 0.68 & 0.94 \\
 &  & w & 0.03 & 0.77 & 0.77 & 0.75 & 0.95 & 0.01 & 0.71 & 0.71 & 0.71 & 0.95 \\
 & $\hat\tau_{adj}^*$ & w & 0.02 & 0.77 & 0.77 & 0.75 & 0.94 & 0.01 & 0.72 & 0.72 & 0.71 & 0.95 \\
 \midrule
$|\mathcal S_n|=50,n=1500$ & $\hat\tau$ & w/o & 0.00 & 0.40 & 0.40 & 0.38 & 0.94 & 0.00 & 0.39 & 0.39 & 0.37 & 0.94 \\
 &  & w & 0.00 & 0.40 & 0.40 & 0.40 & 0.95 & 0.00 & 0.39 & 0.39 & 0.38 & 0.94 \\
 & $\hat\tau_{adj}$ & w/o & -0.01 & 0.36 & 0.36 & 0.34 & 0.94 & -0.01 & 0.36 & 0.36 & 0.33 & 0.94 \\
 &  & w & -0.01 & 0.36 & 0.36 & 0.36 & 0.95 & -0.01 & 0.36 & 0.36 & 0.34 & 0.95 \\
 & $\hat\tau_{adj}^*$ & w & -0.01 & 0.36 & 0.36 & 0.36 & 0.95 & -0.01 & 0.36 & 0.36 & 0.34 & 0.95 \\
 \midrule
$|\mathcal S_n|=100,n=4000$ & $\hat\tau$ & w/o & 0.00 & 0.23 & 0.23 & 0.22 & 0.94 & 0.00 & 0.22 & 0.22 & 0.21 & 0.95 \\
 &  & w & 0.00 & 0.23 & 0.23 & 0.22 & 0.94 & 0.00 & 0.22 & 0.22 & 0.22 & 0.95 \\
 & $\hat\tau_{adj}$ & w/o & 0.00 & 0.22 & 0.22 & 0.21 & 0.94 & 0.00 & 0.21 & 0.21 & 0.21 & 0.94 \\
 &  & w & 0.00 & 0.22 & 0.22 & 0.22 & 0.95 & 0.00 & 0.21 & 0.21 & 0.21 & 0.95 \\
 & $\hat\tau_{adj}^*$ & w & 0.00 & 0.22 & 0.22 & 0.22 & 0.95 & 0.00 & 0.21 & 0.21 & 0.21 & 0.95 \\ 
 \bottomrule
    \end{tabular}
    \begin{tablenotes}
    \item SR, (stratified) simple randomization; SBR, stratified block randomization; MIN, minimization; CP, coverage probability; RMSE, root mean squared error; SD, standard deviation; SE, standard error; Est, estimator. D.F. w/o: without adjustment for degrees of freedom; D.F. w: with adjustment for degrees of freedom.
    \end{tablenotes}
    \end{threeparttable}
    \end{adjustbox}
\end{table}

\begin{table}
    \centering
    \caption{Simulation results for Model 2 under same assignment probability ($\pi=0.5$).}
    \begin{adjustbox}{width=1\textwidth}
    \begin{threeparttable}
    \begin{tabular}{lllcccccccccccccccc}
    \toprule
    & & & \multicolumn{5}{c}{SR} & 
    \multicolumn{5}{c}{MIN} & \multicolumn{5}{c}{SBR} &
    \\ \cmidrule(lr){4-8}\cmidrule(lr){9-13}\cmidrule(lr){14-18}
    Setting & Est & D.F. & \multicolumn{1}{c}{Bias} & \multicolumn{1}{c}{SD} & \multicolumn{1}{c}{RMSE} & \multicolumn{1}{c}{SE} & \multicolumn{1}{c}{CP} & \multicolumn{1}{c}{Bias} & \multicolumn{1}{c}{SD} & \multicolumn{1}{c}{RMSE} & \multicolumn{1}{c}{SE} & \multicolumn{1}{c}{CP} & \multicolumn{1}{c}{Bias} & \multicolumn{1}{c}{SD} & \multicolumn{1}{c}{RMSE} & \multicolumn{1}{c}{SE} & \multicolumn{1}{c}{CP} \\ \midrule
    $|\mathcal S_n|=25,n=500$ & $\hat\tau$ & w/o & -0.03 & 2.05 & 2.05 & 2.00 & 0.95 & 0.08 & 2.09 & 2.09 & 1.98 & 0.94 & 0.08 & 2.03 & 2.03 & 1.97 & 0.94 \\
    &  & w & -0.03 & 2.05 & 2.05 & 2.07 & 0.95 & 0.08 & 2.09 & 2.09 & 2.05 & 0.94 & 0.08 & 2.03 & 2.03 & 2.02 & 0.95 \\
    & $\hat\tau_{adj}$ & w/o & 0.00 & 0.87 & 0.87 & 0.84 & 0.94 & 0.01 & 0.88 & 0.88 & 0.84 & 0.93 & -0.01 & 0.85 & 0.85 & 0.82 & 0.94 \\
    &  & w & 0.00 & 0.87 & 0.87 & 0.87 & 0.95 & 0.01 & 0.88 & 0.88 & 0.86 & 0.94 & -0.01 & 0.85 & 0.85 & 0.84 & 0.95 \\
    & $\hat\tau_{adj}^*$ & w & 0.00 & 0.87 & 0.87 & 0.86 & 0.95 & 0.01 & 0.88 & 0.88 & 0.86 & 0.94 & -0.02 & 0.86 & 0.86 & 0.84 & 0.95 \\
    \midrule
    $|\mathcal S_n|=50,n=1500$ & $\hat\tau$ & w/o & 0.04 & 1.19 & 1.19 & 1.15 & 0.94 & 0.04 & 1.14 & 1.14 & 1.15 & 0.95 & -0.02 & 1.17 & 1.17 & 1.14 & 0.94 \\
    & & w & 0.04 & 1.19 & 1.19 & 1.18 & 0.94 & 0.04 & 1.14 & 1.14 & 1.17 & 0.96 & -0.02 & 1.17 & 1.17 & 1.16 & 0.95 \\
    & $\hat\tau_{adj}$ & w/o &  0.00 & 0.48 & 0.48 & 0.47 & 0.94 & -0.01 & 0.48 & 0.47 & 0.46 & 0.94 & 0.00 & 0.46 & 0.46 & 0.46 & 0.95 \\
    &  & w & 0.00 & 0.48 & 0.48 & 0.47 & 0.95 & -0.01 & 0.48 & 0.48 & 0.47 & 0.94 & 0.00 & 0.46 & 0.46 & 0.46 & 0.95 \\
    & $\hat\tau_{adj}^*$ & w & 0.00 & 0.48 & 0.48 & 0.47 & 0.95 & -0.01 & 0.48 & 0.48 & 0.47 & 0.94 & 0.00 & 0.47 & 0.47 & 0.46 & 0.95 \\
    \midrule
    $|\mathcal S_n|=100,n=4000$ & $\hat\tau$ & w/o & 0.02 & 0.74 & 0.74 & 0.71 & 0.94 & 0.01 & 0.72 & 0.72 & 0.71 & 0.95 & 0.01 & 0.69 & 0.69 & 0.70 & 0.95 \\
    &  & w & 0.02 & 0.74 & 0.74 & 0.72 & 0.94 & 0.01 & 0.72 & 0.72 & 0.72 & 0.95 & 0.01 & 0.69 & 0.69 & 0.71 & 0.96 \\
    & $\hat\tau_{adj}$ & w/o & 0.00 & 0.29 & 0.29 & 0.29 & 0.95 & 0.00 & 0.29 & 0.29 & 0.29 & 0.94 & 0.00 & 0.29 & 0.29 & 0.29 & 0.95 \\
    & & w & 0.00 & 0.29 & 0.29 & 0.29 & 0.95 & 0.00 & 0.29 & 0.29 & 0.29 & 0.95 & 0.00 & 0.29 & 0.29 & 0.29 & 0.96 \\
    & $\hat\tau_{adj}^*$ & w & 0.00 & 0.29 & 0.29 & 0.29 & 0.95 & 0.00 & 0.29 & 0.29 & 0.29 & 0.95 & 0.00 & 0.29 & 0.29 & 0.29 & 0.96 \\
    \bottomrule
    \end{tabular}
    \begin{tablenotes}
    \item SR, (stratified) simple randomization; SBR, stratified block randomization; MIN, minimization; CP, coverage probability; RMSE, root mean squared error; SD, standard deviation; SE, standard error; Est, estimator. D.F. w/o: without adjustment for degrees of freedom; D.F. w: with adjustment for degrees of freedom.
    \end{tablenotes}
    \end{threeparttable}
    \end{adjustbox}
\end{table}

\begin{table}
    \centering
    \caption{Simulation results for Model 2 under varying assignment probability ($\pi\in\{0.2,0.8\}$).}
    \begin{adjustbox}{width=1\textwidth}
    \begin{threeparttable}
    \begin{tabular}{lllccccccccccccccc}
    \toprule
    & & & \multicolumn{5}{c}{SR} & \multicolumn{5}{c}{SBR} &
    \\ \cmidrule(lr){4-8}\cmidrule(lr){9-13}
    Setting & Est & D.F. & \multicolumn{1}{c}{Bias} & \multicolumn{1}{c}{SD} & \multicolumn{1}{c}{RMSE} & \multicolumn{1}{c}{SE} & \multicolumn{1}{c}{CP} & \multicolumn{1}{c}{Bias} & \multicolumn{1}{c}{SD} & \multicolumn{1}{c}{RMSE} & \multicolumn{1}{c}{SE} & \multicolumn{1}{c}{CP} \\ \midrule
    $|\mathcal S_n|=25,n=500$ & $\hat\tau$ & w/o & 0.00 & 3.00 & 3.00 & 2.57 & 0.90 & -0.08 & 2.76 & 2.76 & 2.45 & 0.91 \\
     &  & w & 0.00 & 3.00 & 3.00 & 2.89 & 0.94 & -0.08 & 2.76 & 2.76 & 2.70 & 0.94 \\
     & $\hat\tau_{adj}$ & w/o & -0.01 & 1.20 & 1.20 & 1.06 & 0.92 & -0.03 & 1.07 & 1.07 & 0.97 & 0.92 \\
     &  & w & -0.01 & 1.00 & 1.00 & 0.98 & 0.95 & -0.02 & 0.88 & 0.88 & 0.87 & 0.95 \\
     & $\hat\tau_{adj}^*$ & w & -0.01 & 0.91 & 0.91 & 0.93 & 0.95 & 0.00 & 0.81 & 0.81 & 0.80 & 0.94 \\ \midrule
    $|\mathcal S_n|=50,n=1500$ & $\hat\tau$ & w/o & -0.02 & 1.70 & 1.70 & 1.53 & 0.92 & -0.03 & 1.58 & 1.58 & 1.47 & 0.94 \\
     &  & w & -0.02 & 1.70 & 1.70 & 1.68 & 0.95 & -0.03 & 1.58 & 1.58 & 1.57 & 0.95 \\
     & $\hat\tau_{adj}$ & w/o & -0.01 & 0.62 & 0.62 & 0.57 & 0.92 & -0.01 & 0.58 & 0.58 & 0.54 & 0.93 \\
     &  & w & -0.01 & 0.49 & 0.49 & 0.49 & 0.94 & -0.01 & 0.46 & 0.46 & 0.45 & 0.95 \\
     & $\hat\tau_{adj}^*$ & w & -0.01 & 0.42 & 0.42 & 0.44 & 0.95 & 0.00 & 0.41 & 0.41 & 0.41 & 0.95 \\ \midrule
    $|\mathcal S_n|=100,n=4000$ & $\hat\tau$ & w/o & 0.05 & 1.02 & 1.02 & 0.95 & 0.94 & 0.01 & 0.97 & 0.97 & 0.92 & 0.94 \\
     &  & w & 0.05 & 1.02 & 1.02 & 1.01 & 0.95 & 0.01 & 0.97 & 0.97 & 0.96 & 0.95 \\
     & $\hat\tau_{adj}$ & w/o & 0.01 & 0.37 & 0.37 & 0.35 & 0.94 & 0.00 & 0.35 & 0.35 & 0.34 & 0.94 \\
     &  & w & 0.01 & 0.29 & 0.29 & 0.29 & 0.96 & 0.00 & 0.28 & 0.28 & 0.28 & 0.96 \\
     & $\hat\tau_{adj}^*$ & w & 0.00 & 0.26 & 0.26 & 0.26 & 0.96 & 0.00 & 0.25 & 0.25 & 0.26 & 0.95\\
     \bottomrule
    \end{tabular}
    \begin{tablenotes}
    \item SR, (stratified) simple randomization; SBR, stratified block randomization; MIN, minimization; CP, coverage probability; RMSE, root mean squared error; SD, standard deviation; SE, standard error; Est, estimator. D.F. w/o: without adjustment for degrees of freedom; D.F. w: with adjustment for degrees of freedom.
    \end{tablenotes}
    \end{threeparttable}
    \end{adjustbox}

\end{table}

\begin{table}
    \centering
    \caption{Simulation results for Model 3 under same assignment probability ($\pi=0.5$).}
    \begin{adjustbox}{width=1\textwidth}
    \begin{threeparttable}
    \begin{tabular}{lllcccccccccccccccc}
    \toprule
    & & & \multicolumn{5}{c}{SR} & 
    \multicolumn{5}{c}{MIN} & \multicolumn{5}{c}{SBR} &
    \\ \cmidrule(lr){4-8}\cmidrule(lr){9-13}\cmidrule(lr){14-18}
    Setting & Est & D.F. & \multicolumn{1}{c}{Bias} & \multicolumn{1}{c}{SD} & \multicolumn{1}{c}{RMSE} & \multicolumn{1}{c}{SE} & \multicolumn{1}{c}{CP} & \multicolumn{1}{c}{Bias} & \multicolumn{1}{c}{SD} & \multicolumn{1}{c}{RMSE} & \multicolumn{1}{c}{SE} & \multicolumn{1}{c}{CP} & \multicolumn{1}{c}{Bias} & \multicolumn{1}{c}{SD} & \multicolumn{1}{c}{RMSE} & \multicolumn{1}{c}{SE} & \multicolumn{1}{c}{CP} \\ \midrule
    $|\mathcal S_n|=25,n=500$ & $\hat\tau$ & w/o & -0.03 & 3.06 & 3.06 & 2.97 & 0.94 & -0.01 & 3.07 & 3.06 & 2.95 & 0.94 & -0.06 & 3.01 & 3.01 & 2.95 & 0.94 \\
    &  & w & -0.03 & 3.06 & 3.06 & 3.04 & 0.95 & -0.01 & 3.07 & 3.06 & 3.02 & 0.94 & -0.06 & 3.01 & 3.01 & 3.01 & 0.95 \\
    & $\hat\tau_{adj}$ & w/o & 0.05 & 2.49 & 2.49 & 2.44 & 0.94 & 0.02 & 2.51 & 2.51 & 2.43 & 0.95 & 0.00 & 2.50 & 2.50 & 2.42 & 0.94 \\
    &  & w & 0.05 & 2.49 & 2.49 & 2.49 & 0.95 & 0.02 & 2.51 & 2.51 & 2.47 & 0.95 & 0.00 & 2.50 & 2.50 & 2.46 & 0.95 \\
    & $\hat\tau_{adj}^*$ & w & 0.05 & 2.50 & 2.50 & 2.48 & 0.94 & 0.01 & 2.52 & 2.52 & 2.47 & 0.95 & 0.00 & 2.52 & 2.51 & 2.46 & 0.95 \\ 
    \midrule
    $|\mathcal S_n|=50,n=1500$ & $\hat\tau$ & w/o & 0.03 & 1.48 & 1.48 & 1.49 & 0.95 & 0.00 & 1.48 & 1.48 & 1.49 & 0.96 & 0.00 & 1.51 & 1.51 & 1.49 & 0.96 \\
    &  & w & 0.03 & 1.48 & 1.48 & 1.51 & 0.95 & 0.00 & 1.48 & 1.48 & 1.51 & 0.96 & 0.00 & 1.51 & 1.51 & 1.50 & 0.96 \\
    & $\hat\tau_{adj}$ & w/o & 0.03 & 1.36 & 1.36 & 1.37 & 0.95 & 0.01 & 1.36 & 1.36 & 1.37 & 0.96 & 0.00 & 1.37 & 1.37 & 1.37 & 0.95 \\
    &  & w & 0.03 & 1.36 & 1.36 & 1.38 & 0.96 & 0.01 & 1.36 & 1.36 & 1.38 & 0.96 & 0.00 & 1.37 & 1.37 & 1.38 & 0.95 \\
    & $\hat\tau_{adj}^*$ & w & 0.03 & 1.36 & 1.36 & 1.38 & 0.95 & 0.01 & 1.36 & 1.36 & 1.38 & 0.96 & 0.00 & 1.37 & 1.37 & 1.38 & 0.95 \\
    \midrule
    $|\mathcal S_n|=100,n=4000$ & $\hat\tau$ & w/o & 0.00 & 0.32 & 0.32 & 0.32 & 0.96 & 0.00 & 0.32 & 0.32 & 0.32 & 0.95 & 0.00 & 0.32 & 0.32 & 0.32 & 0.95 \\
    &  & w & 0.00 & 0.32 & 0.32 & 0.32 & 0.96 & 0.00 & 0.32 & 0.32 & 0.32 & 0.96 & 0.00 & 0.32 & 0.32 & 0.32 & 0.95 \\
    & $\hat\tau_{adj}$ & w/o & 0.00 & 0.30 & 0.30 & 0.31 & 0.95 & 0.00 & 0.31 & 0.31 & 0.31 & 0.95 & 0.00 & 0.30 & 0.30 & 0.30 & 0.95 \\
    & & w & 0.00 & 0.30 & 0.30 & 0.31 & 0.96 & 0.00 & 0.31 & 0.31 & 0.31 & 0.95 & 0.00 & 0.30 & 0.30 & 0.31 & 0.95 \\
    & $\hat\tau_{adj}^*$ & w & 0.00 & 0.30 & 0.30 & 0.31 & 0.96 & 0.00 & 0.31 & 0.31 & 0.31 & 0.95 & 0.00 & 0.30 & 0.30 & 0.31 & 0.96 \\
    \bottomrule    
    \end{tabular}
    
    \begin{tablenotes}
    \item SR, (stratified) simple randomization; SBR, stratified block randomization; MIN, minimization; CP, coverage probability; RMSE, root mean squared error; SD, standard deviation; SE, standard error; Est, estimator. D.F. w/o: without adjustment for degrees of freedom; D.F. w: with adjustment for degrees of freedom.
    \end{tablenotes}
        
    \end{threeparttable}
    \end{adjustbox}

\end{table}

\begin{table}
    \centering
    \caption{Simulation results for Model 3 under varying assignment probability ($\pi\in[0.2,0.8]$).}
    \begin{adjustbox}{width=1\textwidth}
    \begin{threeparttable}
    \begin{tabular}{lllccccccccccccccc}
    \toprule
    & & & \multicolumn{5}{c}{SR} & \multicolumn{5}{c}{SBR} &
    \\ \cmidrule(lr){4-8}\cmidrule(lr){9-13}
    Setting & Est & D.F. & \multicolumn{1}{c}{Bias} & \multicolumn{1}{c}{SD} & \multicolumn{1}{c}{RMSE} & \multicolumn{1}{c}{SE} & \multicolumn{1}{c}{CP} & \multicolumn{1}{c}{Bias} & \multicolumn{1}{c}{SD} & \multicolumn{1}{c}{RMSE} & \multicolumn{1}{c}{SE} & \multicolumn{1}{c}{CP} \\ \midrule
    $|\mathcal S_n|=25,n=500$ & $\hat\tau$ & w/o & -0.02 & 2.84 & 2.83 & 2.85 & 0.95 & -0.03 & 2.86 & 2.86 & 2.80 & 0.94 \\
 &  & w & -0.02 & 2.84 & 2.83 & 2.91 & 0.96 & -0.03 & 2.86 & 2.86 & 2.85 & 0.95 \\
 & $\hat\tau_{adj}$ & w/o & 0.03 & 2.51 & 2.51 & 2.43 & 0.94 & 0.03 & 2.44 & 2.44 & 2.38 & 0.94 \\
 &  & w & 0.03 & 2.43 & 2.43 & 2.44 & 0.95 & 0.03 & 2.39 & 2.39 & 2.38 & 0.95 \\
 & $\hat\tau_{adj}^*$ & w & 0.01 & 2.40 & 2.40 & 2.42 & 0.95 & 0.02 & 2.38 & 2.38 & 2.36 & 0.95 \\
 \midrule
$|\mathcal S_n|=50,n=1500$ & $\hat\tau$ & w/o & -0.05 & 1.49 & 1.49 & 1.46 & 0.94 & 0.02 & 1.48 & 1.48 & 1.45 & 0.94 \\
 &  & w & -0.05 & 1.49 & 1.49 & 1.47 & 0.95 & 0.02 & 1.48 & 1.48 & 1.46 & 0.94 \\
 & $\hat\tau_{adj}$ & w/o & 0.01 & 1.40 & 1.40 & 1.36 & 0.94 & 0.02 & 1.36 & 1.36 & 1.35 & 0.95 \\
 &  & w & 0.00 & 1.39 & 1.39 & 1.36 & 0.95 & 0.02 & 1.36 & 1.36 & 1.35 & 0.95 \\
 & $\hat\tau_{adj}^*$ & w & -0.01 & 1.38 & 1.38 & 1.36 & 0.95 & 0.02 & 1.36 & 1.36 & 1.35 & 0.95 \\
 \midrule
$|\mathcal S_n|=100,n=4000$ & $\hat\tau$ & w/o & 0.00 & 0.35 & 0.35 & 0.33 & 0.94 & 0.00 & 0.34 & 0.34 & 0.33 & 0.94 \\
 &  & w & 0.00 & 0.35 & 0.35 & 0.33 & 0.94 & 0.00 & 0.34 & 0.34 & 0.33 & 0.94 \\
 & $\hat\tau_{adj}$ & w/o & 0.00 & 0.33 & 0.33 & 0.31 & 0.94 & 0.00 & 0.32 & 0.32 & 0.31 & 0.94 \\
 &  & w & 0.00 & 0.33 & 0.33 & 0.32 & 0.94 & 0.00 & 0.32 & 0.32 & 0.31 & 0.94 \\
 & $\hat\tau_{adj}^*$ & w & 0.00 & 0.33 & 0.33 & 0.32 & 0.94 & 0.00 & 0.32 & 0.32 & 0.31 & 0.95\\
 \bottomrule
    \end{tabular}
    \begin{tablenotes}
    \item SR, (stratified) simple randomization; SBR, stratified block randomization; MIN, minimization; CP, coverage probability; RMSE, root mean squared error; SD, standard deviation; SE, standard error; Est, estimator. D.F. w/o: without adjustment for degrees of freedom; D.F. w: with adjustment for degrees of freedom.
    \end{tablenotes}
    \end{threeparttable}
    \end{adjustbox}
\end{table}

\subsection{Simulation with an extremely large number of strata}
In this section, we will evaluate the empirical performance of two algorithms proposed in Section 7. We consider number of strata from $10$ to $100$ with fixed sample size $500$.

For $a\in\{0,1\}$ and $1\leq i\leq n$, the potential outcomes are generated using the equation:
\[
Y_i(a)=\mu_a+g_a(B_i,X_i)+\sigma_a(B_i,X_i)\v_i(a),
\]
where $\mu_a$, $g_a(B_i,X_i)$, $\sigma_a(B_i,X_i)$, and $\v_i(a)$ are specified as follows. $\{B_i,X_i,\v_i(0),\v_i(1)\}_{i=1}^n$ are independent and identically distributed (i.i.d.), and $\v_i(a)$ follows a standard normal distribution.

Model: $\boldsymbol{X}_i$ is a five-dimensional vector, for $a\in\{0,1\}$,
$$
g_a\left(\boldsymbol{X}_i\right)=\sum_{j=1}^5 \beta_{aj} X_{i j},
$$
where $X_{i 1} \sim \operatorname{Beta}(2,2), X_{i 2}$ takes values in $\{1,2\}$ with probabilities $\{0.7,0.3\}$, $X_{i 3} \sim$ Unif $[-2,3], X_{i 4}$ takes values in $\{1,2,3,4,5\}$ with equal probability, $X_{i 5} \sim \mathcal{N}(0,1)$, and they are all independent of each other. $\sigma_0\left(\boldsymbol{X}_i\right)=1, \sigma_1\left(\boldsymbol{X}_i\right)=2, \beta_0=(2,8,10,3,6)^{\mathrm{T}}$ and  $\beta_1=(6,3,10,8,2)^{\mathrm{T}}$.
$X_{i 2},X_{i 4}$ and $X_{i 5}$ are used for randomization, and $X_{i 1}$ and $X_{i 3}$ are used as the additional covariates. 

$X_{i5}$ is further discretized according to the uniform quantiles of the normal distribution as $X_{i5S}$. $X_{i5S}$ is considered as a site variable, and $|X_{i5S}|$ represents the number of sites.

The total sample size is $500$, with the total stratum number being $2\times5\times|X_{i 5S}|$, where $|X_{i 5S}|$ varies from $1$ to $10$. We conducted 1000 replicates. Figure 1 illustrates $\mathrm{Median}(n(s))$ and $\pr\{n(s)\ge4\}$. In Figure 1(a), the trend resembles a $x\sim1/x$ curve, while Figure 1(b) reveals that when the number of strata is 50, $\pr\{n(s)\ge4\}$ is approximately $0.92$, indicating that Assumption (B3') is violated with a non-ignorable probability.

Figure 2 presents the bias, standard deviation, and coverage probability. Concerning bias, the imputation case exhibits relatively small bias compared to all randomization schemes and estimators, while the complete case shows increasing bias as the number of strata increases. Stratified block randomization maintains ignorable bias since it achieves strong balance, but the other two randomization methods result in large bias. Adjusted estimators demonstrate greater stability compared to unadjusted estimators. Regarding standard deviation, unadjusted estimators exhibit no clear trend. However, the SDs of adjusted estimators initially decrease but then show an upward trend especially in complete case, indicating the benefit of stratification and the risk of over-stratification.

In terms of coverage probability, unadjusted estimators are more stable due to their larger SDs. The imputation case consistently maintains coverage probability near $0.95$, whereas the coverage probabilities under the complete case notably decrease as the number of strata increases, especially under minimization and simple randomization. This highlights the advantages of stratified block randomization and the imputation case.

\begin{figure}[htbp]
    \centering
    \begin{subfigure}{0.46\textwidth}
        \centering
        \includegraphics[width=\textwidth]{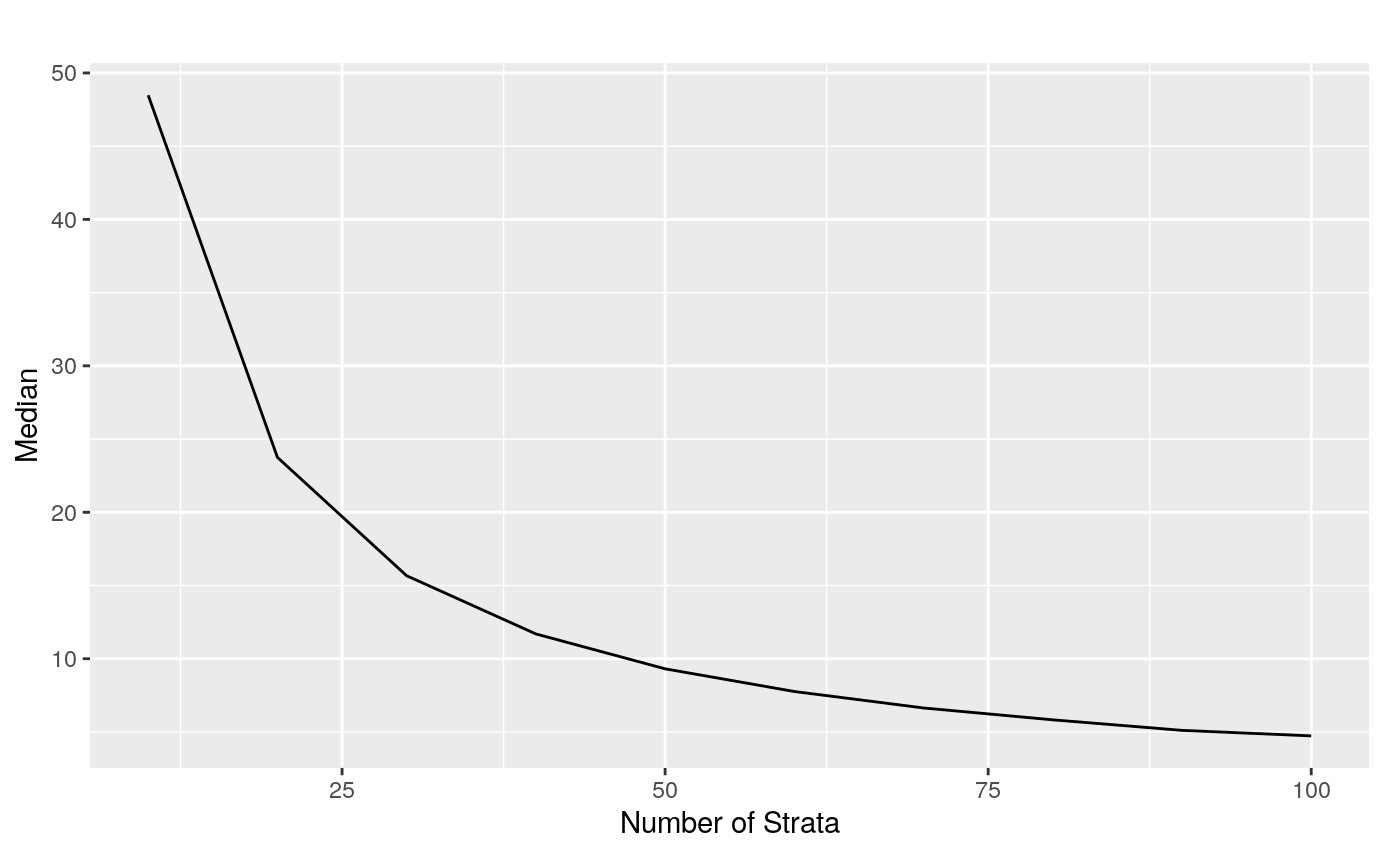}
        \caption{Median of stratum size.}
    \end{subfigure}
    \hfill
    \begin{subfigure}{0.46\textwidth}
        \centering
        \includegraphics[width=\textwidth]{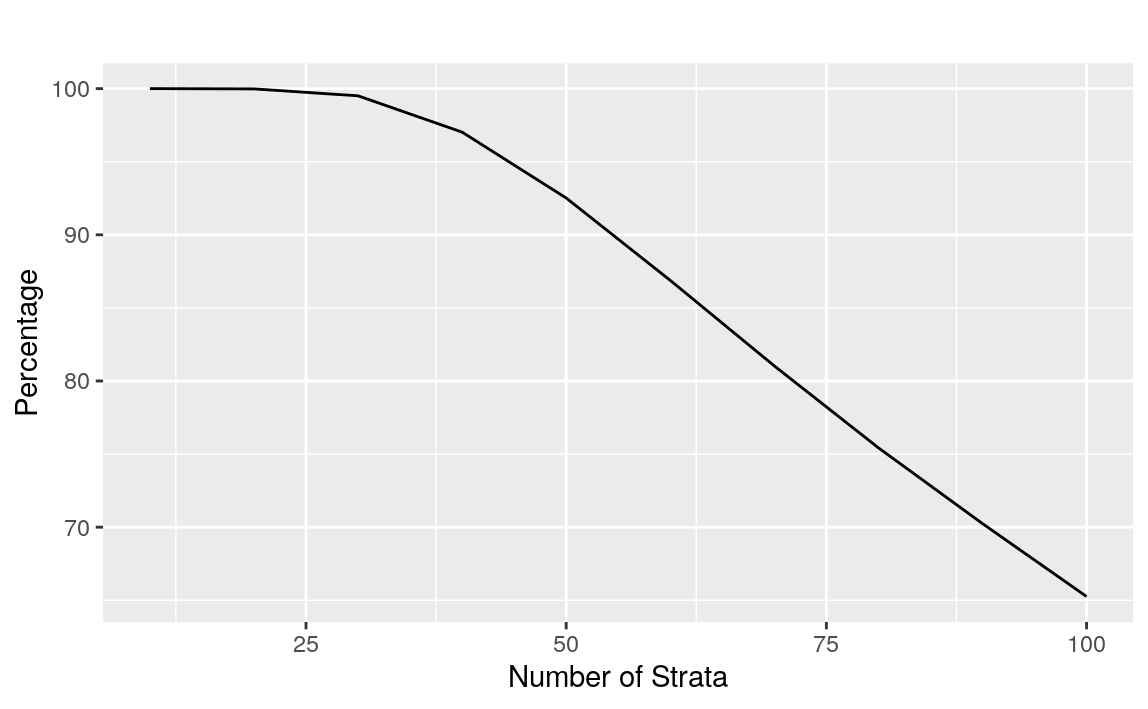}
        \caption{Percentage of stratum size $\ge4$.}
    \end{subfigure}
    \caption{Median stratum size and the percentage of strata with at least four observations in cases of extremely large numbers of strata.}
    \label{fig:subfigures}
\end{figure}

\begin{figure}[htbp]
    \centering
    \begin{subtable}{1\textwidth}
        \centering
        \includegraphics[width=0.66\textwidth]{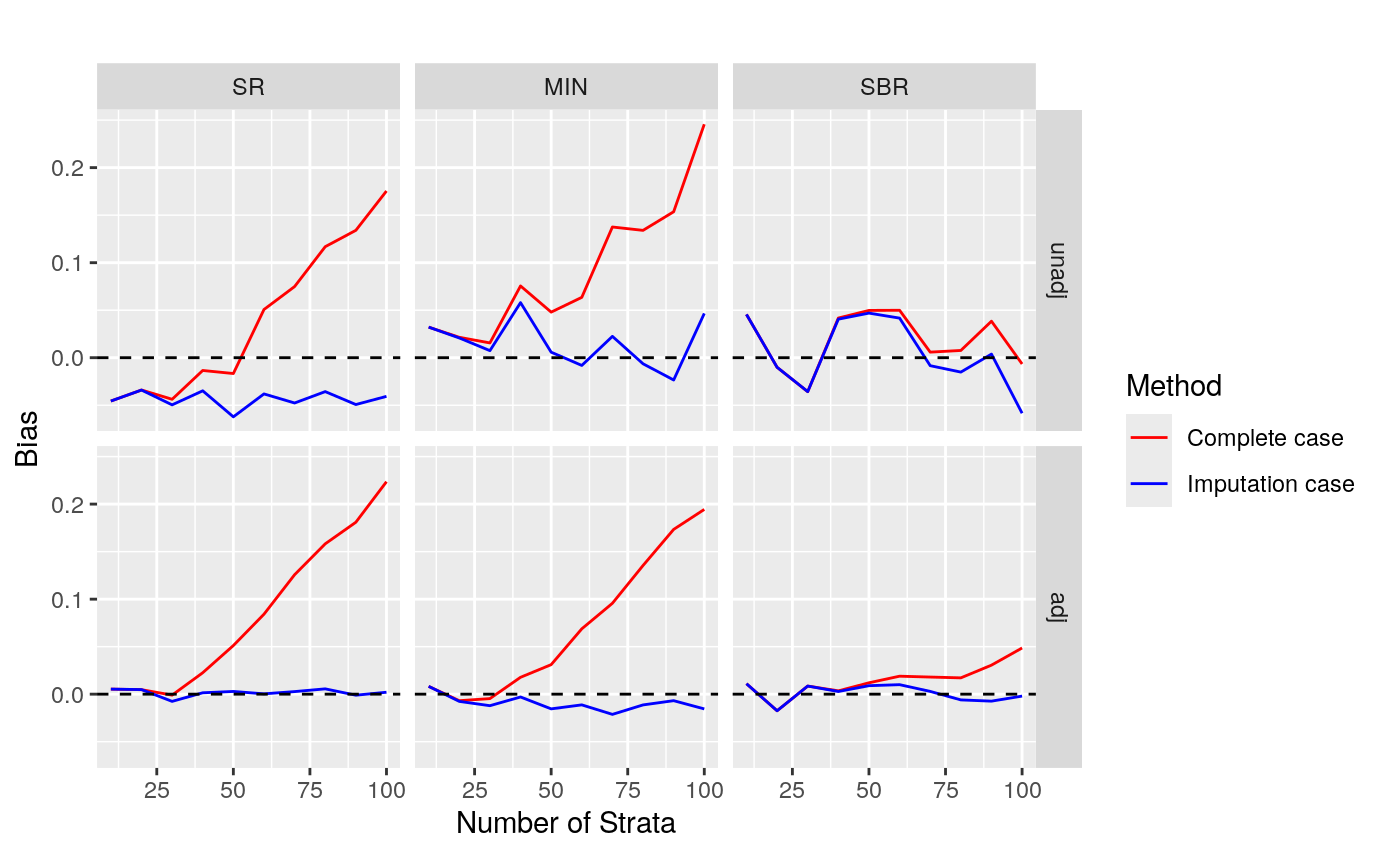}
        \caption{Bias of $\hat\tau$ and $\hat\tau_{adj}^*$.}
        \label{fig:subfiga}
    \end{subtable}
    \begin{subtable}{1\textwidth}
        \centering
        \includegraphics[width=0.66\textwidth]{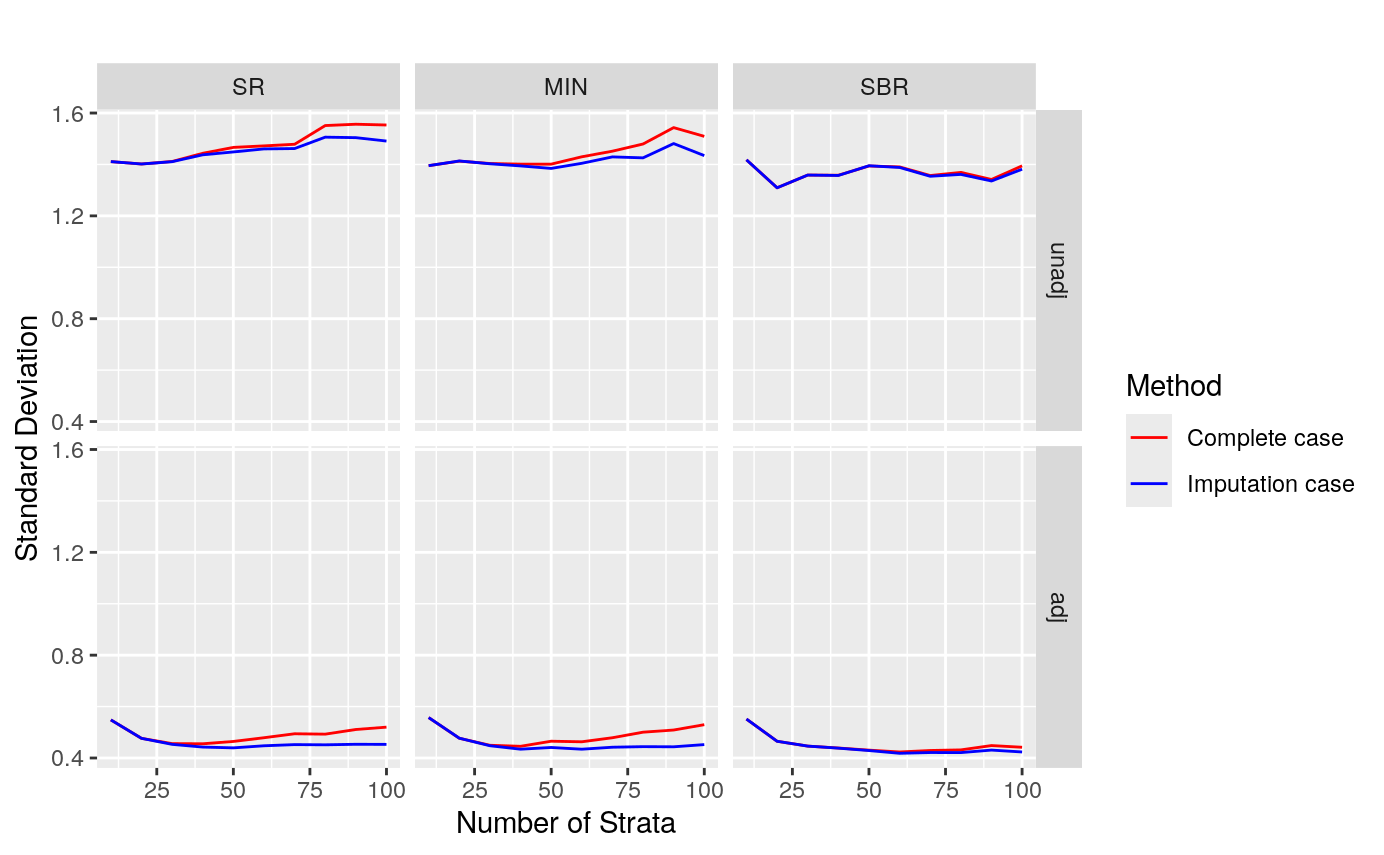}
        \caption{Standard deviation of $\hat\tau$ and $\hat\tau_{adj}^*$.}
        \label{fig:subfigb}
    \end{subtable}
    \begin{subtable}{1\textwidth}
        \centering
        \includegraphics[width=0.66\textwidth]{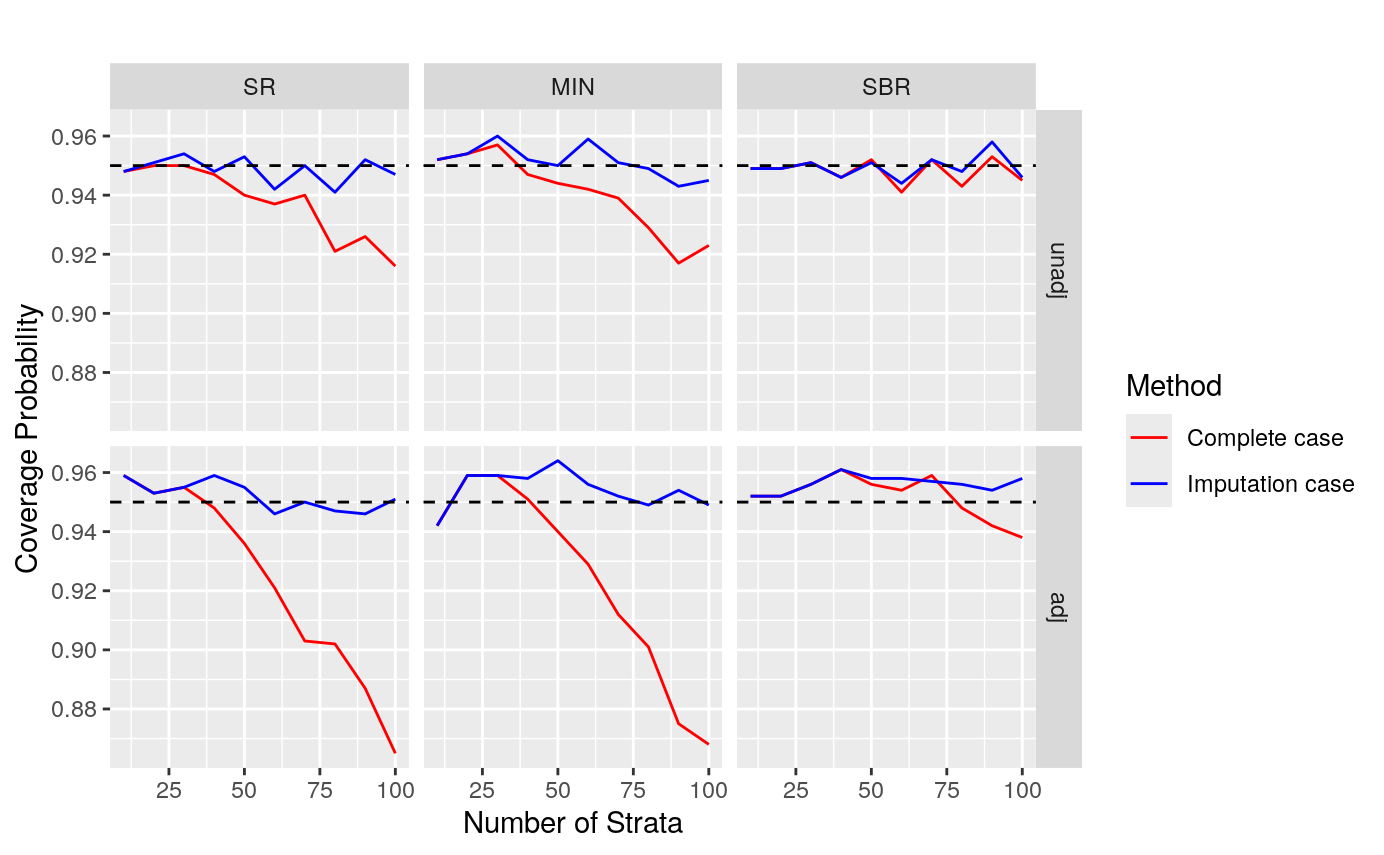}
        \caption{Coverage probability of $\hat\tau$ and $\hat\tau_{adj}^*$.}
        \label{fig:subfigc}
    \end{subtable}
    \caption{Bias, standard deviation and coverage probability with extremely large numbers of strata.}
\end{figure}

\subsection{Clinical trial example}
Finally, we consider a clinical trial example based on the synthetic data of the Nefazodone CBASP trial, which was conducted to compare the efficacy of three treatments for chronic depression \citep{keller2000comparison}.  Here we focus on two of the treatments, Nefazodone and the combination of Nefazodone and the cognitive behavioral-analysis system of psychotherapy (CBASP). To generate the synthetic data, we fit a non-parametric spline using the function ``bigssa" in the $\mathrm{R}$ package ``bigspline" with six selected covariates: AGE, HAMD17, HAMD24, HAMD\_COGIN, GENDER, and TreatPD. The detailed variable description is as follows: AGE, Age in years at screening;
        HAMD17, Total HAMD-17 score;
        HAMD24, Total HAMD-24 score;
        HAMD\_COGIN, HAMD cognitive disturbance score;
        GENDER, 1 female and 0 male;
        TreatPD, Treated past depression (1 yes and 0 no).

The trail with two treatment arms contains 440 patients after removing missing values. Then, we implement simple randomization, minimization and stratified block randomization to obtain the treatment assignments for the equal $(\pi=1 / 2)$ allocation. Among the six covariates, we use the stratified AGE, HAMD17, GENDER and TreatPD in stratified block randomization, where the stratified AGE is determined by the relative values of 40 and the stratified HAMD17 is determined by the relative values of 18. There are 16 strata in total. Once the treatment assignments are produced, we generate patient outcomes through the fitted model. For data analysis, AGE, $\text{AGE}^2$, HAMD17, HAMD24 and HAMD\_COGIN are used as the additional covariates. The analysis was performed by $\hat\tau$, $\hat\tau_{adj}$ and $\hat\tau_{adj}^*$, and the results are shown in Table \ref{tab:real data}.

\begin{table}[]
    \centering
    \caption{Estimates, 95\% confidence intervals, 95\% confidence interval lengths, and variance reductions under simple randomization, minimization and stratified block randomization
    for synthetic Nefazodone CBASP trial data.}
    \begin{threeparttable}
    \begin{tabular}{cccccccc}
    \toprule
        Randomization & Estimator & \multicolumn{1}{c}{Estimate} & \multicolumn{1}{c}{95\% CI} & \multicolumn{1}{c}{CI length} & \multicolumn{1}{c}{VR(\%)}\\
        \midrule
        SR & $\hat\tau$ & 4.69 & (4.16, 5.23) & 1.07 & - \\
        & $\hat\tau_{adj}$ & 4.62 & (4.12, 5.12) & 1.01 & 11.48 \\
        & $\hat\tau_{adj}^*$ & 4.62 & (4.12, 5.12) & 1.01 & 11.48 \\
        \midrule
        MIN & $\hat\tau$ & 4.63 & (4.08, 5.18) & 1.10 & -  \\
        & $\hat\tau_{adj}$ & 4.70 & (4.19, 5.22) & 1.04 & 11.10 \\
        & $\hat\tau_{adj}^*$ & 4.72 & (4.21, 5.24) & 1.03 & 11.21\\
        \midrule
        SBR & $\hat\tau$ &  4.69 & (4.17, 5.21) & 1.04 & -  \\
         & $\hat\tau_{adj}$ & 4.82 & (4.33, 5.32) & 0.99 & 9.62 \\
         & $\hat\tau_{adj}^*$ & 4.81 & (4.32, 5.31) & 0.99 & 9.70 \\
        \bottomrule
    \end{tabular}
    \begin{tablenotes}
        \item CI: Confidence interval. VR: Variance reduction.
    \end{tablenotes}
        
    \end{threeparttable}
    \label{tab:real data}
\end{table}
Among the three randomization procedures, incorporating adjustments for stratification and additional baseline covariates significantly enhances estimation efficiency. Specifically, compared to $\hat{\tau}$, the variance reductions for $\hat{\tau}_{adj}$ amount to approximately 10\%, under equal allocation. Given the assignment probability is set at 0.5, $\hat{\tau}_{adj}$ and $\hat{\tau}_{adj}^*$ behave almost identically in numerical terms.

In conclusion, the performance of each recommended estimator remains comparable across different randomization methods, with stratified block randomization demonstrating a slight advantage.

\section{Conclusion}

The literature on covariate-adaptive randomization typically assumes a fixed number of strata. However, this framework often conflicts with experiments in practice, where the number of strata is not relatively small compared to the total sample size. This paper develops a framework that accommodates a diverging number of strata under covariate-adaptive randomization. Through this framework, we demonstrate the robustness of the stratified difference-in-means estimator and regression-adjusted estimators. In particular, we propose a novel weighted regression adjustment to guarantee efficiency improvement.

Furthermore, this paper investigates the scenario with an extremely large number of strata. We recommend flexible imputation for addressing extreme inadequacy of sample size within strata. More detailed theoretical investigations into scenarios with an extremely large number of strata would be both interesting and valuable.

Non-asymptotic statistics is utilized to show that some important randomization examples satisfy the assumptions required by the developed theory. We rediscover the importance of design balance by linking it to inference robustness. Stratified block randomization is preferable to simple randomization as it allows finer stratification and exhibits more robust empirical performance.  


In summary, this paper provides a comprehensive landscape of inference under covariate-adaptive randomization, spanning fixed, diverging, and extremely large numbers of strata. We believe that our framework and related techniques may contribute to solving other problems in this field. Also, investigating the non-asymptotic properties of covariate-adaptive randomization would be promising for future work.

\bibliographystyle{apalike}
\bibliography{refs}

\newpage

\appendix

\section{Proof of Main Results}

\subsection{Proof of Proposition~\ref{prop:sr}}
\begin{proof}
Randomization naturally guarantees Assumption (B1). We only to verify Assumption (B3).

We prove a stronger result:
For $a\in\{0,1\}$ and any fixed positive number $C$, we have that
    \[
    \sup_{s\in\mathcal S_n}\mid \hat\pi_{na}(s)-\pi_{na}(s)\mid =o_P(1),
    \]
    and
    \[
\lim_{n\to\infty}\pr(\inf_{s\in\mathcal S_n}n_a(s)\ge C)=1.
    \]
Without loss of generality, we assume $|\mathcal S_n|$ is non-decreasing and $\inf_s p_n(s)$ is non-increasing as $n\to\infty$.

\textbf{Case 1.}  It does not hold that $\inf_{s}p_n(s)\to0$, and hence $|\mathcal S_n|$ is bounded away from $\infty$ and $\inf_{s}p_n(s)$ is bounded away from $0$. 

Discuss separately in  every stratum and the results can be checked. 

\textbf{Case 2.} $\inf_{s}p_n(s)\to0$.

First, we prove that $\sup_{s\in\mathcal S_n}\mid \hat\pi_{na}(s)-\pi_{na}(s)\mid =o_P(1)$.

    For any $\v>0$,
    \begin{align*}
        &\pr\{\sup_{s}|\hat\pi_{na}(s)-\pi_{na}(s)|\ge\v\mid S^{(n)} \}\\
        =&\pr\{\sup_{s}|n_a(s)-\pi_{na}(s)n(s)|\ge\v n(s)\mid S^{(n)} \}\\
        \leq&\sum_{s\in\mathcal S_n}\pr\{|n_a(s)-\pi_{na}(s)n(s)|\ge\v n(s)\mid S^{(n)} \}.
    \end{align*}
    Stratified simple randomization is independent within every stratum.  Given $n(s)$, $n_a(s)$ is the independent sum of $n(s)$ $\{0,1\}$-valued random variables. Hence conditional on $n(s)$, $n_a(s)$ is sub-Gaussian with sub-Gaussian parameter $\zeta_a^2(s)=n(s)$ (see, e.g., Example 2.4 and Proposition 2.5 (page 24) in \cite{wainwright2019high}). By the tail bound of sub-Gaussian random variables,
    \begin{align*}
        \pr\{|n_a(s)-\pi_{na}(s)n(s)|\ge\v n(s)\mid S^{(n)} \}
        \leq&e^{-\frac{\v^2n^2(s)}{2n(s)}}\\
        =&e^{-\frac{\v^2n(s)}{2}}.
    \end{align*}
    Plugging into the previous inequality, we have
    \begin{align*}
        \pr\{\sup_{s}|\hat\pi_{na}(s)-\pi_{na}(s)|\ge\v\mid n(s) \}
        \leq&\sum_{s\in\mathcal S_n}e^{-\frac{\v^2n(s)}{2}}\\
        \leq&|\mathcal S_n|e^{-\frac{\v^2\inf_{s\in\mathcal S_n}n(s)}{2}}\\
        =&e^{-\frac{\v^2\inf_{s\in\mathcal S_n}n(s)}{2}+\log(|\mathcal S_n|)}.
    \end{align*}
    
    Then we will show $\frac{\v^2\inf_{s\in\mathcal S_n}n(s)}{2}-\log(|\mathcal S_n|)\xrightarrow{P}\infty$. For any fixed $C>0$, 
    \begin{align*}
        &\pr\{\frac{\v^2\inf_{s\in\mathcal S_n}n(s)}{2}-\log(|\mathcal S_n|)\leq C\}\\
        =&\pr\{{\inf_{s\in\mathcal S_n}n(s)}\leq \frac{2\{C+\log(|\mathcal S_n|)\}}{\v^2} \}\\
        \leq&\sum_{s\in\mathcal S_n}\pr\{n(s)\leq \frac{2\{C+\log(|\mathcal S_n|)\}}{\v^2} \}.
    \end{align*}
    Because $n(s)$ has the distribution $Binoimal(n,p_n(s))$, Theorem 2.1 in \cite{buldygin2013sub} shows that the exact sub-Gaussian parameter of $n(s)$ is \[\zeta^2(s)=\frac{n\{2p_n(s)-1\}}{2[\log\{p_n(s)\}-\log\{1-p_n(s)\}]},\] 
    which implies that
    \begin{align*}
        \pr\{n(s)-np_n(s)\leq -C\}\leq& \exp\left(-\frac{2C^2[\log\{p_n(s)\}-\log\{1-p_n(s)\}]}{n\{2p_n(s)-1\}}\right),
    \end{align*}
    equivalently,
    \begin{align}
    \label{n(s)tail}
        \pr\{n(s)\leq C\}=&\pr\{n(s)-np_n(s)\leq C-np_n(s)\}\nonumber\\ 
        \leq&\exp\left(-\frac{2\{np_n(s)-C\}^2[\log\{p_n(s)\}-\log\{1-p_n(s)\}]}{n\{2p_n(s)-1\}}\right).
    \end{align}
Plug into the previous inequality,

    \begin{align*}
        &\pr\{\frac{\v^2\inf_{s\in\mathcal S_n}n(s)}{2}-\log(|\mathcal S_n|)\leq C\}\\
        \leq&\sum_{s\in\mathcal S_n}\pr\{n(s)\leq \frac{2\{C+\log(|\mathcal S_n|)\}}{\v^2} \}\\
        \leq&|\mathcal S_n|\pr\{ n(s_0)\leq \frac{2\{C+\log(|\mathcal S_n|)\}}{\v^2}|s_0=\mathrm{argmin}_s p_n(s)\}\\
        \leq&\exp\left(-\frac{2\left[n\inf_s p_n(s)-\frac{2\{C+\log(|\mathcal S_n|)\}}{\v^2}\right]^2\left[\log\{\inf_s p_n(s)\}-\log\{1-\inf_s p_n(s)\}\right]}{n\{2\inf_s p_n(s)-1\}}+\log|\mathcal S_n|\right).
    \end{align*}
    We need the exponential term goes to $-\infty$ for any $\v$ and $C$. Lemma 2.1 (K7) in \cite{buldygin2013sub} shows that  $\lim_{p_n(s)\to0}2|\log\{p_n(s)\}|\zeta^2(s)=1$. If $\inf_{s}p_n(s)\to0$, we have
    \begin{align*}
        &\frac{2\left[n\inf_s p_n(s)-\frac{2\{C+\log(|\mathcal S_n|)\}}{\v^2}\right]^2\left[\log\{\inf_s p_n(s)\}-\log\{1-\inf_s p_n(s)\}\right]}{n\{2\inf_s p_n(s)-1\}}\\
        =&O\left(\frac{\left[n\inf_s p_n(s)-\frac{\log(|\mathcal S_n|)}{\v^2}\right]^2\log\{|\inf_s p_n(s)|\}}{n}\right)\\
        =&O\left(\frac{\left\{n\inf_s p_n(s)\right\}^2\log\{|\inf_s p_n(s)|\}}{n}\right)\\
        \to&\infty.
    \end{align*}
    The last equality follows because \[\log(|\mathcal S_n|)=o\left(n \inf_s p^2_n(s)\log\{1/\inf_s p_n(s)\} \right),\] where $\inf_s p_n(s)\log\{1/\inf_s p_n(s)\}=o(1)$.
    
    Since $\log(|\mathcal S_n|)=o\left(n \inf_s p^2_n(s)\log\{|\inf_s p_n(s)|\} \right)$, we have the exponential term goes to $-\infty$. Then \[\frac{\v^2\inf_{s\in\mathcal S_n}n(s)}{2}-\log(|\mathcal S_n|)\xrightarrow{P}\infty,\]
    and hence
    \[\pr\{\sup_{s}|\hat\pi_{na}(s)-\pi_{na}(s)|\ge\v\mid S^{(n)} \}\xrightarrow{P}0. \]

    By dominated convergence theorem we have
    \[\pr\{\sup_{s}|\hat\pi_{na}(s)-\pi_{na}(s)|\ge\v \}\to0. \]

    Second, we prove that $\lim_{n\to\infty}\pr(\inf_{s\in\mathcal S_n}n_a(s)\ge C)=1$.
    In the previous step we have obtained for any fixed $C>0$ and $\v>0$ 
    \[\pr\{\frac{\v^2\inf_{s\in\mathcal S_n}n(s)}{2}-\log(|\mathcal S_n|)\leq C\}\to0,\] which implies that
    \[
    \pr\{\inf_{s\in\mathcal S_n}n(s)\leq C\}\to0.
    \]

    The previous proved result is that
    \[\pr\{\sup_{s}|n_a(s)-\pi_{na}(s)n(s)|\ge\v n(s)|S^{(n)}\}\xrightarrow{P}0.\]
    Take $\v=\inf_{s}\pi_{na}(s)/2$ and we have
    \begin{align*}
        &\pr\{\sup_{s}|n_a(s)-\pi_{na}(s)n(s)|\ge\inf_{s}\pi_{na}(s)n(s)/2 |S^{(n)}\}\xrightarrow{P}0,
    \end{align*}
    which implies that
    \begin{align*}
        &\pr\{\inf_{s}n_a(s) \leq \inf_{s}\pi_{na}(s)n(s)/2|S^{(n)}\}\\
        \leq&\pr\{\sup_{s}|n_a(s)-\pi_{na}(s)n(s)|\ge\inf_{s}\pi_{na}(s)\cdot n(s)/2|S^{(n)}\}\\
        \xrightarrow{P}&0.
    \end{align*}
    Finally, denoting the event  $E_1=\{\inf_{s\in\mathcal S_n}n(s)> 2C/\inf_{s}\pi_{na}(s)\}$,
    \begin{align*}
        &\pr\{\inf_s n_a(s)<C\}\\
        =&\pr\{\inf_s n_a(s)<C\cap E_1\}+\pr(E_1^c)\\
        \leq&\pr\{\inf_{s}n_a(s) \leq \inf_{s}\pi_{na}(s)\cdot n(s)/2 \}+\pr(E_1^c)\\
        =&E[\pr\{\inf_{s}n_a(s) \leq \inf_{s}\pi_{na}(s)\cdot n(s)/2| S^{(n)}\}]+\pr(E_1^c)\\
        \to&0.
    \end{align*}

\end{proof}

\subsection{Proof of Proposition~\ref{prop:sbr}}
\begin{proof}
Randomization naturally guarantees Assumption (B1). We only to verify Assumption (B3).

We prove a stronger result:
For $a\in\{0,1\}$ and any fixed positive number $C$, we have that
    \[
    \sup_{s\in\mathcal S_n}\mid \hat\pi_{na}(s)-\pi_{na}(s)\mid =o_P(1),
    \]
    and
    \[
\lim_{n\to\infty}\pr(\inf_{s\in\mathcal S_n}n_a(s)\ge C)=1.
    \]
    Since $n_1(s)=\lfloor \pi_{n1}(s)n(s) \rfloor$, we have
    \[
    |n_a(s)-\pi_{na}(s)n(s)|\leq1,
    \]
    and hence
    \begin{align*}
        \sup_s|\hat\pi_{na}(s)-\pi_{na}(s)|=&\sup_s|\{n_a(s)-\pi_{na}(s)n(s)\}/n(s)|\\
        \leq&\frac{1}{\inf_s n(s)}.
    \end{align*}
    Then for any fixed $C>0$,
    \begin{align*}
        \pr\{\sup_s|\hat\pi_{na}(s)-\pi_{na}(s)|\leq 1/C\}\geq&\pr\{1/\inf_s n(s)\leq 1/C\}\\
        =&\pr\{\inf_s n(s)\geq C\}.
    \end{align*}
    We only need to prove that for any fixed $C>0$, $\pr\{\inf_s n(s)\geq C\}\to1$. Without loss of generality, we assume $|\mathcal S_n|$ is non-decreasing and $\inf_s p_n(s)$ is non-increasing as $n\to\infty$.

    \textbf{Case 1.} It does not hold that $\inf_{s}p_n(s)\to0$, and hence $|\mathcal S_n|$ is bounded away from $\infty$ and $\inf_{s}p_n(s)$ is bounded away from $0$. This case degenerates into the case where the number of strata is limited and the result can be checked.

    \textbf{Case 2.} $\inf_{s}p_n(s)\to0$.
    \begin{align*}
        \pr\{\inf_s n(s)> C\}&=\prod_{s} \pr\{n(s)> C\}\\
        &\geq  [\min_s \pr\{n(s)> C\}]^{|\mathcal S_n|}\\
        &=[\pr\{n(\mathrm{argmin}_s p_n(s))> C\}]^{|\mathcal S_n|}.
    \end{align*}
By the tail bound of $n(s)$ (\ref{n(s)tail}),
    \[\pr\{n(s)\leq C\}\leq \exp\left(-\frac{\{np_n(s)-C\}^2}{\zeta^2(s)}\right),\]
    where
    \[\zeta^2(s)=\frac{n\{2p_n(s)-1\}}{2[\log\{p_n(s)\}-\log\{1-p_n(s)\}]},\]
we furthermore have
\begin{align*}
    &\pr\{\inf_s n(s)> C\}\\
    \geq&
    \left\{1-\exp\left(-\frac{\{n\inf_s p_n(s)-C\}^2}{\zeta^2(\mathrm{argmin}_s p_n(s))}\right)\right\}^{|\mathcal S_n|}\\
    =&\left\{1-\exp\left(-\frac{\{n\inf_s p_n(s)-C\}^2}{\zeta^2(\mathrm{argmin}_s p_n(s))}\right)\right\}^{\exp\left(\frac{\{n\inf_s p_n(s)-C\}^2}{\zeta^2(\mathrm{argmin}_s p_n(s))}\right)\cdot {|\mathcal S_n|}\exp\left(-\frac{\{n\inf_s p_n(s)-C\}^2}{\zeta^2(\mathrm{argmin}_s p_n(s))}\right)}.
\end{align*}
    In case 2, $\inf_{s}p_n(s)\to0$. Lemma 2.1 (K7) in \cite{buldygin2013sub} shows that  $\lim_{p_n(s)\to0}2|\log\{p_n(s)\}|\zeta^2(s)=1$. Since $n\inf_s p_n(s)\to\infty$, we have
    \begin{align*}
    \frac{\{n\inf_s p_n(s)-C\}^2}{\zeta^2(\mathrm{argmin}_s p_n(s))}&=O\left(2|\log\{\inf_s p_n(s)\}|\{n\inf_s p_n(s)-C\}^2\right)\\
    &=O\left({2|\log\{\inf_s p_n(s)\}|}{\{n\inf_s p_n(s)\}^2}\right)\to\infty.
    \end{align*}
    Hence we do  equivalent to infinity replacement,
    \begin{align*}
        \left\{1-\exp\left(-\frac{\{n\inf_s p_n(s)-C\}^2}{\zeta^2(\mathrm{argmin}_s p_n(s))}\right)\right\}^{\exp\left(\frac{\{n\inf_s p_n(s)-C\}^2}{\zeta^2(\mathrm{argmin}_s p_n(s))}\right)}\to \exp(-1).
    \end{align*}
    Plug in,
    \begin{align*}
        &\left\{1-\exp\left(-\frac{\{n\inf_s p_n(s)-C\}^2}{\zeta^2(\mathrm{argmin}_s p_n(s))}\right)\right\}^{\exp\left(\frac{\{n\inf_s p_n(s)-C\}^2}{\zeta^2(\mathrm{argmin}_s p_n(s))}\right)\cdot \mathcal{|S|}\exp\left(-\frac{\{n\inf_s p_n(s)-C\}^2}{\zeta^2(\mathrm{argmin}_s p_n(s))}\right)}\\
        \to&\exp\left(-|\mathcal S_n|\exp\left(-{2|\log\{\inf_s p_n(s)\}|}{\{n\inf_s p_n(s)\}^2}\right)\right).
    \end{align*}
    We require the exponential term tends to zero which means that 
    \[
    2{|\log\{\inf_s p_n(s)\}|}{n^2\inf_s p^2_n(s)} -\log|\mathcal S_n|\to\infty.
    \]
\end{proof}

\subsection{Proof of Theorem~~\ref{thm:var}}

\begin{proof}
We first claim that $E\{\sigma^2(Y,a,S_{ni})/\pi_{na}(S_{ni})\}$ and $\text{var}\big[\{Y_i(1)-Y_i(0)|S_{ni}\}\big]$ exist.
The former exists since $\sup_{n,s}\sigma^2(Y,a,s)<\infty$ and $\inf_{n,s}\{\pi_{na}(s)\}>0$. The latter exists since $\sup_{n,s}\sigma^2(Y,a,s)<\infty$.

Similar as \cite{Bugni2019}, we decompose $\hat\tau$ as follows
\begin{equation}
    \begin{aligned}
    \sqrt{n}\left(\hat\tau-\tau\right)= & \sqrt n \Bigg(\sum_{s \in \mathcal S_n} \frac{n(s)}{n}\left\{\frac{1}{n_1(s)} \sum_{i\in[1,s]} \tilde{Y}_i(1)\right\} -\sum_{s \in \mathcal S_n} \frac{n(s)}{n}\left\{\frac{1}{n_0(s)} \sum_{i\in[0,s]}\tilde{Y}_i(0)\right\} \\ 
    & +\sum_{i=1}^n \frac{1}{n} \left[E\left\{Y_i(1)-Y_i(0)|S_{ni}\right\}-E\left\{Y_i(1)-Y_i(0)\right\}\right]\Bigg),
    \end{aligned}
\end{equation}
where the first two terms can be dealt with Lemma \ref{crly:s d-i-m} and the last term can be dealt with triangular-array central limit theorem.

Directly from Lemma~\ref{crly:s d-i-m}, we obtain
\begin{equation}
    \pr\Bigg\{\frac{\sqrt n \left[\sum_{s \in \mathcal S_n} \frac{n(s)}{n}\hat\mu(\tilde Y,1,s)-\sum_{s \in \mathcal S_n} \frac{n(s)}{n}\hat\mu(\tilde Y,0,s)\right]}{\sqrt{\frac{1}{n}\sum_{s\in\mathcal S_n}\left\{\frac{n^2(s)\sigma^2(Y,0,s)}{n_0(s)}+\frac{n^2(s)\sigma^2(Y,1,s)}{n_1(s)}\right\}}}\leq u|(S^{(n)},A^{(n)}) \Bigg\}\xrightarrow{P} \Phi(u).
\end{equation}

The remaining last term is i.i.d. with finite variance per row. We use the central limit theorem for triangular arrays 
\begin{equation}
\begin{aligned}
    &\frac{(\sqrt n)^{-1}\sum_{i=1}^n  [E\left\{Y_i(1)-Y_i(0)|S_{ni}\right\}-E\left\{Y_i(1)-Y_i(0)\right\}]
    }{\sqrt{\text{var}\big[E\left\{Y_i(1)-Y_i(0)|S_{ni}\right\}\big]}}\\
    =&\frac{(\sqrt n)^{-1}\sum_{i=1}^n  [E\left\{Y_i(1)-Y_i(0)|S_{ni}\right\}-E\left\{Y_i(1)-Y_i(0)\right\}]
    }{\sqrt{V_{nB}(Y)}}
    \\
    \xrightarrow{d}& \mathcal N\left(0,1\right).
\end{aligned}
\end{equation}

Notice that ${\sigma^2(Y,a,S_{ni})}/{\pi_{na}(S_{ni})}$ is bounded since $0<\inf_{n,s}\pi_{na}(s)$ and  $\sup_{n,s}\sigma^2(f,a,s)<\infty$. Furthermore we have
\begin{equation}
\begin{aligned}
    \frac{1}{n}\sum_{s\in\mathcal S_n}\left\{\frac{n^2(s)\sigma^2(Y,0,s)}{n_0(s)}+\frac{n^2(s)\sigma^2(Y,1,s)}{n_1(s)}\right\}=&\frac{1}{n}\sum_{s\in\mathcal S_n}\left\{\frac{n(s)\sigma^2(Y,0,s)}{\pi_{n0}(s)}+\frac{n(s)\sigma^2(Y,1,s)}{\pi_{n1}(s)}\right\}+o_P(1)\\
    = &\frac{1}{n}\left\{\sum_{i=1}^n\frac{\sigma^2(Y,0,S_{ni})}{\pi_{n0}(S_{ni})}+\sum_{i=1}^n\frac{\sigma^2(Y,1,S_{ni})}{\pi_{n1}(S_{ni})}\right\}+o_P(1)\\
    =& E\left\{\frac{\sigma^2(Y,0,S_{ni})}{\pi_{n0}(S_{ni})}+\frac{\sigma^2(Y,1,S_{ni})}{\pi_{n1}(S_{ni})}\right\}+o_P(1)\\
    =&V_{nW}(Y,0)+V_{nW}(Y,1)+o_P(1)\\
    =&V_W(Y,0)+V_W(Y,1)+o_P(1),
\end{aligned}
\end{equation}
where the first equality is from uniform convergence of ${n(s)}/{n_a(s)}$, the third equality is from the weak law of large numbers for triangular-arrays and the last equality is from the assumption that limits exist.

With (3), (5) and Lemma~\ref{lem:shao},
\begin{equation}
    \frac{\frac{1}{\sqrt n}\left[\sum_{s \in \mathcal S_n} \frac{n(s)}{n}\hat\mu(\tilde Y,1,s)-\sum_{s \in \mathcal S_n} \frac{n(s)}{n}\hat\mu(\tilde Y,0,s)\right]
    }{\sqrt{V_W(Y,0)+V_W(Y,1)}}
    \xrightarrow{d} \mathcal N\left(0,1\right).
\end{equation}

From the assumption that $\lim_{n\to\infty}V_{nB}(Y)=V_B(Y)$ and (4), we use Slutsky Theorem,
\begin{equation}
\frac{(\sqrt n)^{-1}\sum_{i=1}^n  [E\left\{Y_i(1)-Y_i(0)|S_{ni}\right\}-E\left\{Y_i(1)-Y_i(0)\right\}]
    }{\sqrt{V_{B}(Y)}}
    \xrightarrow{d}\mathcal N\left(0,1\right).
\end{equation}

Finally note that the last term $1/n \sum_{i=1}^n  [E\left\{Y_i(1)-Y_i(0)|S_{ni}\right\}-E\left\{Y_i(1)-Y_i(0)\right\}]$ is function of $(S^{(n)},A^{(n)})$. With (6) and (7), Lemma~\ref{lem:bai} implies the result.
\end{proof}

\subsection{Proof of Theorem~\ref{thm:var2}}

\begin{proof}
\textbf{Step 1.} We show that $\hat\beta_n-\beta_n\xrightarrow{P}0$, i.e., $\{\pi_1\hat\beta_n(0)+\pi_0\hat\beta_n(1)\}-\{\pi_1\beta_n(0)+\pi_0\beta_n(1)\}\xrightarrow{P}0$. It suffices to show $\hat\Sigma_{\tilde X(a)\tilde X(a)}\xrightarrow{P}\Sigma_{\tilde X\tilde X}$ and $\hat\Sigma_{\tilde X(a)\tilde Y(a)}\xrightarrow{P}\Sigma_{\tilde X\tilde Y(a)}$ for $a\in\{0,1\}$.

\textbf{Step 1(a).} We show the consistency of $\hat\Sigma_{\tilde X(a) \tilde X(a)}$. 

We will use the conditional Markov inequality to get the conditional convergence in probability. We  prove the conditional variance converges in probability to zero. Without loss of generality, we consider the set $\{\inf_{a,s}n_a(s)\geq2\}$ in the probability space. By Assumption (B3'), $\pr\{\inf_{a,s}n_a(s)\geq2\}\to1$.
\begin{align*}
    &\mathrm{var}[\{\hat\Sigma_{\tilde X(a) \tilde X(a)}\}_{kl}\mid (S^{(n)},A^{(n)})]\\
    =&\mathrm{var}\left[\sum_{s\in\mathcal S_n} \frac{n(s)}{n}\{\hat\sigma^2(X,a,s)\}_{kl}\mid (S^{(n)},A^{(n)})\right]\\
    =&\sum_{s\in\mathcal S_n} \frac{n^2(s)}{n^2}\mathrm{var}\left[\{\hat\sigma^2(X,a,s)\}_{kl}\mid (S^{(n)},A^{(n)})\right]\\
    =&\sum_{s\in\mathcal S_n} \frac{n^2(s)}{n^2}\left[\frac{\sigma_{kkll}(X,\cdot,s)}{n_a(s)}-\frac{\sigma^2_{kl}(X,\cdot,s)\{n_a(s)-2\}}{n_a(s)\{n_a(s)-1\}}+\frac{\sigma_{kk}(X,\cdot,s)\sigma_{ll}(X,\cdot,s)}{n_a(s)\{n_a(s)-1\}}\right]\\
    \leq&\sum_{s\in\mathcal S_n} \frac{n^2(s)}{n^2}\left[\frac{\sigma^{1/2}_{kkkk}(X,\cdot,s)\sigma^{1/2}_{llll}(X,\cdot,s)}{n_a(s)}+\frac{\sigma_{kk}(X,\cdot,s)\sigma_{ll}(X,\cdot,s)\{n_a(s)-2\}}{n_a(s)\{n_a(s)-1\}}+\frac{\sigma_{kk}(X,\cdot,s)\sigma_{ll}(X,\cdot,s)}{n_a(s)\{n_a(s)-1\}}\right]\\
    \leq&\sum_{s\in\mathcal S_n} \frac{n^2(s)}{n^2}\left[\frac{\sigma^{1/2}_{kkkk}(X,\cdot,s)\sigma^{1/2}_{llll}(X,\cdot,s)}{n_a(s)}+\frac{\sigma_{kk}(X,\cdot,s)\sigma_{ll}(X,\cdot,s)}{n_a(s)}\right]\\
    \leq& \frac{\sup_{k,n,s}E(\tilde X^4_{ik}\mid S_{ni}=s)}{n}\sum_{s\in\mathcal S_n} \frac{n(s)}{n} \frac{n(s)}{n_a(s)}\\
    =&\frac{\sup_{k,n,s}E(\tilde X^4_{ik}\mid S_{ni}=s)}{n}\sum_{s\in\mathcal S_n} \frac{n(s)}{n} \frac{1}{\pi_{na}(s)}+o_P(1)\\
    \leq&\frac{\sup_{k,n,s}E(\tilde X^4_{ik}\mid S_{ni}=s)}{n}\left\{\sum_{s\in\mathcal S_n} \frac{n(s)}{n}\right\}\frac{1}{\inf_{n,s}\pi_{na}(s)}+o_P(1)\\
    =&o(1)+o_P(1)=o_P(1),
\end{align*}
where the third equality is from Lemma~\ref{lem:var_var}, the fourth equality is from the uniform convergence in probability of $n_a(s)/n(s)$. Here $\sigma_{kl}(X,\cdot,s)=E\{(X_k-\mu(X_k,\cdot,s))(X_l-\mu(X_l,\cdot,s))\}$ and $\sigma_{kkll}(X,\cdot,s)=E\{(X_k-\mu(X_k,\cdot,s))^2(X_l-\mu(X_l,\cdot,s))^2\}$.

Hence from conditional Markov inequality (Lemma~\ref{lem:cond_markov}), we have for any $\v>0$,
\begin{align*}
    &\pr\left(|\{\hat\Sigma_{\tilde X(a) \tilde X(a)}\}_{kl} - E[\{\hat\Sigma_{\tilde X(a) \tilde X(a)}\}_{kl}\mid (S^{(n)},A^{(n)})] | \geq\v|(S^{(n)},A^{(n)})\right)\\
    \leq&\frac{\mathrm{var}[\{\hat\Sigma_{\tilde X(a) \tilde X(a)}\}_{kl}\mid (S^{(n)},A^{(n)})]}{\v^2}\xrightarrow{P}0.
\end{align*}

Further from Lemma~\ref{lem:uncond-conv-in-prob} we obtain that for any $\v>0$,
\begin{align*}
  \pr\left(|\{\hat\Sigma_{\tilde X(a) \tilde X(a)}\}_{kl} - E[\hat\Sigma_{\tilde X(a) \tilde X(a)}\}_{kl}\mid (S^{(n)},A^{(n)})]|\geq \v\right)\xrightarrow{P}0,
\end{align*} 
which means that $\{\hat\Sigma_{\tilde X(a) \tilde X(a)}\}_{kl} - E[\{\hat\Sigma_{\tilde X(a) \tilde X(a)}\}_{kl}\mid (S^{(n)},A^{(n)})]\xrightarrow{P}0$.

Next we prove that $E[\hat\Sigma_{\tilde X(a) \tilde X(a)}\}_{kl}\mid (S^{(n)},A^{(n)})]\xrightarrow{P}\{\Sigma_{\tilde X\tilde X}\}_{kl}$. From Assumption A3 we have $\sup_{k,l,n,s}|\mathrm{cov}(X_{ik},X_{il}\mid S_{ni}=s)|<\infty$ for any element $(k,l)$. Then we have
\begin{align*}
    E\left[\left\{\hat\Sigma_{\tilde X(a) \tilde X(a)}\right\}_{kl}\mid (S^{(n)},A^{(n)})\right]
    =&\sum_{s\in\mathcal S_n} \frac{n(s)}{n}E\left\{\hat\sigma^2(X,a,s)\mid (S^{(n)},A^{(n)})\right\}_{kl}\\
    =&\sum_{s\in\mathcal S_n} \frac{n(s)}{n}\mathrm{cov}(X_{ik},X_{il}\mid S_{ni}=s)\\
    =&\sum_{i=1}^n  \frac{1}{n}\mathrm{cov}(X_{ik},X_{il}\mid S_{ni})\\
    \xrightarrow{P}&E\{\mathrm{cov}(X_{ik},X_{il}\mid S_{ni})\}=\left\{\Sigma_{\tilde X\tilde X}\right\}_{kl},
\end{align*}
where the convergence from the weak law of large number (Lemma~\ref{lem:wlln}).

We conclude that $\{\hat\Sigma_{\tilde X(a) \tilde X(a)}\}_{kl}\xrightarrow{P}\{\Sigma_{\tilde X \tilde X}\}_{kl}$.

\textbf{Step 1(b).} Similarly, we can use the conditional variance and conditional expectation to prove that $\{\hat\Sigma_{\tilde X(a) \tilde Y(a)}\}_{k}\xrightarrow{P}\{\Sigma_{\tilde X \tilde Y(a)}\}_{k}$. We use $\widehat{\mathrm{cov}}(\cdot)$ to denote the sample covariance.

\begin{align*}
    &\mathrm{var}[\{\hat\Sigma_{\tilde X(a) \tilde Y(a)}\}_{k}\mid (S^{(n)},A^{(n)})]\\
    =&\mathrm{var}\left[\sum_{s\in\mathcal S_n} \frac{n(s)}{n}\widehat{\mathrm{cov}}\{X_{ik}(a),Y_i(a)\mid (S^{(n)},A^{(n)}\}\right]\\
    =&\sum_{s\in\mathcal S_n} \frac{n^2(s)}{n^2}\mathrm{var}\left[\widehat{\mathrm{cov}}\{X_{ik}(a),Y_i(a)\mid (S^{(n)},A^{(n)}\}\right]\\
    \leq&\sum_{s\in\mathcal S_n} \frac{n^2(s)}{n^2}\left(\frac{\{E(\tilde X^4_{ik}\mid S_{ni}=s)\}^{1/2}[E\{\tilde Y^4_i(a)\mid S_{ni}=s\}]^{1/2}}{n_a(s)}+\frac{E(\tilde X^2_{ik}\mid S_{ni}=s)E(\tilde Y^2_i(a)\mid S_{ni}=s)}{n_a(s)}\right)\\
    \leq& \frac{\max\left\{\sup_{k,n,s}E(\tilde X^4_{ik}\mid S_{ni}=s),\sup_{n,s}E\{\tilde Y^4_i(a)\mid S_{ni}=s\}\right\}}{n}\sum_{s\in\mathcal S_n} \frac{n(s)}{n} \frac{n(s)}{n_a(s)}\\
    =&\frac{\max\left\{\sup_{k,n,s}E(\tilde X^4_{ik}\mid S_{ni}=s),\sup_{n,s}E\{\tilde Y^4_i(a)\mid S_{ni}=s\}\right\}}{n}\sum_{s\in\mathcal S_n} \frac{n(s)}{n} \frac{1}{\pi_{na}(s)}+o_P(1)\\
    \leq&\frac{\max\left\{\sup_{k,n,s}E(\tilde X^4_{ik}\mid S_{ni}=s),\sup_{n,s}E\{\tilde Y^4_i(a)\mid S_{ni}=s\}\right\}}{n}\left\{\sum_{s\in\mathcal S_n} \frac{n(s)}{n}\right\}\frac{1}{\inf_{n,s}\pi_{na}(s)}+o_P(1)\\
    =&o(1)+o_P(1)=o_P(1).
\end{align*}
\begin{align*}
    E\left[\left\{\hat\Sigma_{\tilde X(a) \tilde Y(a)}\right\}_{k}\mid (S^{(n)},A^{(n)})\right]
    =&\sum_{s\in\mathcal S_n} \frac{n(s)}{n}E\left[\widehat{\mathrm{cov}}\{X_{ik}(a),Y_i(a)\mid (S^{(n)},A^{(n)})\}\right]\\
    =&\sum_{s\in\mathcal S_n} \frac{n(s)}{n}\mathrm{cov}\{X_{ik},Y_i(a)\mid S_{ni}=s\}\\
    =&\sum_{i=1}^n  \frac{1}{n}\mathrm{cov}\{X_{ik},Y_i(a)\mid S_{ni}\}\\
    \xrightarrow{P}&E[\mathrm{cov}\{X_{ik},Y_i(a)\mid S_{ni}\}]=\left\{\Sigma_{\tilde X\tilde Y(a)}\right\}_{k}.
\end{align*}

Up to now we obtain $\hat\beta_n(a)-\beta_n(a)\xrightarrow{P}0$ and $\hat\beta_n-\beta_n\xrightarrow{P}0$. 

\textbf{Step 2.} We study the asymptotic normality of $\hat\tau_{adj}$ as follows.

By the definition of $\hat\tau_{adj}$, we have
\begin{align*}
\hat{\tau}_{adj} = & \sum_{s\in\mathcal S_n} \frac{n(s)}{n} \{\hat\mu(\hat r,1,s)-\hat\mu(\hat r,0,s)\}\\
= & \sum_{s\in\mathcal S_n} \frac{n(s)}{n} \{\hat\mu(Y,1,s) - \hat\mu(X,1,s)^\T\hat \beta_n -\hat\mu(Y,0,s) + \hat\mu(X,0,s)^\T\hat \beta_n \}\\
= & \sum_{s\in\mathcal S_n} \frac{n(s)}{n}\left[\hat\mu(Y,1,s) -\hat\mu(Y,0,s) - \{\hat\mu(X,1,s) -\hat\mu(X,0,s)\}^\T\hat \beta_n \right]\\
= & \sum_{s\in\mathcal S_n} \frac{n(s)}{n}\left[\hat\mu(Y,1,s) -\hat\mu(Y,0,s) - \{\hat\mu(X,1,s) -\hat\mu(X,0,s)\}^\T \beta_n \right] \\
& -\sum_{s\in\mathcal S_n} \frac{n(s)}{n}\{\hat\mu(X,1,s) -\hat\mu(X,0,s)\}^\T\{\hat \beta_n-\beta_n\}\\
:= & R_1+R_2.
\end{align*}
Then we will prove that $\sqrt n R_2=o_P(1)$ so that $\hat\tau_{adj}$ has the same asymptotic distribution as $R_1$. Since $X_i(1)=X_i(0)$, we have the conditional expectation 
\begin{align*}
&E\left[\sum_{s\in\mathcal S_n} \frac{n(s)}{n}\left\{\hat\mu(X,1,s)-\hat\mu(X,0,s)\right\}\mid (S^{(n)},A^{(n)})\right]\\
=&\sum_{s\in\mathcal S_n} \frac{n(s)}{n} E\left\{\hat\mu(X,1,s)-\hat\mu(X,0,s)\mid (S^{(n)},A^{(n)})\right\}\\
=&0.
\end{align*}

Similarly, we have the $(k,l)$ element  of conditional variance
\begin{align*}
    &\mathrm{var}\left[\sum_{s\in\mathcal S_n} \frac{n(s)}{n}\left\{\hat\mu(X,1,s)-\hat\mu(X,0,s)\right\}\mid (S^{(n)},A^{(n)})\right]_{kl}\\
    =&\sum_{s\in\mathcal S_n} \frac{n^2(s)}{n^2} \mathrm{var}\left[\{\hat\mu(X,1,s)-\hat\mu(X,0,s)\}\mid (S^{(n)},A^{(n)})\right]_{kl}\\
    =&\sum_{s\in\mathcal S_n} \frac{n^2(s)}{n^2}\left[\mathrm{var}\{\hat\mu(X,1,s)\mid (S^{(n)},A^{(n)})\}_{kl}+\mathrm{var}\{\hat\mu(X,0,s)\mid (S^{(n)},A^{(n)})\}_{kl}\right]\\
    =&\sum_{s\in\mathcal S_n} \frac{n^2(s)}{n^2}\left[\frac{\{\sigma^2(X,1,s)\}_{kl}}{n_1(s)}+\frac{\{\sigma^2(X,0,s)\}_{kl}}{n_0(s)}\right]\\
    \leq&\sum_{s\in\mathcal S_n} \frac{n^2(s)}{n^2}\left[\frac{\sigma(X_k,\cdot,s)\sigma(X_l,\cdot,s)}{n_1(s)}+\frac{\sigma(X_k,\cdot,s)\sigma(X_l,\cdot,s)}{n_0(s)}\right]\\
    \leq&\sum_{s\in\mathcal S_n} \frac{n^2(s)}{n^2}\left[\frac{\sup_{k,n,s}\sigma^2(X_k,\cdot,s)}{n_1(s)}+\frac{\sup_{k,n,s}\sigma^2(X_k,\cdot,s)}{n_0(s)}\right]\\
    =&\frac{1}{n}\sum_{s\in\mathcal S_n} \frac{n(s)}{n}\sum_{a\in\{0,1\}}\left[\sup_{k,n,s}\sigma^2(X_k,\cdot,s)\left\{\frac{1}{\pi_{na}(s)}+\left|\frac{n(s)}{n_a(s)}-\frac{1}{\pi_{na}(s)}\right|\right\}\right]\\
    \leq&\frac{1}{n}\left\{\frac{2\sup_{k,n,s}\sigma^2(X_k,\cdot,s)}{\inf_{a,n,s}\pi_{na}(s)}+o_P(1)\right\}\\
    =&O_P(1/n).
\end{align*}
Hence by Lemma~\ref{lem:cond_markov} and the property of probability,
\begin{align*}
    \pr\left\{|\sqrt n\sum_{s\in\mathcal S_n} {n(s)}/{n}\left\{\hat\mu(X,1,s)-\hat\mu(X,0,s)\right\}|\ge M \Big|(S^{(n)},A^{(n)})\right\}\leq \min\left\{1,\frac{\frac{2\sup_{k,n,s}\sigma^2(X_k,\cdot,s)}{\inf_{a,n,s}\pi_{na}(s)}+o_P(1)}{M^2}\right\}.
\end{align*}
Then take the expectation
\begin{align*}
    &\pr\left\{|\sqrt n\sum_{s\in\mathcal S_n} {n(s)}/{n}\left\{\hat\mu(X,1,s)-\hat\mu(X,0,s)\right\}|\ge M \right\}\\
    =&E\left[\pr\left\{|\sqrt n\sum_{s\in\mathcal S_n} {n(s)}/{n}\left\{\hat\mu(X,1,s)-\hat\mu(X,0,s)\right\}|\ge M \Big|(S^{(n)},A^{(n)})\right\}\right]\\
    \leq&E\left[\min\left\{1,\frac{\frac{2\sup_{k,n,s}\sigma^2(X_k,\cdot,s)}{\inf_{a,n,s}\pi_{na}(s)}+o_P(1)}{M^2}\right\}\right]\\
    \leq&\frac{\frac{2\sup_{k,n,s}\sigma^2(X_k,\cdot,s)}{\inf_{a,n,s}\pi_{na}(s)}}{M^2}+E\left[\min\left\{1-\frac{\frac{2\sup_{k,n,s}\sigma^2(X_k,\cdot,s)}{\inf_{a,n,s}\pi_{na}(s)}}{M^2},\frac{o_P(1)}{M^2}\right\}\right]\\
    \leq& \frac{\frac{2\sup_{k,n,s}\sigma^2(X_k,\cdot,s)}{\inf_{a,n,s}\pi_{na}(s)}}{M^2} + o(1),
\end{align*}
where $M$ is some sufficiently large constant and the last inequality is from the dominated convergence theorem. 

Hence we have $\sqrt n\sum_{s\in\mathcal S_n} {n(s)}/{n}\left\{\hat\mu(X,1,s)-\hat\mu(X,0,s)\right\}=O_P(1)$. Combining with the consistency of $\hat\beta_n$, we have
\begin{align*}
    \sqrt n R_2=-\sqrt n\sum_{s\in\mathcal S_n} \frac{n(s)}{n}\{\hat\mu(X,1,s) -\hat\mu(X,0,s)\}^\T\{\hat \beta_n-\beta_n\}=O_P(1)\cdot o_P(1)=o_P(1).
\end{align*}

Therefore, $\sqrt n \hat{\tau}_{adj}$ has the same asymptotic distribution as
\begin{align*}
    \sqrt n\sum_{s\in\mathcal S_n} \frac{n(s)}{n}\left[\hat\mu(Y,1,s) -\hat\mu(Y,0,s) - \{\hat\mu(X,1,s) -\hat\mu(X,0,s)\}^\T \beta_n \right]
\end{align*}
which is the stratified difference-in-means estimator applied to the transformed outcomes $r_{ni}(a)=Y_i(a)-{X}_i^\T\beta_n$. We verify the conditions of Theorem~\ref{thm:var}. The main requirement is the moment condition similar as Assumption (A1).

We need for $a\in\{0,1\}$, for some $\v>0$,
    \[
    \sup_{n\ge1,s\in\mathcal S_n}E\{\mid \tilde r_{ni}(a)\mid^{2+\v}\mid S_{ni}=s\}<\infty.
    \]
Assumption (C2) guarantees that $\beta_n\to\beta$ and hence $\sup_n|\beta_n|<\infty$. Assumption (A1') and (A3) guarantee that $\sup_{n\ge1,s\in\mathcal S_n}E\{\mid \tilde Y_i(a)\mid^{2+\v}\mid S_{ni}=s\}<\infty$ and $\sup_{n\ge1,s\in\mathcal S_n}E\{\mid \tilde X_{ik}\mid^{2+\v}\mid S_{ni}=s\}<\infty$.

We have 
\begin{align*}
    &\sup_{n\ge1,s\in\mathcal S_n}E\{\mid \tilde X_i^\T\beta_n\mid^{2+\v}\mid S_{ni}=s\}\\
    \leq&p\cdot\sup_n|\beta_n|^{2+\v}\cdot\sup_{n\ge1,s\in\mathcal S_n,k\in\{1,\cdots,p\}}E\{\mid \tilde X_{ik}\mid^{2+\v}\mid S_{ni}=s\}\\
    <&\infty.
\end{align*}
Then by Lemma~\ref{lem:add mom} we have
\[
    \sup_{n\ge1,s\in\mathcal S_n}E\{\mid \tilde Y_i(a)-X_i^\T\beta_n\mid^{2+\v}\mid S_{ni}=s\}<\infty.
    \]

\end{proof}

\subsection{Proof of Theorem~~\ref{thm:var3}}
To prove the asymptotic normality, we follow the similar steps in the proof of Theorem~~\ref{thm:var2}. 

We then prove the efficiency gain.
By the above definitions we have $E\{r^*_{ni}(1)-r^*_{ni}(0)\mid S_{ni}\}=E\{Y_i(1)-Y_i(0)\mid S_{ni}\}$.  Hence $E\{r^*_{ni}(1)-r^*_{ni}(0)\}=E\{Y_i(1)-Y_i(0)\}$ and $V_{nB}(Y)=V_{nB}(r^*)$. The asymptotic variance difference is only between $V_{nW}(r^*,a)$ and $V_{nW}(Y,a)$.

We have
\begin{align*}
    &\{V_{nW}(r^*,0) + V_{nW}(r^*,1)\} -\{V_{nW}(Y,0) + V_{nW}(Y,1)\}\\
    =&E\left[\frac{\mathrm{var}\{ r^*_{ni}(0)\mid S_{ni}\}}{\pi_{n0}(S_{ni})}-\frac{\mathrm{var}(\{Y_i(0)\mid S_{ni}\}}{\pi_{n0}(S_{ni})}\right]+E\left[\frac{\mathrm{var}\{ r^*_{ni}(1)\mid S_{ni}\}}{\pi_{n1}(S_{ni})}-\frac{\mathrm{var}(\{Y_i(1)\mid S_{ni}\}}{\pi_{n1}(S_{ni})}\right]\\
    =&E[\mathrm{var}\{\check r^*_{ni}(0)\mid S_{ni}\}-\mathrm{var}\{\check Y_i(0)\mid S_{ni}\}]+E[\mathrm{var}\{\check r^*_{ni}(1)\mid S_{ni}\}-\mathrm{var}\{\check Y_i(1)\mid S_{ni}\}].
\end{align*}
Recall that $\check{f}_{ni}(a) = \tilde{f}_{ni}(a)/\sqrt{\pi_{na}(S_{ni})}$ and then we have
\begin{align*}
    E[\mathrm{var}\{\check r^*_{ni}(a)\mid S_{ni}\}]
    &=E[\mathrm{var}\{\check Y_i(a)-\check X_i(a)^\T\beta^*_n\mid S_{ni}\}]\\
    &=E[\mathrm{var}\{\check Y_i(a)\mid S_{ni}\}]+E\{\beta_n^{*\T}\check X_i(a)\check X_i(a)^\T\beta^*_n\mid S_{ni}\}-2E\{\beta_n^{*\T}\check X_i(a)\check Y_i(a)\mid S_{ni}\}]\\
    &=E[\mathrm{var}\{\check Y_i(a)\mid S_{ni}\}+\beta_n^{*\T} \mathrm{cov}\{\check X_i(a),\check X_i(a)\mid S_{ni}\}\beta^*_n-2\beta^{*\T}\mathrm{cov}\{\check X_i(a),\check Y_i(a)\mid S_{ni}\}]\\
    &=E[\mathrm{var}\{\check Y_i(a)\mid S_{ni}\}]+\beta_n^{*\T} \Sigma_{\check X(a),\check X(a)}\beta^*_n-2\beta_n^{*\T}\Sigma_{\check X(a)\check Y(a)}.
\end{align*}

Hence 
\begin{align*}
    &\{V_{nW}(r^*,0) + V_{nW}(r^*,1)\} -\{V_{nW}(Y,0) + V_{nW}(Y,1)\}\\
    =&\beta^{*\T}_n \Sigma_{\check X(0)\check X(0)}\beta^*_n-2\beta^{*\T}_n\Sigma_{\check X(0)\check Y(0)}+\beta^{*\T}_n \Sigma_{\check X(1)\check X(1)}\beta^*_n-2\beta^{*\T}_n\Sigma_{\check X(1)\check Y(1)}\\
    =&\beta^{*\T}_n \{\Sigma_{\check X(0)\check X(0)}+\Sigma_{\check X(1)\check X(1)}\}\beta^*_n-2\beta^{*\T}_n\{\Sigma_{\check X(0)\check Y(0)}+\Sigma_{\check X(1)\check Y(1)}\}\\
    =&\beta^{*\T}_n \{\Sigma_{\check X(0)\check X(0)}+\Sigma_{\check X(1)\check X(1)}\}\beta^{*}_n-2\beta^{*\T}_n\{\Sigma_{\check X(0)\check X(0)}+\Sigma_{\check X(1)\check X(1)}\}\beta^{*}_n\\
    =&-\beta^{*\T}_n\{\Sigma_{\check X(0)\check X(0)}+\Sigma_{\check X(1)\check X(1)}\}\beta^{*}_n\\
    \leq&0.
\end{align*}

So we have
\begin{align*}
    V_{\hat\tau^*_{adj}}-V_{\hat\tau}&=\{V_W(r^*,0) + V_W(r^*,1)+V_B(r^*)\} -\{V_W(Y,0)+V_W(Y,1)+V_B(Y)\}\\
    &=\{V_W(r^*,0) + V_W(r^*,1)\} -\{V_W(Y,0)+V_W(Y,1)\}\\
    &=\lim_{n\to\infty}\{V_{nW}(r^*,0) + V_{nW}(r^*,1)\} -\{V_{nW}(Y,0)+V_{nW}(Y,1)\}\\
    &\leq0.
\end{align*}

\subsection{Proof of Theorem~~\ref{thm:est}}

\begin{proof}
First, we prove that $\hat V_{nW}(Y,a)-V_{nW}(Y,a)\xrightarrow{P}0$.
\begin{align*}
\hat V_{nW}(Y)=&\hat V_{nW}(Y,0)+\hat V_{nW}(Y,1)\\
=&\frac{1}{n}\sum_{s\in\mathcal S_n}\left\{\frac{n^2(s)\hat\sigma^2(Y,0,s)}{n_0(s)}+\frac{n^2(s)\hat\sigma^2(Y,1,s)}{n_1(s)}\right\}.
\end{align*}
Then by Lemma~\ref{crly:s d-i-m}, we have for any $\v>0$,
\[
\pr\{\mid\hat V_{nW}(Y,a)/V_{nW}(Y,a)-1\mid >\v\mid (S^{(n)},A^{(n)})\}\xrightarrow{P}0.
\]
Furthermore by Lemma~\ref{lem:uncond-conv-in-prob}, we have for any $\v>0$,
\[
\pr\{\mid\hat V_{nW}(Y,a)/V_{nW}(Y,a)-1\mid >\v\}\to0,
\]
which means that $\hat V_{nW}(Y,a)-V_{nW}(Y,a)\xrightarrow{P}0$. Hence 
\begin{align*}
    \hat V_{nW}(Y,a)&= V_{nW}(Y,a)+o_P(1)\\
    &= V_{W}(Y,a)+o(1)+o_P(1)\\
    &=V_{W}(Y,a)+o_P(1).
\end{align*}

Second, we prove that $\hat V_{nB}(Y)-V_{nB}(Y)\xrightarrow{P}0$.
    \begin{align*}
    V_{nB}(Y) &= \mathrm{var}\big[E\left\{Y_i(1)-Y_i(0)|S_{ni}\right\}\big]\\
    &=E\left([E\left\{Y_i(1)\mid S_{ni}\right\}-E\left\{Y_i(0)\mid S_{ni}\right\}]^2\right)-
    [E\{Y_i(1)\}-E\{Y_i(0)\}]^2\\
    &=E([E\{Y_i(1)\mid S_{ni}\}]^2-2E\{Y_i(1)\mid S_{ni}\}E\{Y_i(0)\mid S_{ni}\}+[E\left\{Y_i(0)\mid S_{ni}\right\}]^2)\\
    &\quad-
    [E\{Y_i(1)\}-E\{Y_i(0)\}]^2\\
    &=E[-\mathrm{var}\{Y_i(1)\mid S_{ni}\}+E\{Y_i^2(1)\mid S_{ni}\}-2E\{Y_i(1)\mid S_{ni}\}E\{Y_i(0)\mid S_{ni}\}\\
    &\quad-\mathrm{var}\{Y_i(0)\mid S_{ni}\}+E\{Y_i^2(0)\mid S_{ni}\}]-
    [E\{Y_i(1)\}-E\{Y_i(0)\}]^2\\
    &=E[-\mathrm{var}\{Y_i(1)\mid S_{ni}\}-2E\{Y_i(1)\mid S_{ni}\}E\{Y_i(0)\mid S_{ni}\}-\mathrm{var}\{Y_i(0)\mid S_{ni}\}]\\
    &\quad+[E\{Y^2_i(1)\}+E\{Y^2_i(0)\}]-[E\{Y_i(1)\}-E\{Y_i(0)\}]^2.
    \end{align*}
Hence, we need to estimate four types of population quantities: $E[\text{var}\{Y_i(a)\mid S_{ni}\}]$, $E[E\{Y_i(1)\mid S_{ni}\}E\{Y_i(0)\mid S_{ni}\}]$, $E\{Y_i^2(a)\}$ and $E\{Y_i(a)\}$. We will prove that all the corresponding estimators are consistent: $\sum_{s\in\mathcal S_n}\{n(s)/n\} \hat\sigma^2(Y,a,s)$, $\sum_{s\in\mathcal S_n}\{n(s)/n\}\hat\mu(Y,0,s)\hat\mu(Y,1,s)$, $\sum_{s\in\mathcal S_n}\{n(s)/n\}\hat \mu_2(Y,a,s)$ and $\sum_{s\in\mathcal S_n}\{n(s)/n\}\hat\mu(Y,a,s)$.

We first show that $\sum_{s\in\mathcal S_n}\{n(s)/n\} \hat\sigma^2(f,a,s)\xrightarrow{P}E[var\{Y_i(a)\mid S_{ni}\}]$. Taking 
\begin{align*}
    c_a(s)&=\frac{\sqrt{\{n(s)n_a(s)\}}}{n},\\
    \tau_a^2&=\sum_{s\in\mathcal S_n}\frac{ c^2_a(s)}{n_a(s)}\sigma^2(Y,a,s)=\sum_{s\in\mathcal S_n}\frac{n(s)\sigma^2(Y,a,s)}{n^2},\\
    \hat\tau_a^2&=\sum_{s\in\mathcal S_n}\frac{ c^2_a(s)}{n_a(s)}\hat\sigma^2(Y,a,s)=\sum_{s\in\mathcal S_n}\frac{n(s)\hat\sigma^2(Y,a,s)}{n^2},\\
    w^2_a(s)&=\frac{c^2_a(s)\sigma^2(Y,a,s)}{n_a(s)\tau_a^2}=n(s)\sigma^2(Y,a,s)\left\{\sum_{s\in\mathcal S_n} n(s)\sigma^2(Y,a,s)\right\}^{-1},
\end{align*} we verify the conditions of Lemma~\ref{lem:relax} similarly as Lemma~\ref{crly:d-i-m} and Lemma~\ref{crly:s d-i-m}.
\begin{align*}
    &\sup_{s\in\mathcal S_n}\frac{w^2_a(s)}{n_a(s)}\\
    =&\sup_{s\in\mathcal S_n}\frac{n(s)\sigma^2(Y,a,s)}{n_a(s)\sum_{s\in\mathcal S_n} n(s)\sigma^2(Y,a,s)}\\
    \leq&\sup_{s\in\mathcal S_n}\left\{\left|\frac{n(s)}{n_a(s)}-\frac{1}{\pi_{na}(s)}\right|\frac{\sigma^2(Y,a,s)}{\sum_{s\in\mathcal S_n} n(s)\sigma^2(Y,a,s)}\right\}
    +\sup_{s\in\mathcal S_n}\left\{\frac{1}{\pi_{na}(s)}\frac{\sigma^2(Y,a,s)}{\sum_{s\in\mathcal S_n} n(s)\sigma^2(Y,a,s)}\right\}\\
    \leq&\sup_{s\in\mathcal S_n}\left|\frac{n(s)}{n_a(s)}-\frac{1}{\pi_{na}(s)}\right|\sup_{s\in\mathcal S_n}\left\{\frac{\sigma^2(Y,a,s)}{\sum_{s\in\mathcal S_n} n(s)\sigma^2(Y,a,s)}\right\}
    +\sup_{s\in\mathcal S_n}\frac{1}{\pi_{na}(s)}\sup_{s\in\mathcal S_n}\left\{\frac{\sigma^2(Y,a,s)}{\sum_{s\in\mathcal S_n} n(s)\sigma^2(Y,a,s)}\right\}\\
\xrightarrow{P}&0.
\end{align*}

So we get the conditional convergence in probability \[
\pr\left\{\left|\sum_{s\in\mathcal S_n}\frac{n(s)}{n} \hat\sigma^2(Y,a,s) - \sum_{s\in\mathcal S_n}\frac{n(s)}{n}  \sigma^2(Y,a,s)\right|>\varepsilon|(S^{(n)},A^{(n)})\right\}\xrightarrow{P} 0
\]
for any $\v>0$. Then Lemma~\ref{lem:uncond-conv-in-prob} implies the unconditional convergence in probability
\[
\sum_{s\in\mathcal S_n}\frac{n(s)}{n} \hat\sigma^2(Y,a,s) - \sum_{s\in\mathcal S_n}\frac{n(s)}{n}  \sigma^2(Y,a,s)\xrightarrow{P}0.
\]
Furthermore we have
\begin{align*}
    \sum_{s\in\mathcal S_n}\frac{n(s)}{n}  \sigma^2(Y,a,s)=\frac{1}{n}\sum_{i=1}^n  \sigma^2(Y,a,S_{ni})\xrightarrow{P} E\{\text{var}(Y_i(a)\mid S_{ni})\},
\end{align*}
where the convergence follows from WLLN for triangular arrays (Lemma~\ref{lem:wlln}). The WLLN holds because of the boundedness of $\sup_{s\in\mathcal S_n}{\sigma^2(Y,a,s)}$.

Next, we will show that $\sum_{s\in\mathcal S_n}\{n(s)/n\}\hat\mu(Y,0,s)\hat\mu(Y,1,s)\xrightarrow{P}E[E\{Y_i(1)\mid S_{ni}\}E\{Y_i(0)\mid S_{ni}\}]$ with analysis of conditional expectation and conditional variance.

We analyze the conditional variance and prove it tending to zero.
\begin{align*}
    &\mathrm{var}\left\{\sum_{s\in\mathcal S_n}\frac{n(s)}{n}\hat\mu(Y,0,s)\hat\mu(Y,1,s)\mid(S^{(n)},A^{(n)})\right\}\\
    =&\sum_{s\in\mathcal S_n}\frac{n^2(s)}{n^2}\mathrm{var}\left\{\hat\mu(Y,0,s)\hat\mu(Y,1,s)\mid(S^{(n)},A^{(n)})\right\}\\
    =&\sum_{s\in\mathcal S_n}\frac{n^2(s)}{n^2}\left(E\left\{\hat\mu^2(Y,0,s)\hat\mu^2(Y,1,s)\mid(S^{(n)},A^{(n)})\right\}-\left[E\left\{\hat\mu(Y,0,s)\hat\mu(Y,1,s)\mid(S^{(n)},A^{(n)})\right\}\right]^2\right)\\
    =&\sum_{s\in\mathcal S_n}\frac{n^2(s)}{n^2}\Bigg(E\left\{\hat\mu^2(Y,0,s)\mid(S^{(n)},A^{(n)})\right\}E\left\{\hat\mu^2(Y,1,s)\mid(S^{(n)},A^{(n)})\right\}\\
    &\quad\quad\quad-\left[E\left\{\hat\mu(Y,0,s)\mid(S^{(n)},A^{(n)})\right\}\right]^2\left[E\left\{\hat\mu(Y,1,s)\mid(S^{(n)},A^{(n)})\right\}\right]^2\Bigg)\\
    =&\sum_{s\in\mathcal S_n}\frac{n^2(s)}{n^2}\Bigg[E\left\{\hat\mu^2(Y,0,s)\mid(S^{(n)},A^{(n)})\right\}E\left\{\hat\mu^2(Y,1,s)\mid(S^{(n)},A^{(n)})\right\}-\mu^2(Y,0,s)\mu^2(Y,1,s)\Bigg].\\
\end{align*}
From Lemma~\ref{lem:var_sqr_mean} we have
$E\left\{\hat\mu^2(Y,a,s)\mid(S^{(n)},A^{(n)})\right\}=\mu^2(Y,a,s)+\frac{\sigma^2(Y,a,s)}{n_a(s)}$, then plug in,

    \begin{align*}
    &\mathrm{var}\left\{\sum_{s\in\mathcal S_n}\frac{n(s)}{n}\hat\mu(Y,0,s)\hat\mu(Y,1,s)\mid(S^{(n)},A^{(n)})\right\}\\
    =&\sum_{s\in\mathcal S_n}\frac{n^2(s)}{n^2}\left\{\frac{\mu^2(Y,1,s)\sigma^2(Y,0,s)}{n_0(s)}+\frac{\mu^2(Y,0,s)\sigma^2(Y,1,s)}{n_1(s)}+\frac{\sigma^2(Y,0,s)\sigma^2(Y,1,s)}{n_0(s)n_1(s)}\right\}
\\
    \leq&\mathrm{Const}\cdot
    \sum_{s\in\mathcal S_n}\frac{n^2(s)}{n^2}\frac{n_1(s)+n_0(s)}{n_1(s)n_0(s)}\\
    =&\mathrm{Const}\cdot
    \frac{1}{n}\left[\sum_{s\in\mathcal S_n}\frac{n(s)}{n}\left\{\frac{1}{\pi_{n0}(s)}+\frac{1}{\pi_{n1}(s)}\right\}+o_P(1)\right]\\
    =&O\left(1/n\right)+O\left(1/n\right)\cdot o_P(1)= O_P(1/n),
\end{align*}
where $\mathrm{Const}$ is some constant only depending on $\sup_{a,s}\mu(Y,a,s)$ and $\sup_{a,s}\sigma^2(Y,a,s)$.

Then by conditional Markov inequality (Lemma~\ref{lem:cond_markov}) we have the conditional convergence in probability, conditional on $(S^{(n)},A^{(n)})$, $\sum_{s\in\mathcal S_n}\{n(s)/n\}\hat\mu(Y,0,s)\hat\mu(Y,1,s)$ converges in probability to its conditional expectation $E\left\{\sum_{s\in\mathcal S_n}\frac{n(s)}{n}\hat\mu(Y,0,s)\hat\mu(Y,1,s)\mid (S^{(n)},A^{(n)})\right\}$. By Lemma~\ref{lem:uncond-conv-in-prob}, we furthermore have the unconditional convergence in probability.

The conditional expectation is as follows:
\begin{align*}
    &E\left\{\sum_{s\in\mathcal S_n}\frac{n(s)}{n}\hat\mu(Y,0,s)\hat\mu(Y,1,s)\mid (S^{(n)},A^{(n)})\right\}\\
    &=\sum_{s\in\mathcal S_n}\frac{n(s)}{n}E\{\hat\mu(Y,0,s)\hat\mu(Y,1,s)\mid(S^{(n)},A^{(n)})\}\\
    &=\sum_{s\in\mathcal S_n}\frac{n(s)}{n}E\{\hat\mu(Y,0,s)\mid(S^{(n)},A^{(n)})\}E\{\hat\mu(Y,1,s)\mid(S^{(n)},A^{(n)})\}\\
    &=\sum_{s\in\mathcal S_n}\frac{n(s)}{n}\mu(Y,0,s)\mu(Y,1,s)\\
    &=\frac{1}{n}\sum_{i=1}^n\mu(Y,0,S_{ni})\mu(Y,1,S_{ni})\\
    &\xrightarrow{P}E[E\{Y_i(1)\mid S_{ni}\}E\{Y_i(0)\mid S_{ni}\}]<\infty,
\end{align*}
where the convergence follows from weak law of large numbers (WLLN) for triangular arrays (Lemma~\ref{lem:wlln}). Since $\sup_{s\in\mathcal S_n}\mid\mu(Y,a,s)\mid<\infty$, the conditional expectation is bounded by $\sup_{s\in\mathcal S_n}\mid\mu(Y,0,s)\mid\sup_{s\in\mathcal S_n}\mid\mu(Y,1,s)\mid<\infty$.

Hence we obtain that
\[\sum_{s\in\mathcal S_n}\{n(s)/n\}\hat\mu(Y,0,s)\hat\mu(Y,1,s)\xrightarrow{P}E[E\{Y_i(1)\mid S_{ni}\}E\{Y_i(0)\mid S_{ni}\}].\]

The remaining part is to estimate $E\{Y_i^2(a)\}$ and $E\{Y_i(a)\}$ with $\sum_{s}\{n(s)/n\}\hat\mu_2(Y,a,s)$ and $\sum_{s}\{n(s)/n\}\hat\mu(Y,a,s)$, respectively. We can use the similar conditional convergence in probability argument to prove and hence is omitted. The consistency of $\sum_{s}\{n(s)/n\}\hat\mu_2(Y,a,s)$ and $\sum_{s}\{n(s)/n\}\hat\mu(Y,a,s)$ needs $\sup_{n,s} E\{Y_i^4(a)\mid S_{ni}=s\}<\infty$ and $\sup_{n,s} E\{Y_i^2(a)\mid S_{ni}=s\}<\infty$, respectively.
\end{proof}

\subsection{Proof of Theorem~~\ref{thm:est2}}

\begin{proof}
We state some useful results.
\begin{align*}
    \hat r_{ni}(a)&=Y_i(a)-X_i^\T\hat\beta_n\\
    &=Y_i(a)-X_i^\T\beta_n+X_i^\T(\beta_n-\hat\beta_n)\\
    &=r_{ni}(a)+X_i^\T (\beta_n-\hat\beta_n),\\
    \hat\mu(\hat r,a,s)&=\frac{1}{n_a(s)}\Sigma_{i\in[a,s]}\hat r_{ni}(a)\\
    &=\frac{1}{n_a(s)}\Sigma_{i\in[a,s]}\{ r_{ni}(a)+\hat r_{ni}(a)-r_{ni}(a)\}\\
    &=\hat\mu(r,a,s)+\frac{1}{n_a(s)}\Sigma_{i\in[a,s]}X_i^\T(\beta_n-\hat\beta_n)\\
    &=\hat\mu(r,a,s)+\hat\mu(X,a,s)^\T (\beta_n-\hat\beta_n),\\
    \hat r_{ni}(a)-\hat\mu(\hat r,a,s)&= \{r_{ni}(a)-\hat\mu(r,a,s)\}+\{\hat r_{ni}(a)-r_{ni}(a)\} - \{\hat\mu(\hat r,a,s)-\hat\mu(r,a,s)\}\\
    &=\{r_{ni}(a)-\hat\mu(r,a,s)\}+\{X_i-\hat\mu(X,a,s)\}^\T(\beta_n-\hat\beta_n).
\end{align*}
Recall that
\begin{align*}
    \hat V_{nB}(\hat r)&=\sum_{s\in\mathcal S_n} \frac{n(s)}{n}\sum_{a\in\{0,1\}}\{\hat \mu_2(\hat r,a,s) - \hat\sigma^2(\hat r,a,s)\}-2\sum_{s\in\mathcal S_n}\frac{n(s)}{n}\left\{\hat\mu(\hat r,0,s)\hat\mu(\hat r,1,s)\right\}- \hat\tau_{adj}^2,
\end{align*}
and
\begin{align*}
\hat V_{nW}(\hat r,a)=\sum_{s\in\mathcal S_n}\frac{n(s)}{n}\left\{\frac{n(s)}{n_a(s)}\hat\sigma^2(\hat r,a,s)\right\}.
\end{align*}
We will prove the consistency of the following four parts: $\sum_{s\in\mathcal S_n} n(s)/n\cdot\{\hat\mu(\hat r,1,s)-\hat\mu(\hat r,0,s)\}$, $\sum_{s\in\mathcal S_n}n(s)/n \cdot \hat \mu_2(\hat r,a,s)$, $\sum_{s\in\mathcal S_n}\{n(s)/n\}/\{n(s)/n_a(s)\}\cdot\hat \sigma^2(\hat r,a,s)$ and $\sum_{s\in\mathcal S_n}n(s)/n\cdot\hat \mu(\hat r,0,s)\hat \mu(\hat r,1,s)$. More specifically, we will prove that all of above $\hat r$-version parts converges to $r$-version parts, respectively.

First,
\begin{align*}
    \sum_{s\in\mathcal S_n}\{\hat\mu(\hat r,1,s)-\hat\mu(\hat r,0,s)\}
    =\sum_{s\in\mathcal S_n}\{\hat\mu(r,1,s)-\hat\mu(r,0,s)\}+\sum_{s\in\mathcal S_n}\{\hat\mu(X,1,s)-\hat\mu(X,0,s)\}^\T(\beta_n-\hat\beta_n).
\end{align*}
Because $X_i(0)=X_i(1)$, with similar arguments (or conditional convergence in probability) as the analysis of $\sum_{s\in\mathcal S_n}\frac{n(s)}{n}\{\hat\mu(Y,1,s)-\hat\mu(Y,0,s)\}$, $\sum_{s\in\mathcal S_n}\frac{n(s)}{n}\{\hat\mu(X,1,s)-\hat\mu(X,0,s)\}=O_P(n^{-1/2})$. Additional on $\hat\beta_n-\beta_n\xrightarrow{P}0$, we have $\sum_{s\in\mathcal S_n}\frac{n(s)}{n}\{\hat\mu(X,1,s)-\hat\mu(X,0,s)\}^\T(\beta_n-\hat\beta_n)=o_P(1)$ and
then
\[
\sum_{s\in\mathcal S_n}\{\hat\mu(\hat r,1,s)-\hat\mu(\hat r,0,s)\}-\sum_{s\in\mathcal S_n}\{\hat\mu(r,1,s)-\hat\mu(r,0,s)\}\xrightarrow{P}0.
\]

Second,
\begin{align*}
    \sum_{s\in\mathcal S_n}\frac{n(s)}{n}\hat \mu_2(\hat r,a,s)&=\sum_{s\in\mathcal S_n}\frac{n(s)}{n}\frac{1}{n_a(s)}\sum_{i\in[a,s]}\hat r^2_{ni}(a)\\
    &=\sum_{s\in\mathcal S_n}\frac{n(s)}{n}\frac{1}{n_a(s)}\sum_{i\in[a,s]}\left\{r_{ni}(a)+X_i^\T(\hat\beta_n-\beta_n)\right\}^2\\
    &=\sum_{s\in\mathcal S_n}\frac{n(s)}{n}\frac{1}{n_a(s)}\sum_{i\in[a,s]}\Big\{r^2_{ni}(a) + (\hat\beta_n-\beta_n)^\T X_i X_i^\T(\hat\beta_n-\beta_n)+ 2r_{ni}(a)X_i^\T(\hat\beta_n-\beta_n)\Big\}\\
    :&=A_2+B_2+C_2.
\end{align*}
Note that \begin{align*}
    A_2=&\sum_{s\in\mathcal S_n}\frac{n(s)}{n}\frac{1}{n_a(s)}\sum_{i\in[a,s]}r^2_{ni}(a)\\
    =&\sum_{s\in\mathcal S_n}\frac{n(s)}{n}\hat \mu_2(r,a,s),
\end{align*} which is the term we want. Hence only focus on the term $B_2$ and $C_2$.
\begin{align*}
    B_2=&(\hat\beta_n-\beta_n)^\T\sum_{s\in\mathcal S_n}\frac{n(s)}{n}\frac{1}{n_a(s)}\sum_{i\in[a,s]}( X_i X_i^\T)(\hat\beta_n-\beta_n)\\
    =&(\hat\beta_n-\beta_n)^\T\sum_{s\in\mathcal S_n}\frac{n(s)}{n}\hat\mu(XX^\T,a,s)(\hat\beta_n-\beta_n).
\end{align*}
Because $X_i$ has uniformly bounded fourth moment within each stratum, we have $\sup_{n,s}E(X_{ij}^2X_{ik}^2\mid S_{ni}=s)<\infty$ for all $j,k$. Then with the argument of conditional convergence in probability in the proof of Theorem~\ref{thm:var2}, we have that
\[
\sum_{s\in\mathcal S_n}\frac{n(s)}{n}\hat\mu(XX^\T,a,s)\xrightarrow{P}E(XX^\T)=O(1).
\]
With $\hat\beta_n-\beta_n\xrightarrow{P}0$, we have
\[
B_2=o_P(1).
\]
For $C_2$, we use Cauchy-Schwarz inequality,
\begin{align*}
    C_2=&2\sum_{s\in\mathcal S_n}\sum_{i\in[a,s]}\left\{\frac{n(s)}{n}\frac{1}{n_a(s)}r_{ni}(a)X_i^\T(\hat\beta_n-\beta_n)\right\}\\
    \leq&2\sqrt{\sum_{s\in\mathcal S_n}\sum_{i\in[a,s]}\left\{\frac{n(s)}{n}\frac{1}{n_a(s)}r_{ni}^2(a)\right\}\sum_{s\in\mathcal S_n}\sum_{i\in[a,s]}\left[\frac{n(s)}{n}\frac{1}{n_a(s)}\{X_i^\T(\hat\beta_n-\beta_n)\}^2\right]}\\
    =&2\sqrt{A_2B_2}\\
    =&2\sqrt{O_P(1)\cdot o_P(1)}\\
    =&o_P(1).
\end{align*}
Hence \[\sum_{s\in\mathcal S_n}\frac{n(s)}{n}\hat \mu_2(\hat r,a,s)-\sum_{s\in\mathcal S_n}\frac{n(s)}{n}\hat \mu_2(r,a,s)\xrightarrow{P}0.\]

Third,
\begin{align*}
    \sum_{s\in\mathcal S_n}\frac{n(s)}{n}\hat\sigma^2(\hat r,a,s)&=\sum_{s\in\mathcal S_n}\frac{n(s)}{n}\frac{n(s)}{n_a(s)}\frac{1}{n_a(s)-1}\sum_{i\in[a,s]}\left\{\hat r_{ni}(a)-\hat\mu(\hat r,a,s)\right\}^2\\
    &=\sum_{s\in\mathcal S_n}\frac{n(s)}{n}\frac{n(s)}{n_a(s)}\frac{1}{n_a(s)-1}\sum_{i\in[a,s]}\left[r_{ni}(a)-\hat\mu(r,a,s)+\{X_i-\hat\mu(X,a,s)\}^\T(\beta_n-\hat\beta_n)\right]^2\\
    &=\sum_{s\in\mathcal S_n}\frac{n(s)}{n}\frac{n(s)}{n_a(s)}\frac{1}{n_a(s)-1}\sum_{i\in[a,s]}\Big[\{r_{ni}(a)-\hat\mu(r,a,s)\}^2\\
    &\quad+(\beta_n-\hat\beta_n)^\T\{X_i-\hat\mu(X,a,s)\}\{X_i-\hat\mu(X,a,s)\}^\T(\beta_n-\hat\beta_n)\\
    &\quad+2\{r_{ni}(a)-\hat\mu(r,a,s)\}\{X_i-\hat\mu(X,a,s)\}^\T(\beta_n-\hat\beta_n)\Big]\\
    :&=A_3+B_3+C_3.
\end{align*}
The following proof is quite similar as the second part. Notice that $A_3=\sum_{s\in\mathcal S_n}\frac{n(s)}{n}\frac{n(s)}{n_a(s)}\hat\sigma^2(\hat r,a,s)$ and we focus on \begin{align*}B_3&=(\beta_n-\hat\beta_n)^\T\sum_{s\in\mathcal S_n}\frac{n(s)}{n}\frac{n(s)}{n_a(s)}\frac{1}{n_a(s)-1}\sum_{i\in[a,s]}\{X_i-\hat\mu(X,a,s)\}\{X_i-\hat\mu(X,a,s)\}^\T(\beta_n-\hat\beta_n)\\
:&=(\beta_n-\hat\beta_n)^\T D_3 (\beta_n-\hat\beta_n).\end{align*}
We have $\hat\beta_n-\beta_n\xrightarrow{P}0$ and hence only need to prove the middle term $D_3$ is $O_P(1)$. We show it with conditional Markov inequality.
\begin{align*}
    &E\left[\sum_{s\in\mathcal S_n}\frac{n(s)}{n}\frac{n(s)}{n_a(s)}\frac{1}{n_a(s)-1}\sum_{i\in[a,s]}\{X_i-\hat\mu(X,a,s)\}\{X_i-\hat\mu(X,a,s)\}^\T\mid (S^{(n)},A^{(n)}\right]\\
    =&\sum_{s\in\mathcal S_n}\frac{n(s)}{n}\frac{n(s)}{n_a(s)}E\left[\frac{1}{n_a(s)-1}\sum_{i\in[a,s]}\{X_i-\hat\mu(X,a,s)\}\{X_i-\hat\mu(X,a,s)\}^\T\mid (S^{(n)},A^{(n)})\right]\\
    =&\sum_{s\in\mathcal S_n}\frac{n(s)}{n}\frac{n(s)}{n_a(s)}\mathrm{var}(X_i\mid S_{ni}=s)\\
    =&\frac{1}{n}\sum_{i=1}^n\frac{n(s)}{n_a(s)}\mathrm{var}(X_i\mid S_{ni})\\
    =&\frac{1}{n}\left\{\sum_{i=1}^n\frac{1}{\pi_{na}(S_{ni})}\mathrm{var}(X_i\mid S_{ni})+\sup_{n,s}\mathrm{var}(X_i\mid S_{ni})\cdot o_P(1)\right\}\\
    \xrightarrow{P}&E\left\{\frac{\mathrm{var}(X_i\mid S_{ni})}{\pi_{na}(S_{ni})}\right\}=O(1),
\end{align*}
where the convergence in probability is from  WLLN  for triangular arrays (Lemma~\ref{lem:wlln}) and uniform boundedness of $\mathrm{var}(X_i\mid S_{ni}=s)$.
\begin{align*}
    &\left|\mathrm{var}\left\{D_3\mid (S^{(n)},A^{(n)})\right\}\right|\\
    =&\sum_{s\in\mathcal S_n}\frac{n^2(s)}{n^2}\frac{n^2(s)}{n^2_a(s)}\left|\mathrm{var}\left[\frac{1}{n_a(s)-1}\sum_{i\in[a,s]}\{X_i-\hat\mu(X,a,s)\}\{X_i-\hat\mu(X,a,s)\}^\T\mid (S^{(n)},A^{(n)})\right]\right|\\
    \leq&\frac{1}{n}\sum_{s\in\mathcal S_n}\frac{n^2(s)}{n}\frac{n^2(s)}{n^2_a(s)}\frac{\mathrm{Const}}{n_a(s)}\\
    =&\frac{1}{n}\left\{\sum_{s\in\mathcal S_n}\frac{n(s)}{n}\frac{\mathrm{Const}}{\pi^3_{na}(s)}+o_P(1)\right\}\\
    =&o_P(1/n),
\end{align*}
where $\mathrm{Const}$ only depends on $\sup_{n,s,k} E(X_{ik}^4\mid S_{ni}=s)$ and the inequality is from Lemma~\ref{lem:var_var} and the uniform boundedness of $E(X^4_{ik}\mid S_{ni}=s)$.

Then by conditional Markov inequality, for any $\varepsilon>0$,
\begin{align*}
    &\pr\left[\mid D_3-E\left\{D_3\mid(S^{(n)},A^{(n)})\right\}\mid>\varepsilon\mid(S^{(n)},A^{(n)})\right]\\
    \leq&\frac{\mathrm{var}\left\{D_3\mid (S^{(n)},A^{(n)})\right\}}{\varepsilon^2}\\
    \xrightarrow{P}&0.
\end{align*}
Furthermore by the dominated convergence theorem, 
\begin{align*}
    &\pr\left[\mid D_3-E\left\{D_3\mid(S^{(n)},A^{(n)})\right\}\mid>\varepsilon\right]\\
    =&E\left(\pr\left[\mid D_3-E\left\{D_3\mid(S^{(n)},A^{(n)})\right\}\mid>\varepsilon\mid(S^{(n)},A^{(n)})\right]\right)\\
    \to&E(0)=0.
\end{align*}

Hence $D_3=O_P(1)$ and $B_3=o_P(1)$.

For $C_3$, we again use Cauchy-Schwarz inequality,
\[|C_3|\leq\sqrt{A_3 B_3}=o_P(1).\]
Hence \[
\sum_{s\in\mathcal S_n}\frac{n(s)}{n}\frac{n(s)}{n_a(s)}\hat\sigma^2(\hat r,a,s)-\sum_{s\in\mathcal S_n}\frac{n(s)}{n}\frac{n(s)}{n_a(s)}\hat\sigma^2(r,a,s)\xrightarrow{P}0.
\]
Fourth,
\begin{align*}
    &\sum_{s\in\mathcal S_n}\frac{n(s)}{n}\hat\mu(\hat r,0,s)\hat\mu(\hat r,1,s)\\
    =&\sum_{s\in\mathcal S_n}\frac{n(s)}{n}\frac{1}{n_0(s)}\sum_{i\in[0,s]}\hat r_{ni}(0)\frac{1}{n_1(s)}\sum_{j\in[1,s]}\hat r_{nj}(1)\\
    =&\sum_{s\in\mathcal S_n}\frac{n(s)}{n}\frac{1}{n_0(s)}\sum_{i\in[0,s]}\left\{r_{ni}(0)+(\beta_n-\hat\beta_n)^\T X_i\right\}\frac{1}{n_1(s)}\sum_{j\in[1,s]}\left\{r_{nj}(1)+(\beta_n-\hat\beta_n)^\T X_j\right\}\\
    =&\sum_{s\in\mathcal S_n}\frac{n(s)}{n}\frac{1}{n_0(s)}\frac{1}{n_1(s)}\sum_{i\in[0,s]}\sum_{j\in[1,s]}\Big\{r_{ni}(0)r_{nj}(1)+r_{ni}(0)(\beta_n-\hat\beta_n)^\T X_j\\
    &\quad+r_{nj}(1)(\beta_n-\hat\beta_n)^\T X_i+(\beta_n-\hat\beta_n)^\T X_iX_j^\T(\beta_n-\hat\beta_n)\Big\}\\
    :=&A_4+B_4+C_4+D_4.
\end{align*}
Note that $A_4=\sum_{s\in\mathcal S_n}\frac{n(s)}{n}\hat\mu(r,0,s)\hat\mu(r,1,s)$ and we will prove the remaining 3 terms $B_4,C_4$ and $D_4$ are all $o_P(1)$. Recall that $r_{ni}(a)=Y_i(a)-X_i^\T \beta_n$ and $\sup_{n,s}E\{Y^2_i(a)\mid S_{ni}=s\}<\infty$, $\sup_{n,s}E\{X^2_{ik}(a)\mid S_{ni}=s\}<\infty$ for all $k$. Hence as the regression residual, $\sup_{n,s}E\{r^2_{ni}(a)\mid S_{ni}=s\}<\infty$. 

\begin{align*}
B_4=&\sum_{s\in\mathcal S_n}\frac{n(s)}{n}\frac{1}{n_0(s)}\frac{1}{n_1(s)}\sum_{i\in[0,s]}\sum_{j\in[1,s]}\left\{r_{ni}(0)X_j^\T \right\}(\hat\beta_n-\beta_n)\\
:=&E_4^\T (\hat\beta_n-\beta_n).
\end{align*}
With conditional independence and no overlap between $[0,s]$ and $[1,s]$, we use triangular-array WLLN,
\begin{align*}
    &E\left\{E_4\mid(S^{(n)},A^{(n)})\right\}\\
    =&\sum_{s\in\mathcal S_n}\frac{n(s)}{n}\frac{1}{n_0(s)}\frac{1}{n_1(s)}\sum_{i\in[0,s]}\sum_{j\in[1,s]}E\left\{r_{ni}(0)X_j \mid(S^{(n)},A^{(n)})\right\}\\
    =&\sum_{s\in\mathcal S_n}\frac{n(s)}{n}E\left\{r_{ni}(0)X_j\mid S_{ni}=S_{nj}=s,i\ne j\right\}\\
    =&\sum_{s\in\mathcal S_n}\frac{n(s)}{n}E\{r_{ni}(0)\mid S_{ni}=s\}E(X_j\mid S_{nj}=s)\\
    =&\sum_{i=1}^n\frac{1}{n}E\{r_{ni}(0)\mid S_{ni}\}E(X_i\mid S_{ni})\\
    \xrightarrow{P}&E[E\{r_{ni}(0)\mid S_{ni}\}E(X_i\mid S_{ni})],
\end{align*}
where the condition of WLLN is guaranteed because $\sup_{n,s}E\{|r_{ni}(0)|\mid S_{ni}=s\}<\infty$ and $\sup_{n,s}E(|X_i|\mid S_{ni}=s)<\infty$.

By the symmetry between $a=0$ and $a=1$, $C_4=B_4$.

For $D_4$,
\begin{align*}
    D_4=&(\beta_n-\hat\beta_n)^\T\sum_{s\in\mathcal S_n}\frac{n(s)}{n}\frac{1}{n_0(s)}\frac{1}{n_1(s)}\sum_{i\in[0,s]}\sum_{j\in[1,s]}X_i X_j^\T(\beta_n-\hat\beta_n)
\\
:=&(\beta_n-\hat\beta_n)^\T F_4 (\beta_n-\hat\beta_n).
\end{align*}
We also use conditional Markov inequality and triangular-array WLLN for $F_4$.
\begin{align*}
    &E\left\{F_4\mid(S^{(n)},A^{(n)})\right\}\\
    =&\sum_{s\in\mathcal S_n}\frac{n(s)}{n}\frac{1}{n_0(s)}\frac{1}{n_1(s)}\sum_{i\in[0,s]}\sum_{j\in[1,s]}
    E\left\{X_i X_j^\T\mid(S^{(n)},A^{(n)})\right\}\\
    =&\sum_{s\in\mathcal S_n}\frac{n(s)}{n}\frac{1}{n_0(s)}\frac{1}{n_1(s)}\sum_{i\in[0,s]}\sum_{j\in[1,s]}
    E(X_i X_j^\T\mid S_{ni}=S_{nj}=s,i\ne j )\\
    =&\sum_{s\in\mathcal S_n}\frac{n(s)}{n}
    E(X_i\mid S_{ni}=s )E(X_j^\T\mid S_{nj}=s )\\
    =&\sum_{i=1}^n\frac{1}{n}
    E(X_i\mid S_{ni}=s )E(X_i^\T\mid S_{ni}=s )\\
    \xrightarrow{P}&E\{E(X_i\mid S_{ni})E(X_i\mid S_{ni})^\T\}.
\end{align*}
Conditions of WLLN are guaranteed because $\sup_{n,s}E\{X^2_{ik}\mid S_{ni}=s\}<\infty$ for all $k$.

Then consider the $(k,l)$ element of $F_4$. Similar as $\mathrm{var}\left\{\sum_{s\in\mathcal S_n}\frac{n(s)}{n}\hat\mu(Y,0,s)\hat\mu(Y,1,s)\mid\allowbreak(S^{(n)},A^{(n)})\right\}$ in the proof of Theorem~\ref{thm:est},
\begin{align*}
    &\mathrm{var}\left\{F_{4,kl}\mid(S^{(n)},A^{(n)})\right\}\\
    =&\mathrm{var}\left\{\sum_{s\in\mathcal S_n}\frac{n(s)}{n}\frac{1}{n_0(s)}\frac{1}{n_1(s)}\sum_{i\in[0,s]}\sum_{j\in[1,s]}
    X_{ik} X_{jl}\mid(S^{(n)},A^{(n)})\right\}\\
    =&\sum_{s\in\mathcal S_n}\frac{n^2(s)}{n^2}\Bigg[E\left\{\hat\mu^2(X_{ik},0,s)\mid(S^{(n)},A^{(n)})\right\}E\left\{\hat\mu^2(X_{jl},1,s)\mid(S^{(n)},A^{(n)})\right\}-\mu^2(X_{ik},0,s)\mu^2(X_{jl},1,s)\Bigg]\\
    =&\sum_{s\in\mathcal S_n}\frac{n^2(s)}{n^2}\Bigg[\left\{\mu^2(X_{ik},0,s)+\frac{\sigma^2(X_{ik},0,s)}{n_0(s)}\right\}\left\{\mu^2(X_{jl},1,s)+\frac{\sigma^2(X_{jl},1,s)}{n_1(s)}\right\}-\mu^2(X_{ik},0,s)\mu^2(X_{jl},1,s)\Bigg]\\
    =&\sum_{s\in\mathcal S_n}\frac{n^2(s)}{n^2}\left\{\frac{\mu^2(X_{jl},1,s)\sigma^2(X_{ik},0,s)}{n_0(s)}+\frac{\mu^2(X_{ik},0,s)\sigma^2(X_{jl},1,s)}{n_1(s)}+\frac{\sigma^2(X_{ik},0,s)\sigma^2(X_{jl},1,s)}{n_0(s)n_1(s)}\right\}
\\
    \leq&\text{Const}\cdot
    \sum_{s\in\mathcal S_n}\frac{n^2(s)}{n^2}\frac{n_1(s)+n_0(s)}{n_1(s)n_0(s)}\\
    =&\text{Const}\cdot
    \frac{1}{n}\left[\sum_{s\in\mathcal S_n}\frac{n(s)}{n}\left\{\frac{1}{\pi_{n0}(s)}+\frac{1}{\pi_{n1}(s)}\right\}+o_P(1)\right]\\
    =&O_P\left(1/n\right),
\end{align*}
where $\mathrm{Const}$ only depends on $\sup_{n,s,k} E(X_{ik}^4\mid S_{ni}=s)$.

Therefore by conditional Markov inequality (Lemma~\ref{lem:cond_markov}) we have the conditional convergence in probability. Then with Lemma~\ref{lem:uncond-conv-in-prob}, we can derive the unconditional convergence in probability which means $F_4=O_P(1)$ and hence $D_4=o_P(1)$.

So we have \[\sum_{s\in\mathcal S_n}\frac{n(s)}{n}\hat\mu(\hat r,0,s)\hat\mu(\hat r,1,s)-\sum_{s\in\mathcal S_n}\frac{n(s)}{n}\hat\mu(r,0,s)\hat\mu(r,1,s)\xrightarrow{P}0.\]

We have proved the  $\hat r-$version variance estimator converges in probability to the $r-$version variance estimator. Finally we need to prove that  the $r-$version variance estimator is consistent.

Applying Theorem~\ref{thm:est} with respect to $r$ requires $\sup_{n,s}E\{r^4_{ni}(a)\mid S_{ni}=s\}<\infty$. This condition is guaranteed because $\sup_{n,s}E\{Y^4_i(a)\mid S_{ni}=s\}<\infty$, $\sup_{n,s}E\{X^4_{ik}(a)\mid S_{ni}=s\}<\infty$ for all $k$ and $\beta_n\to\beta$.
\begin{align*}
    \sup_{n,s}E\{r^4_{ni}(a)\mid S_{ni}=s\}
    &=\sup_{n,s}E\left(\left\{Y_i(a)-X_i^\T\beta_n\right\}^4\mid S_{ni}=s\right)\\
    &=\sup_{n,s}E\left\{\sum_{m=1}^4\binom{4}{m}Y^m_i(a)\left(X_i^\T\beta_n\right)^{4-m}\mid S_{ni}=s\right\}\\
    &\leq\sum_{m=1}^4\binom{4}{m}\sup_{n,s}E\left\{|Y^m_i(a)\left(X_i^\T\beta_n\right)^{4-m}|\mid S_{ni}=s\right\}\\
    &\leq\sum_{m=1}^4\binom{4}{m}\sup_{n,s}E\left\{Y^4_i(a)\mid S_{ni}=s\right\}^{\frac{m}{4}}\sup_{n,s}E\left\{\left(X_i^\T\beta_n\right)^{4}\mid S_{ni}=s\right\}^{\frac{4-m}{4}}\\
    &<\infty.
\end{align*}

We finish the proof.
\end{proof}

\subsection{Proof of Theorem~\ref{thm:est3}}
In the proof of Theorem~\ref{thm:var3}, we have proved that $\hat\beta_n^*-\beta_n^*\xrightarrow{P}0$.

Compared with Theorem~\ref{thm:est2}, the only difference of Theorem~\ref{thm:est3} is to replace $r$ with $r^*$. Recall that $r_{ni}(a)=Y_i(a)-X_i^\T\beta_n$ and $r^*_{ni}(a)=Y_i(a)-X_i^\T\beta^*_n$. Apply Theorem~\ref{thm:est2} and we get the result. 

\section{Auxiliary Results}
\subsection{Stratified sampling conditional on $(S^{(n)},A^{(n)})$}
For any potential outcome $f_{ni}(a)$ with $a\in\{0,1\}$, we denote the weighted mean with the stratum weight $c_a(s)$ as \[\gamma_a=\sum_{s\in\mathcal S_n} c_a(s)\mu(f,a,s).\]
Letting $c_a(s)={n_a(s)}/{n_a}$ or ${n(s)}/{n}$, $\gamma_a$ can be mean or stratified mean, respectively.
The parameter of interest is $\gamma_1-\gamma_0$ with the plug-in estimator $\hat\gamma_1-\hat\gamma_0$, where
\[
\hat\gamma_a=\sum_{s\in\mathcal S_n}c_a(s)\hat\mu(f,a,s).\] Furthermore let the variance of $\hat\gamma_a$ and its estimator be
\[\tau_a^2=\sum_{s\in\mathcal S_n}\frac{c^2_a(s)}{n_a(s)}\sigma^2(f,a,s)\]
and
\[\hat\tau_a^2=\sum_{s\in\mathcal S_n}\frac{c^2_a(s)}{n_a(s)}{\hat\sigma^2(f,a,s)},\]respectively.

Using the decoupling technique in  \cite{bugni2018inference} and \cite{Rafi2023}, we will construct the stratified sampling framework conditional on $(S^{(n)},A^{(n)})$. Independently for each $s \in \mathcal S_n$ and independently of $(S^{(n)}, A^{(n)})$, let $\left\{\left(f_i^{(s)}(1), f_i^{(s)}(0)\right)\right\}_{i=1}^n$ be i.i.d. with marginal distribution equal to the conditional distribution of $\left(f_i(1), f_i(0)\right)\mid S_{ni}=s$. Conditional on $(A^{(n)},S^{(n)})$, we view the estimation of $\gamma_a$ in the stratified sampling framework in \citet{Bickel1984}.

\begin{condition}[Naive Condition]
\label{cond:naive}
    For $a\in\{0,1\},$ $$ \lim_{n\to\infty}\pr\left(\inf_{s\in\mathcal S_n}n_a(s)\geq 2\right)=1$$ and $$\sup_{n\ge1,s\in\mathcal S_n}\sigma^2(f,a,s)<\infty.$$
\end{condition}

\begin{condition}[Lindeberg Condition]
\label{cond:lindeberg}
For $a\in\{0,1\}$ and any $\varepsilon>0$,
\begin{equation}
\label{eq:lind}
\tau_a^{-2} \sum_{s\in\mathcal S_n}n_a(s)^{-1} c_a^2(s)  E\left[\phi^2\Big\{f_i(a)-\mu(f,a,s), \varepsilon n_a(s) \tau_a|c_a(s)|^{-1}\Big\}| S_{ni}=s\right]\xrightarrow{P}0,
\end{equation}
where the function $
\phi(u,\varepsilon)=uI\{|u|\geq \varepsilon\}.$
\end{condition}

Furthermore, we denote stratum-standardized potential outcome as  \[V_i(a)=\frac{f_i(a)-\mu(f,a,S_{ni})}{\sigma(f,a,S_{ni})}\] and stratum variance weight as \[w_a^2(s)=\mathrm{var}\big\{\frac{c_a(s)\hat\mu(f,a,s)}{\tau_a}\big\}=\frac{c^2_a(s)\sigma^2(f,a,s)}{n_a(s)\tau_a^2}.\]
\begin{condition}[Equivalent Lindeberg Condition]
\label{cond:lind_equi}
    For $a\in\{0,1\}$ and any $\varepsilon>0$,
\[
\sum_{s\in\mathcal S_n} E\Big[\phi^2\Big\{w_a(s)V_i(a), \varepsilon \sqrt{n_a(s)}\Big\}\mid S_{ni}=s\Big]\xrightarrow{P} 0.
\]
\end{condition}

\begin{condition}[Sufficient Condition]
\label{cond:lind_suff}
$$\sup_{n\ge1,s\in\mathcal S_n} E[|V_i(a)|^3\mid S_{ni}=s]<\infty$$
and
$$\sup_{s\in\mathcal S_n}{w^2_a(s)}/{n_a(s)}\xrightarrow{P}0$$ for $a\in\{0,1\}$.
\end{condition}

\begin{condition}[Weak Sufficient Condition]
\label{cond:lind_weak}
For some $\v>0$, 
$\sup_{n\ge1,s\in\mathcal S_n} E[|V_i(a)|^{2+\v}\mid S_{ni}=s]<\infty$
and
$\operatorname{sup}_{s\in\mathcal S_n}\frac{w^2_a(s)}{n_a(s)}\xrightarrow{P}0$ for $a\in\{0,1\}$.
\end{condition}

We summarize some of super population results in \cite{Bickel1984} as the following theorem.
\begin{theorem}[Bickel and Freedman, 1984]
\label{thm:1984}
Condition~\ref{cond:lind_suff} can deduce Condition~\ref{cond:lind_equi} and Condition~\ref{cond:lind_equi} is equivalent to Condition~\ref{cond:lindeberg}.

If Condition~\ref{cond:naive} and Condition~\ref{cond:lind_suff} hold, or Condition~\ref{cond:naive} and Condition~\ref{cond:lind_equi} hold, then for $a\in\{0,1\}$, 
\[
\pr\left\{(\hat\gamma_a-\gamma_a)/{\tau_a} \leq u|(S^{(n)},A^{(n)})\right\} \xrightarrow{P} \Phi\left(u\right),
\]
where $\Phi(\cdot)$ is the standard normal cumulative distribution function (c.d.f.) and for any $\varepsilon>0$,
\[
    \pr\{\mid \hat\tau_a/\tau_a -1 \mid\geq\varepsilon\mid (S^{(n)},A^{(n)})\}\xrightarrow{P} 0.
    \]
\end{theorem}
\begin{proof}
    See \cite{Bickel1984} for details.
\end{proof}

We give one lemma to relax Condition~\ref{cond:lind_suff}.
\begin{lemma}
\label{lem:relax} Condition~\ref{cond:lind_weak} deduces Condition~\ref{cond:lind_equi}. Then with Theorem~\ref{thm:1984}, we obtain that if  Condition~\ref{cond:naive} and Condition~\ref{cond:lind_weak} hold, then for $a\in\{0,1\}$, for any $u\in\mathbb R$,
\[
\pr\left\{(\hat\gamma_a-\gamma_a)/{\tau_a} \leq u|(S^{(n)},A^{(n)})\right\} \xrightarrow{P} \Phi\left(u\right),
\]
where $\Phi(\cdot)$ is the standard normal c.d.f. and for any $\varepsilon>0$,
\[
    \pr\{\mid \hat\tau_a/\tau_a -1 \mid\geq\varepsilon\mid (S^{(n)},A^{(n)})\}\xrightarrow{P} 0.
    \]
\end{lemma}
\begin{proof}
    For the former claim, we will prove that Condition~\ref{cond:lind_weak} implies Condition \ref{cond:lind_equi}. The main techniques are Holder's inequality and Markov's inequality.

    Without loss of generality we assume that $E\{V_i(a)\big|S_{ni}=s\}=0$, $E\{V_i^2(a)\big|S_{ni}=s\}=1$, $\sup_{n\ge1,s\in\mathcal S_n}E\{|V_i(a)|^{2+\varepsilon}\big|S_{ni}=s\}\leq M<\infty$ and $\sup_{s\in\mathcal S_n}{w^2_a(s)}/{n_a(s)}\xrightarrow{P}0$.
    
    Without loss of generality, we only consider strata with $w_a(s)>0$. Then we have 
    \begin{align*}
        &\sum_{s\in\mathcal S_n}  E\Big[\phi^2\Big\{w_a(s)V_i(a), \varepsilon \sqrt{n_a(s)}\Big\}\Big| S_{ni}=s\Big]\\
        =&\sum_{s\in\mathcal S_n}  E\Big[w^2_a(s)\phi\Big\{V_i^2(a), \frac{\varepsilon ^2{n_a(s)}}{w^2_a(s)}\Big\}\Big|S_{ni}=s\Big]\\
        =&\sum_{s\in\mathcal S_n} w^2_a(s)  E\Big[V_i^2(a)I\big\{ V_i^2(a)\geq\frac{\varepsilon ^2{n_a(s)}}{w^2_a(s)}\big\}\Big|S_{ni}=s\Big]\\
        \leq&\sum_{s\in\mathcal S_n} w^2_a(s) E\{| V_i(a)|^{2+\varepsilon}\mid S_{ni}=s\}^{\frac{2}{2+\varepsilon}}E\left[I\Big\{ V_i^2(a)\geq\frac{\varepsilon ^2{n_a(s)}}{w^2_a(s)}\Big\}|S_{ni}=s\right]^{\frac{\varepsilon}{2+\varepsilon}}\\
        \leq&\sum_{s\in\mathcal S_n} w^2_a(s) M^{\frac{2}{2+\varepsilon}}\pr\Big\{V_i^2(a)\geq\frac{\varepsilon ^2{n_a(s)}}{w^2_a(s)}\Big|S_{ni}=s\Big\}^\frac{\varepsilon}{2+\varepsilon}\\
        \leq& M^{\frac{2}{2+\varepsilon}}\sum_{s\in\mathcal S_n} w^2_a(s)  \left[\left(\frac{\varepsilon^2{n_a(s)}}{w^2_a(s)}\right)^{-1}{E\{V_i^2(a)|S_{ni}=s\}}\right]^\frac{\varepsilon}{2+\varepsilon}\\
        \leq& \frac{M^{\frac{2}{2+\varepsilon}}}{\varepsilon^\frac{2\varepsilon}{2+\varepsilon}} \left\{\sup_{s\in\mathcal S_n}\frac{w^2_a(s)}{n_a(s)}\right\}^\frac{\varepsilon}{2+\varepsilon}\sum_{s\in\mathcal S_n} w^2_a(s)  E\{V_i^2(a)|S_{ni}=s]\}^\frac{\varepsilon}{2+\varepsilon}\\
        = &\frac{M^{\frac{2}{2+\varepsilon}}}{\varepsilon^\frac{2\varepsilon}{2+\varepsilon}} \left\{\sup_{s\in\mathcal S_n}\frac{w^2_a(s)}{n_a(s)}\right\}^\frac{\varepsilon}{2+\varepsilon}\sum_{s\in\mathcal S_n} w^2_a(s) \\
        = & \frac{M^{\frac{2}{2+\varepsilon}}}{\varepsilon^\frac{2\varepsilon}{2+\varepsilon}}\left\{\sup_{s\in\mathcal S_n}\frac{w^2_a(s)}{n_a(s)}\right\}^\frac{\varepsilon}{2+\varepsilon} \to 0,
    \end{align*}
    where the first inequality uses Holder's inequality, the third inequality is from Markov's inequality and the last equality is from the property of stratum weights.
\end{proof}

Conditional on $(S^{(n)},A^{(n)})$, the independence among both strata and treatment groups will hold and we have lemmas for difference-in-means estimator (Lemma~\ref{crly:d-i-m}) and stratified difference-in-means estimator (Lemma~\ref{crly:s d-i-m}), respectively.
\begin{lemma}[Difference-in-means]
\label{crly:d-i-m}
Let $\hat\gamma_1-\hat\gamma_0$ denote difference-in-means, i.e.,
\begin{align*}
        \hat\gamma_1-\hat\gamma_0=\frac{1}{n_1}\sum_{s\in\mathcal S_n}n_1(s)\hat\mu(f,1,s) - \frac{1}{n_0}\sum_{s\in\mathcal S_n}n_0(s)\hat\mu(f,0,s). 
    \end{align*}
    If Conditions~\ref{cond:naive} and \ref{cond:lind_weak} hold, we have
    $$
\pr\left\{\frac{(\hat\gamma_1-\hat\gamma_0)-(\gamma_1-\gamma_0)}{\sqrt{\tau_0^2+\tau_1^2}} \leq u|(S^{(n)},A^{(n)})\right\} \xrightarrow{P}\Phi\left(u\right),
$$
where $\Phi(\cdot)$ is the standard normal c.d.f.
and
\[
    \pr\{\mid \hat\tau_a/\tau_a -1 \mid\geq\varepsilon\mid (S^{(n)},A^{(n)})\}\xrightarrow{P} 0\ ,
    \]
    for any $\varepsilon>0$ and $a\in\{0,1\}$. 

    Here we have
\begin{align*}
    \hat\gamma_a&=\sum_{s\in\mathcal S_n}\frac{n_a(s)}{n_a}\hat\mu(f,a,s),\\
    c_a(s)&=\frac{n_a(s)}{n_a},\\
    \tau_a^2&=\sum_{s\in\mathcal S_n}\frac{ c^2_a(s)}{n_a(s)}\sigma^2(f,a,s)=\sum_{s\in\mathcal S_n}\frac{n_a(s)\sigma^2(f,a,s)}{n_{a}^2},\\
    \hat\tau_a^2&=\sum_{s\in\mathcal S_n}\frac{ c^2_a(s)}{n_a(s)}\hat\sigma^2(f,a,s)=\sum_{s\in\mathcal S_n}\frac{n_a(s)\hat\sigma^2(f,a,s)}{n_{a}^2},\\
    w^2_a(s)&=\frac{c^2_a(s)\sigma^2(f,a,s)}{n_a(s)\tau_a^2}=n_a(s)\sigma^2(f,a,s)\Big\{\sum_{s\in\mathcal S_n} n_a(s)\sigma^2(f,a,s)\Big\}^{-1}.
\end{align*}
\end{lemma}
\begin{proof}
    $\hat\gamma_1$ and $\hat\gamma_0$ are independent conditional on $(S^{(n)},A^{(n)})$. 
    
We need Condition \ref{cond:naive} and Condition \ref{cond:lind_weak} both hold.
Condition \ref{cond:naive} is intuitive. Condition \ref{cond:lind_weak} requires that for $a\in\{0,1\}$ and some $\varepsilon>0$, \[
\sup_{n\ge1,s\in\mathcal S_n} E\{|V_i(a)|^{2+\v}\mid S_{ni}=s\}<\infty \text{ and } \frac{\sup_{s\in\mathcal S_n}\sigma^2(f,a,s)}{\sum_{s\in\mathcal S_n} n_a(s)\sigma^2(f,a,s)}\xrightarrow{P}0.
\]
Note that
\begin{align*}
    &\frac{1}{n}\sum_{s\in\mathcal S_n} n_a(s)\sigma^2(f,a,s)\\
    =&\frac{1}{n}\sum_{s\in\mathcal S_n} \left\{\frac{n_a(s)}{n(s)}-\pi_{na}(s)\right\}n(s)\sigma^2(f,a,s)+\frac{1}{n}\sum_{s\in\mathcal S_n} n(s)\pi_{na}(s)\sigma^2(f,a,s)\\
    \leq& \frac{1}{n}\sum_{s\in\mathcal S_n} \left|\frac{n_a(s)}{n(s)}-\pi_{na}(s)\right|n(s)\sigma^2(f,a,s)+\sup_{n,s}\pi_{na}(s)\cdot\frac{1}{n}\sum_{s\in\mathcal S_n} n(s)\sigma^2(f,a,s)\\
    \leq& \sup_{s}\left|\frac{n_a(s)}{n(s)}-\pi_{na}(s)\right|\cdot\frac{1}{n}\sum_{i=1}^n \sigma^2(f,a,S_{ni})+\sup_{n,s}\pi_{na}(s)\cdot\frac{1}{n}\sum_{i=1}^n \sigma^2(f,a,S_{ni})\\
    \leq& \sup_{s}\left|\frac{n_a(s)}{n(s)}-\pi_{na}(s)\right|\cdot \sup_{n,s}\{\sigma^2(f,a,S_{ni}=s)\}+\sup_{n,s}\pi_{na}(s)\cdot\frac{1}{n}\sum_{i=1}^n \sigma^2(f,a,S_{ni})\\
    =&o_P(1)+\sup_{n,s}\pi_{na}(s)\cdot\frac{1}{n}\sum_{i=1}^n \sigma^2(f,a,S_{ni})\\
    =&o_P(1)+\sup_{n,s}\pi_{na}(s)\cdot E\{ \sigma^2(f,a,S_{ni})\}+o_P(1),
\end{align*}
where the second-to-last equality is from the uniform convergence of $n_a(s)/n(s)$ and uniform boundedness of $\sup_{n,s}\{\sigma^2(f,a,S_{ni}=s)\}$ and the last equality is from WLLN of triangular arrays (Lemma~\ref{lem:wlln}). Similarly we can derive that \[\frac{1}{n}\sum_{s\in\mathcal S_n} n_a(s)\sigma^2(f,a,s)\geq \inf_{n,s}\pi_{na}(s)\cdot E\{ \sigma^2(f,a,S_{ni})\}+o_P(1),\]
and hence from Assumption (A2) we have
\[
\sum_{s\in\mathcal S_n} n_a(s)\sigma^2(f,a,s)\xrightarrow{P}\infty.
\]

Additional on the assumption that $\sup_{s\in\mathcal S_n}\sigma^2(f,a,s)<\infty$, we have
\[\frac{\sup_{s\in\mathcal S_n}\sigma^2(f,a,s)}{\sum_{s\in\mathcal S_n} n_a(s)\sigma^2(f,a,s)}\xrightarrow{P}0.\]

Hence Condition~\ref{cond:lind_weak} holds. 
Combining with Condition~\ref{cond:naive}, Theorem~\ref{thm:1984} holds.

For $a\in\{0,1\}$, \[
\pr\left\{(\hat\gamma_a-\gamma_a)/{\tau_a} \leq u|(S^{(n)},A^{(n)})\right\} \xrightarrow{P} \Phi\left(u\right),
\]
where $\Phi(\cdot)$ is the standard normal c.d.f.
and
\[
    \pr\{\mid \hat\tau_a/\tau_a -1 \mid\geq\varepsilon\mid (S^{(n)},A^{(n)})\}\xrightarrow{P} 0\ ,
    \]
    for any $\varepsilon>0$.

By Lemma~\ref{lem:lind} and \ref{lem:unif_conv}, we have
\begin{align*}
&E\{e^{it{(\hat\gamma_a-\gamma_a)}/{\sqrt{\tau_0^2+\tau_1^2}}}\mid(S^{(n)},A^{(n)})\}\xrightarrow{P} e^{-\frac{t^2\tau_a^2}{2(\tau_0^2+\tau_1^2)}}.
\end{align*}
Finally with the independence between $\hat\gamma_1$ and $\hat\gamma_0$ conditional on $(S^{(n)},A^{(n)})$ we have
\begin{align*}
&E\{e^{it{(\hat\gamma_1-\hat\gamma_0)-(\gamma_1-\gamma_0)}/{\sqrt{\tau_0^2+\tau_1^2}}}\mid(S^{(n)},A^{(n)})\}\\
=&E\{e^{it{(\hat\gamma_1-\gamma_1)}/{\sqrt{\tau_0^2+\tau_1^2}}}\mid(S^{(n)},A^{(n)})\}\cdot E\{e^{i(-t){(\hat\gamma_0-\gamma_0)}/{\sqrt{\tau_0^2+\tau_1^2}}}\mid(S^{(n)},A^{(n)})\}\\
\xrightarrow{P}& e^{-\frac{t^2\tau_1^2}{2(\tau_0^2+\tau_1^2)}}\cdot e^{-\frac{t^2\tau_0^2}{2(\tau_0^2+\tau_1^2)}}\\
=& e^{-t^2/2}.
\end{align*}
\end{proof}

By Lemma~\ref{lem:lind} we obtain the result.

\begin{lemma}[Stratified difference-in-means]
\label{crly:s d-i-m}
Let $\hat\gamma_1-\hat\gamma_0$ denote stratified difference-in-means, i.e.,
\[
\hat\gamma_1-\hat\gamma_0=\sum_{s\in\mathcal S_n}\frac{n(s)}{n}\big\{\hat\mu(f,1,s)-\hat\mu(f,0,s)\big\}.\]
    If conditions of Conditions~\ref{cond:naive}, \ref{cond:lind_weak} and Assumption~(B2) hold,
    we have 
    $$
\pr\left\{\frac{(\hat\gamma_1-\hat\gamma_0)-(\gamma_1-\gamma_0)}{\sqrt{\tau_0^2+\tau_1^2}} \leq u|(S^{(n)},A^{(n)})\right\} \xrightarrow{P} \Phi\left(u\right),
$$
where $\Phi(\cdot)$ is the standard normal c.d.f.
and
\[
    \pr\{\mid \hat\tau_a/\tau_a -1 \mid\geq\varepsilon\mid (S^{(n)},A^{(n)})\}\xrightarrow{P} 0\ ,
    \]
    for any $\varepsilon>0$ and $a\in\{0,1\}$. 

    Here we have
\begin{align*}
    \hat\gamma_a&=\sum_{s\in\mathcal S_n}\frac{n(s)}{n}\hat\mu(f,a,s),\\
    c_a(s)&=\frac{n(s)}{n},\\
    \tau_a^2&=\sum_{s\in\mathcal S_n}\frac{ c^2_a(s)}{n_a(s)}\sigma^2(f,a,s)=\frac{1}{n^2}\sum_{s\in\mathcal S_n}\frac{n^2(s)\sigma^2(f,a,s)}{n_a(s)},\\
    \hat\tau_a^2&=\sum_{s\in\mathcal S_n}\frac{ c^2_a(s)}{n_a(s)}\hat\sigma^2(f,a,s)=\frac{1}{n^2}\sum_{s\in\mathcal S_n}\frac{n^2(s)\hat\sigma^2(f,a,s)}{n_a(s)},\\
    w^2_a(s)&=\frac{c^2_a(s)\sigma^2(f,a,s)}{n_a(s)\tau_a^2}=\frac{n^2(s)\sigma^2(f,a,s)}{n_a(s)}\Big\{\sum_{s\in\mathcal S_n} \frac{n^2(s)\sigma^2(f,a,s)}{n_a(s)}\Big\}^{-1}.
\end{align*}
\end{lemma}
\begin{proof}
$\hat\gamma_1$ and $\hat\gamma_0$ are independent conditional on $(S^{(n)},A^{(n)})$. 

Condition \ref{cond:naive} is satisfied. Condition \ref{cond:lind_weak} requires that for $a\in\{0,1\}$, \begin{equation}
\label{eq:s-d-in-m1}
\sup_{n\geq1,s\in\mathcal S_n} E[|V_i(a)|^{2+\v}\mid S_{ni}=s]<\infty \text{ and } \sup_{s\in\mathcal S_n}\frac{w^2_a(s)}{n_a(s)}=\frac{\sup_{s\in\mathcal S_n}\frac{n^2(s)}{n_a^2(s)}\sigma^2(f,a,s)}{\sum_{s\in\mathcal S_n} \frac{n^2(s)}{n_a(s)}\sigma^2(f,a,s)}\xrightarrow{P}0.
\end{equation}

Assumption~(B2) implies that $\sup_{s\in\mathcal S_n}|\frac{n^2 (s)}{n_a^2(s)}-\frac{1}{\pi_a^2(s)}|\xrightarrow{P}0$. Then we have
\begin{equation}
\label{eq:s-d-in-m2}
\begin{split}
&\left|\sup_{s\in\mathcal S_n}\frac{n^2(s)}{n_a^2(s)}\sigma^2(f,a,s)-\sup_{s\in\mathcal S_n}\frac{1}{\pi_a^2(s)}\sigma^2(f,a,s)\right|\\
\leq&\sup_{s\in\mathcal S_n}\left|\left\{\frac{n^2 (s)}{n_a^2(s)}-\frac{1}{\pi_a^2(s)}\right\}\sigma^2(f,a,s)\right|\\
\leq&\sup_{s\in\mathcal S_n}\left|\frac{n^2 (s)}{n_a^2(s)}-\frac{1}{\pi_a^2(s)}\right|\cdot\sup_{s\in\mathcal S_n}\sigma^2(f,a,s)\xrightarrow{P}0.
\end{split}
\end{equation}

By the similar argument,
\begin{equation}
\label{eq:s-d-in-m3}
\begin{split}
&\frac{\left|\sum_{s\in\mathcal S_n} \frac{n^2(s)}{n_a(s)}\sigma^2(f,a,s) -
\sum_{s\in\mathcal S_n} \frac{n_a(s)}{\pi_a^2(s)}\sigma^2(f,a,s)\right|}{
\sum_{s\in\mathcal S_n}\frac{n_a(s)}{\pi_a^2(s)}\sigma^2(f,a,s)
}\\
\leq&\frac{\sum_{s\in\mathcal S_n} \left|\frac{n^2(s)}{n_a^2(s)}-\frac{1}{\pi_a^2(s)}\right|n_a(s)\sigma^2(f,a,s)}{
\sum_{s\in\mathcal S_n}\frac{n_a(s)}{\pi_a^2(s)}\sigma^2(f,a,s)}\\
\leq&\sup_{s\in\mathcal S_n}\left| \frac{n^2(s)}{n_a^2(s)}-\frac{1}{\pi_a^2(s)}\right| \frac{\sum_{s\in\mathcal S_n} n_a(s)\sigma^2(f,a,s)}{
\sum_{s\in\mathcal S_n}\frac{n_a(s)}{\pi_a^2(s)}\sigma^2(f,a,s)} \\
\leq&\sup_{s\in\mathcal S_n}\left| \frac{n^2(s)}{n_a^2(s)}-\frac{1}{\pi_a^2(s)}\right| \frac{1}{\inf_{s\in\mathcal S_n}\frac{1}{\pi_a^2(s)}}\\
=&\sup_{s\in\mathcal S_n}\left| \frac{n^2(s)}{n_a^2(s)}-\frac{1}{\pi_a^2(s)}\right| \sup_{s\in\mathcal S_n}\pi_a^2(s)\xrightarrow{P}0.
\end{split}
\end{equation}
With (\ref{eq:s-d-in-m2}) and (\ref{eq:s-d-in-m3}), the former condition in~(\ref{eq:s-d-in-m1}) is equivalent to
\begin{equation}
\label{eq:s-d-in-m4}
\begin{split}
    \sup_{s\in\mathcal S_n}\frac{w^2_a(s)} {n_a(s)}&=\frac{\sup_{s\in\mathcal S_n}\frac{1}{\pi_a^2(s)}\sigma^2(f,a,s)}{\sum_{s\in\mathcal S_n} \frac{n_a(s)}{\pi_a^2(s)}\sigma^2(f,a,s)}\\
    \leq&\Big(\frac{\sup_{s\in\mathcal S_n}\sigma^2(f,a,s)}{\sum_{s\in\mathcal S_n} {n_a(s)}\sigma^2(f,a,s)}\frac{\sup_{s\in\mathcal S_n}\frac{1}{\pi_a^2(s)}}{\inf_{s\in\mathcal S_n}\frac{1}{\pi_a^2(s)}}\Big)\\
    =&\Big(\frac{\sup_{s\in\mathcal S_n}\sigma^2(f,a,s)}{\sum_{s\in\mathcal S_n} {n_a(s)}\sigma^2(f,a,s)}\frac{\sup_{s\in\mathcal S_n}\pi_a^2(s)}{\inf_{s\in\mathcal S_n}\pi_a^2(s)}\Big)\xrightarrow{P}0.
\end{split}
\end{equation}
The rest follows as the proof of Lemma \ref{crly:d-i-m}.

\end{proof}
\subsection{Conditional probability space}
We view the conditional distribution $X\mid Y$ as the two-layer structure $Y(\omega)$ and $(X\mid Y(\omega) )(\omega^*)$, where $\omega$ is in the first random layer $Y$ and $\omega^*$ is in the second random layer $X\mid Y(\omega)$. 

The conditional probability $P(X\leq u\mid Y)$ is equivalent to the conditional expectation $E(I\{X\leq u\}\mid Y)$. For some fixed $\omega$ in the probability space $(\Omega,\mathcal F, P)$, the
conditional probability space $(\Omega^*,\mathcal F^*, P^*)$ is constructed under the conditional c.d.f. $P\{X\leq u\mid Y(\omega)\}$.

\begin{lemma}[Conditional triangular Lindeberg CLT]\label{lem:lind}

For triangular arrays $\{X_{ni}\}_{i=1}^n\in\mathbb R$ and $\sigma-$field $\mathcal F_n$ generated by multi-variate random variable arrays $Y_n\in\mathbb R^{r_n}$, assume that $\{X_{ni}\}_{i=1}^n$ are independent conditional on $Y_n$ and $\mathrm{var}(X_{ni}\mid Y_n)=\sigma_{ni}^2(Y_n)$.
Furthermore define \[
Z_n(Y_n)=\sum_{i=1}^n\{X_{ni}-E(X_{ni}\mid Y_n)\}
\]
and
\[
B_n^2(Y_n)=\sum_{i=1}^n
\sigma_{ni}^2(Y_n).\]
If the conditional Lindeberg Condition is satisfied, which means that 
\[
\frac{1}{B_n^2(Y_n)}\sum_{i=1}^n E\left[\{X_{ni}-E(X_{ni}\mid Y_n)\}^2 I\{\mid X_{ni}-E(X_{ni}\mid Y_n)\mid >\varepsilon B_n(Y_n)\}\mid Y_n\right]\to 0 
\]
a.s. or in probability, then we have 
\[
\sup_{u\in\mathbb R}\mid P\{Z_n(Y_n)/B_n(Y_n)\leq u|Y_n\} - \Phi(u)|\to 0
\]
a.s. or in probability, respectively where $\Phi(\cdot)$ is the standard normal c.d.f..
Equivalently, we will also have that for any $t\in \mathbb R$,
\[
E\left[\exp\{itZ_n(Y_n)/B_n(Y_n)\}\mid Y_n\right]\to\exp(-t^2/2)
\]
a.s. or in probability, respectively.
\end{lemma}
\begin{proof}
We prove almost sure convergence first.

    Denote 
    \begin{align*}
    &\Omega_1=
    \\&\{\omega:\frac{1}{B_n^2(Y_n(\omega))}\sum_{i=1}^n E\left([X_{ni}-E\{X_{ni}\mid Y_n(\omega)\}]^2I\{\mid X_{ni}-E\{X_{ni}\mid Y_n(\omega)\mid >\varepsilon B_n(Y_n(\omega))\}\mid Y_n(\omega)\right)\to 0 \}.
    \end{align*}
    For any $\omega\in\Omega_1$, the conditional Lindeberg condition holds in the probability space conditional on $\{Y_n=Y_n(\omega)\}$. Then in the conditional probability space $\{Y_n=Y_n(\omega)\},\ \omega\in\Omega_1$, we use Lindeberg-Feller Central Limit Theorem,
    \[
    \sup_{u\in\mathbb R}\mid P\{Z_n(Y_n(\omega))/B_n(Y_n(\omega))\leq u|Y_n(\omega)\} - \Phi(u)|\to 0
    \]
    and for any $t\in\mathbb R$
    \[
    E\left[\exp\{itZ_n(Y_n(\omega))/B_n(Y_n(\omega))\}\mid Y_n(\omega)\right]\to \exp(-t^2/2).
    \]
    Since $P(\Omega_1)=1$ then we have the almost sure convergence.

    The convergence in probability of $\{X_n\}$ is equivalent that for any sub-array $\{X_{n_k}\}$ of $\{X_n\}$, there exists some sub-sub-array $\{X_{n_k(i)}\}$ converges almost surely.
    For the convergence in probability, we use the subsequencing argument similar as \cite{Bai2022}.

    Hence for any sub-array $\{n_k\}$, we have the sub-sub-array $\{n_{k(i)}\}$ such that conditional Lindeberg condition holds a.s. . Following the first part proof, we have the corresponding result a.s. for the sub-sub-array $\{n_{k(i)}\}$. 

    With the definition of convergence in probability, we end the proof.
\end{proof}
\begin{lemma}
\label{lem:shao}
    Suppose that Lemma~\ref{lem:lind} holds a.s. or in probability and  additionally  $B_n(Y_n)\to B>0$ a.s. or in probability, respectively.
    Then we have
    \[
\sup_{u\in\mathbb R}\mid P\{Z_n(Y_n)/B\leq u|Y_n\} - \Phi(u)|\to 0
\]
a.s. or in probability, respectively.
\end{lemma}
\begin{proof}
    We prove the almost sure convergence.
    
    Denote 
    \begin{align*}
        &\Omega_1
        =\\
        &\{\omega:\frac{1}{B_n^2(Y_n(\omega))}\sum_{i=1}^n E\left[\{X_{ni}-E(X_{ni}\mid Y_n(\omega))\}^2 I\{\mid X_{ni}-E\{X_{ni}\mid Y_n(\omega)\}\mid >\varepsilon B_n(Y_n(\omega))\}\mid Y_n(\omega)\right]\to 0 \}
    \end{align*}
    and $\Omega_2=\{\omega: B_n(Y_n(\omega))\to B \}$. By the almost sure definition we have $P(\Omega_1\cap\Omega_2)=1$.

By Lemma~\ref{lem:lind}, we have
\[
\sup_{u\in\mathbb R}\mid P\{Z_n(Y_n(\omega))/B_n(Y_n(\omega))\leq u|Y_n(\omega)\} - \Phi(u)|\to 0,
\]
for any $w\in\Omega_1$. Because $w$ is fixed, equivalently, we have
\[
\sup_{u\in\mathbb R}\mid P\{Z_n(Y_n(\omega))\leq u|Y_n(\omega)\} - \Phi(u/B_n(Y_n(\omega)))|\to 0,
\]
for any $w\in\Omega_1$. 

Furthermore we claim that
\[
\sup_{u\in\mathbb R}\mid \Phi(u/B) - \Phi(u/B_n(Y_n(\omega)))|\to 0
\]
for any $w\in\Omega_2$.

Hence for any $\omega\in\Omega_1\cap\Omega_2$
\begin{align*}
&\sup_{u\in\mathbb R}\mid P\{Z_n(Y_n(\omega))\geq u|Y_n(\omega)\} - \Phi(u/B) \mid
\\
\leq&\sup_{u\in\mathbb R}\mid P\{Z_n(Y_n(\omega))\geq u|Y_n(\omega)\} - \Phi(u/B_n(Y_n(\omega)))|+ \sup_{u\in\mathbb R}\mid \Phi(u/B) - \Phi(u/B_n(Y_n(\omega))) \mid\\
\to&0.
\end{align*}
Since $P(\Omega_1\cap\Omega_2)=1$ we get the almost sure convergence.

To prove the claim, we define a series of i.i.d. $W_n\sim \mathcal N(0,1)$ which are independent with the probability space $(\Omega,\mathcal F,P)$.
For any fixed $\omega\in\Omega_2$, $B_n(Y_n(\omega))\to B>0$ and then by Slutsky Theorem we obtain that
\[
B_n(Y_n(\omega))W_n \to \mathcal N(0,B^2),
\]
in distribution
and equivalently,
\[
\sup_{u\in\mathbb R}\mid \Phi(u/B) - \Phi(u/B_n(Y_n(\omega)))|\to 0.
\]
We end the proof with subsequencing argument similar as the proof of Lemma~\ref{lem:lind}.
If Lemma~\ref{lem:lind} holds in probability, for any sub-array, we can take one sub-sub-array converging almost surely. Follow the first part proof again.
\end{proof}

\begin{lemma}[Uniform convergence of characteristic function on compact intervals]
\label{lem:unif_conv}
Assume for any $t\in \mathbb R$,
\[
E\left[\exp\{it W_n(Y_n)\}\mid Y_n\right]\to\exp(-t^2/2)
\]
a.s. or in probability.

Then for any $K>0$,
we have
\[
\sup_{t\in[-K,K]}\mid E\left[\exp\{it W_n(Y_n)\}\mid Y_n\right]-\exp(-t^2/2)\mid \to 0
\]
a.s. or in probability, respectively.
\end{lemma}
\begin{proof}
We prove the almost sure convergence. With the sub-sequencing argument the convergence in probability can be deduced analogously.

Only consider the event $\Omega=\{\omega:E\left[\exp\{it W_n(Y_n)\}\mid Y_n(\omega)\right]\to\exp(-t^2/2)\}$ with $P(\Omega)=1$. For fixed $\omega\in\Omega$, we have the convergence in distribution of $W_n(Y_n(\omega))$ on the conditional probability space. 

Denote $\phi_n(t)=E[\exp\{itW_n(Y_n(\omega))\}]$ and $\phi(t)=\exp(-t^2/2)$. Denote $F_n(x)=P(W_n(Y_n(\omega))\leq x)$ and $F(x)=\Phi(x)$, where $\Phi(\cdot)$ is the standard normal c.d.f.. The following proof is by 3 steps.

\textbf{Step 1.} The convergence in distribution is equivalent to tightness of measure.

For any $\varepsilon>0$ there exists $K>0$ such that $P(W_n(Y_n(\omega))\in[-K,K])\ge1-\varepsilon$.

\textbf{Step 2.} $\phi(t)$ is uniformly continuous.

$\phi'(t)=-t\exp(-t^2/2)\to0$ as $\mid t\mid\to\infty$. Hence $\sup_{t\in\mathbb R}\mid\phi'(t)\mid<\infty$ and then $\phi(t)$ is uniformly continuous.

\textbf{Step 3.} $\phi_n(t)$ is uniformly equi-continuous.

For any $\varepsilon>0$, choose some $K>0$ in Step 1. Because $\exp(it)$ is continuous at $t=0$, we can choose $\delta>0$ such that 
\[
|\exp(i0)-\exp\{ix(\xi-\eta)\}|<\varepsilon
\]
for $x\in[-K,K]$ and $|\xi-\eta|<\delta$.

Then we have
\begin{align*}
|\phi_n(\eta)-\phi_n(\xi)| &\leq \int_{[-K,K]}{|e^{i\xi x}-e^{i\eta x}|}d F_n(x) + \int_{[-K,K]^c} {|e^{i\xi x}-e^{i\eta x}|}d F_n(x) \\ 
&\leq \int_{[-K,K]}{|e^{ix\eta}|\cdot|1-e^{ix (\xi -\eta)}|} d F_n(x) + \int_{[-K,K]^c} {2} \, d F_n(x) \\
&\leq \int_{[-K,K]}\varepsilon d F_n(x) + 2 \varepsilon\\
&\leq 3 \varepsilon. 
\end{align*}

\textbf{Step 4.} $\phi_n$ has local uniform convergence.

Since $\phi(t)$ is uniformly continuous and $\phi_n(t)$ is uniformly equi-continuous, for any $\xi\in\mathbb R$ we choose $\delta$ such that 
\[
|\phi(\xi)-\phi(\eta)|<\v,\ |\phi_n(\xi)-\phi_n(\eta)|<\v
\]for any $\mid\xi-\eta \mid<\delta$.

Since $\phi_n(\xi)\to\phi(\xi)$, we choose $N$ such that
\[
|\phi_n(\xi)-\phi(\xi)|<\v
\]for any $n>N$.

Then we have for $|\xi-\eta|<\delta$ and any $n>N$.
\begin{align*} |\phi_n(\eta)-\phi(\eta)| &\leq {|\phi_n(\eta)-\phi_n(\xi)|} + {|\phi_n(\xi)-\phi(\xi)|} + {|\phi(\xi)-\phi(\eta)|}\\
&\leq 3\v.
\end{align*}

Since every open cover of compact set has a finite subcover, local uniform convergence is equivalent to uniform convergence on compact set. We conclude the proof.
\end{proof}

We give one more general lemma than Lemma S.1.2 in \cite{Bai2022}.
\begin{lemma}
\label{lem:bai}
    For $n\geq 1$, let $U_n$ and $V_n$ be real-valued random variables and $\mathcal F_n$ a $\sigma-$field generated by multivariate random variable $Y_n$. Suppose
    \[
    P\{U_n\leq u\mid \mathcal F_n\}\to \Phi(u/\tau_1)
    \]
    in probability, where $\Phi(\cdot)$ is the standard normal c.d.f. and $\tau_1,\tau_2$ are constants. Further assume $V_n$ is $\mathcal F_n$- measurable and 
    \[
    P\{V_n\leq u\}\to \Phi(u/\tau_2).
    \]
    Then
    \[
    P\{U_n+V_n\leq u\}\to\Phi(u/\sqrt{\tau_1^2+\tau_2^2}).
    \]
\end{lemma}
\begin{proof}
From the convergence in probability, we choose one sub-array $\{n_k\}_{k=1}^\infty$ converging almost surely. Then without loss of generality, 
    we prove the a.s. convergence.

    Denote $\Omega_1=\{\omega:P\{U_n\leq u\mid \mathcal F_n\}\to \Phi(u/\tau_1/2)\}$, where $P(\Omega_1)=1$.
    In the conditional probability space $\{Y_n=Y_n(w)\}$ where $\omega\in\Omega_1$, characteristic function converges too. Hence
    \[
    E\{\exp(itU_n)\mid Y_n\}\to \exp(it\tau_1^2/2) \ \mathrm{a.s.}
    \]
    $V_n$ converges to $\mathcal N(0,\tau_2^2)$. Then we have
    \[
    E\{\exp(itV_n)\}\to \exp(it\tau_2^2/2).
    \]
    Hence
    \begin{align*}
        E[\exp\{it(V_n+U_n)\}]=&E(E[\exp\{it(V_n+U_n)\}\mid \mathcal F_n])\\
        =&E[E\{\exp(itU_n)\mid \mathcal F_n\}\exp(itV_n)]\\
        \to&\exp\{it(\tau_1^2+\tau_2^2)/2\}.
    \end{align*}
    The three lines come from the tower property, $V_n\in\mathcal F_n$ and the dominated convergence theorem, respectively.

\end{proof}

\begin{lemma}[Conditional convergence in probability]
\label{lem:uncond-conv-in-prob}
    For $n\geq1$, let $U_n$ be real-valued random variables and $\{\mathcal F_n\}$ a $\sigma$-field. Suppose
    \[
    P(\mid U_n-c\mid\geq\varepsilon\mid \mathcal F_n)\to 0\ \text{in probability}
    \]
    for any $\varepsilon>0$ and some constant $c$. Then, 
    \[
    P(\mid U_n-c\mid\geq \v)\to 0 
    \]
    for any $\v>0$ and the same constant $c$.
\end{lemma}
\begin{proof}
    \begin{align*}
        P(\mid U_n-c\mid\geq\v)&=
        E\{I(\mid U_n-c\mid\geq\v)\}
        \\&=E[E\{I(\mid U_n-c\mid\geq\v)\mid \mathcal F_n\}]\\
        &=E[P(\mid U_n-c\mid\geq\v|\mathcal F_n)]\\
        &\to E(0)=0.
    \end{align*}
    The last convergence is from the dominated convergence theorem.
\end{proof}

\begin{lemma}[Conditional Markov's inequality]
\label{lem:cond_markov}
    Assume $X,S$ are two random variables. Then
    \[
    P\{\mid X-E(X\mid S)\mid \geq \v\mid S\}\leq \frac{\text{var}(X\mid S)}{\v^2}.
    \]
\end{lemma}
\begin{proof}
Since $\{ X-E(X\mid S)\}^2\geq0$ we have
    \begin{align*}
        \v^2P[\{ X-E(X\mid S)\}^2 \geq \v^2\mid S]\leq E[\{ X-E(X\mid S)\}^2\mid S]=\text{var}(X\mid S).
    \end{align*}
    Then by algebra
\[
P\{\mid X-E(X\mid S)\mid \geq \v\mid S\}=P[\{ X-E(X\mid S)\}^2 \geq \v^2\mid S],
\]
we obtain the result.
\end{proof}

\subsection{Other Lemmas}

\begin{lemma}[WLLN for triangular arrays]
\label{lem:wlln}
Let $\left\{U_{n, i}\right\}$ be a row-wise i.i.d. triangular array of random variables. If $\frac{\sup _n \text{var} (U_{n, i})}{n} \rightarrow 0$, then
$$
n^{-1} \sum_{i=1}^n\left(U_{n i}-E U_{n i}\right) \xrightarrow{P} 0.
$$
\end{lemma}
\begin{proof}The proof simply applies the Chebyshev inequality:
$$
\begin{aligned}
\pr\left\{\left|n^{-1} \sum_{i=1}^n\left(U_{n i}-E U_{n i}\right)\right|>\varepsilon\right\} & \leq \frac{E\left\{(n^{-1} \sum_{i=1}^n\left(U_{n i}-E U_{n i}\right)\right\}^2}{\varepsilon^2} \\
& =\frac{E\left(U_{n i}-E U_{n i}\right)^2}{n \varepsilon^2} \\
& \leq \frac{\sup _n \text{var} (U_{n, i})}{n \varepsilon^2} \\
& \rightarrow 0 \text { as } n \rightarrow \infty.
\end{aligned}
$$
The above equation holds for any $\varepsilon>0$. Thus, the lemma holds.
\end{proof}

\begin{lemma}[Variance of sample variance]
\label{lem:var_var}
    Assume $X_1,\cdots,X_n$ are i.i.d. $p-$dimensional random variables. Denote $\bar X=\frac{1}{n}\sum_{i=1}^n X_i$, $\sigma_{kkll}=E\left[(X_{ik}-\mu_k)^2(X_{il}-\mu_l)^2\right]$, $\sigma_{kl}=E\left[(X_{ik}-\mu_k)(X_{il}-\mu_l)\right]$ and $\mu_k=E(X_{ik})$.
    Then
    \begin{align*}
    E\left\{\frac{1}{n-1}\sum_{i=1}^n (X_i-\bar X)(X_i-\bar X)^\T\right\}&=\mathrm{var}(X),\\
    \mathrm{var}\left[\left\{\frac{1}{n-1}\sum_{i=1}^n (X_i-\bar X)(X_i-\bar X)^\T\right\}_{kl}\right]&=\frac{\sigma_{kkll}}{n}-\frac{\sigma_{kl}^2(n-2)}{n(n-1)}+\frac{\sigma_{kk}\sigma_{ll}}{n(n-1)}.
    \end{align*}
\end{lemma}
\begin{proof}
We rewrite the sample variance as the U-statistic form,
    \begin{align*}
        \frac{1}{n-1}\sum_{i=1}^n (X_i-\bar X)(X_i-\bar X)^\T=\frac{1}{\binom{n}{2}}\sum_{\{i,j\}}\frac{1}{2}(X_i-X_j)(X_i-X_j)^\T.
    \end{align*}
    Then we consider the $(k,l)$ element $\frac{1}{\binom{n}{2}}\sum_{\{i,j\}}\frac{1}{2}(X_{ik}-X_{jk})(X_{il}-X_{jl})$. It has expectation $\mathrm{Cov}(X_{ik},X_{il}):=\sigma_{kl}$, so the variance will be
    \[
    E\left[\frac{1}{\binom{n}{2}}\sum_{\{i,j\}}\left\{\frac{1}{2}(X_{ik}-X_{jk})(X_{il}-X_{jl})-\sigma_{kl}\right\}^2\right].
    \]

Expand the outer square and there are 3 types of cross product terms
$\left\{\frac{1}{2}(X_{i_1,k}-X_{j_1,k})(X_{i_1,l}-X_{j_1,l})-\sigma_{kl}\right\}\cdot\allowbreak\left\{\frac{1}{2}(X_{i_2,k}-X_{j_2,k})(X_{i_2,l}-X_{j_2,l})-\sigma_{kl}\right\}$
depending on the size of the intersection $\{i_1,j_1\}\cap\{i_2,j_2\}$.

1. When this intersection is empty, the factors are independent and the expected cross product is zero.

2. There are $n(n-1)(n-2)$ terms where $|\{i_1, j_1\} \cap\{i_2, j_2\}|=1$ and each has an expected cross product of $\left(\sigma_{kkll}-\sigma_{kl}^2\right) / 4$.

3. There are $\binom{n}{2}$ terms where $|\{i_1, j_1\} \cap\{i_2,j_2\}|=2$ and each has an expected cross product of $\left(\sigma_{kkll}+\sigma_{kk}\sigma_{ll}\right) / 2$.

Putting it all together shows that
$$
\operatorname{var}\left(S^2_{kl}\right)=\frac{\sigma_{kkll}}{n}-\frac{\sigma_{kl}^2(n-2)}{n(n-1)}+\frac{\sigma_{kk}\sigma_{ll}}{n(n-1)}.
$$

\end{proof}

\begin{lemma}[Variance of squared mean]
\label{lem:var_sqr_mean}
    Assume $X_1,\cdots,X_n$ are i.i.d. random variables with $\mu=E(X_i)$ and $\sigma^2=\mathrm{var}(X_i)$.
    Denote $\bar X=\frac{1}{n}\sum_{i=1}^n X_i$. Then
    \[
    E(\bar X^2)=\mu^2+\frac{\sigma^2}{n}.
    \]
\end{lemma}
\begin{proof}
\begin{align*}
    E(\bar X^2)=&\{E(\bar X)\}^2+\mathrm{var}(\bar X)\\
    =&\mu^2+\frac{\sigma^2}{n}.
\end{align*}
    
\end{proof}

\begin{lemma}
\label{lem:add mom}
    If for some $\gamma>0$, $ E|U|^\gamma<\infty$ and $E|V|^\gamma<\infty$, then $E|U+V|^\gamma<\infty$.
\end{lemma}
\begin{proof}
    \begin{align*}
        E|U+V|^\gamma\leq&E\{(|U|+|V|)^\gamma\}\\
        \leq&E\{\max(|2U|^\gamma,|2V|^\gamma)\}\\
        \leq&E|2U|^\gamma+E|2V|^\gamma\\
        <&\infty.
    \end{align*}
\end{proof}

\section{Precise Comparison Between New Variance Estimators and Ones Without Degree of Freedom Adjustment in \cite{ma2022regression}\label{section:comparison}}

In this section we denote $\hat V_{\cdot}$ and $\breve V_{\cdot}$ as new variance estimator and that in \cite{ma2022regression}.
We only consider $\hat V_{\mathrm{\hat\tau}}$ and $\breve V_{\hat\tau}$. The performance of the remaining two estimators is similar before and after adjusting the degrees of freedom.

Different variance estimators 
$$\hat V_\mathrm{\hat\tau}=\hat V_\mathrm{nB}(Y)+\hat V_\mathrm{nW}(Y)\text{ and }\breve V_{\hat\tau}=\breve V_{\mathrm{nB}}(Y)+\breve V_{\mathrm{nW}}(Y),$$
where the four estimators are 
\begin{align*}
\hat V_{nB}(Y)&=\sum_{s\in\mathcal S_n} \frac{n(s)}{n}\sum_{a\in\{0,1\}}\{\hat \mu_2(Y,a,s) - \hat\sigma^2(Y,a,s)\}-2\sum_{s\in\mathcal S_n}\frac{n(s)}{n}\left\{\hat\mu(Y,0,s)\hat\mu(Y,1,s)\right\}- \hat\tau^2,\\
\hat V_{nW}(Y)&=\sum_{s\in\mathcal S_n}\sum_{a\in\{0,1\}}\frac{n(s)}{n}\left\{\frac{n(s)}{n_a(s)}\hat\sigma^2(Y,a,s)\right\}.
\end{align*}
and
\begin{align*}
\breve V_{\mathrm{nB}}(Y)&=\sum_{s\in\mathcal S_n} \frac{n(s)}{n}\left\{\hat\mu(Y,1,s)-\hat\mu(Y,0,s)-\hat\tau\right\}^2,\\
\breve V_{\mathrm{nW}}(Y)&=\sum_{s\in\mathcal S_n}\sum_{a\in\{0,1\}}\frac{n(s)}{n}\left[\frac{\{n_a(s)-1\}}{n_a(s)}\frac{n(s)}{n_a(s)}\hat\sigma^2(Y,a,s)\right].
\end{align*}
The following theorem gives the precise difference between $\hat V_\mathrm{\hat\tau}$ and $\hat \breve V_{\hat\tau}$, which is always positive. However, given the stronger assumption
\[
\inf_{s\in\mathcal S_n}n(s)\xrightarrow{P}\infty,
\]
the difference tends to zero in probability. In practice we recommend new estimators to avoid underestimating true variances.

\begin{theorem}
    \[\hat V_\mathrm{nB}(Y)-\breve V_{\mathrm{nB}}(Y)=-\sum_{s\in\mathcal S_n} \frac{n(s)}{n}\left\{\frac{\hat\sigma^2(Y,1,s)}{n_1(s)}+\frac{\hat\sigma^2(Y,0,s)}{n_0(s)}\right\}\]  and \[\hat V_\mathrm{nW}(Y)-\breve V_{\mathrm{nW}}(Y)=\sum_{s\in\mathcal S_n} \frac{n(s)}{n}\left\{\frac{n(s)}{n_1(s)}\frac{\hat\sigma^2(Y,1,s)}{ n_1(s)}+\frac{n(s)}{n_0(s)}\frac{\hat\sigma^2(Y,0,s)}{ n_0(s)}\right\}.\]

    Then $\hat V_\mathrm{\hat\tau}- \breve V_{\hat\tau}=\sum_{s\in\mathcal S_n} \frac{n(s)}{n}\left\{\frac{\hat\sigma^2(Y,1,s)}{ n_1(s)}\frac{n_0(s)}{n_1(s)}+\frac{\hat\sigma^2(Y,0,s)}{n_0(s)}\frac{n_1(s)}{n_0(s)}\right\}$.
\end{theorem}
\begin{proof}
\begin{align*}
    \breve V_{\mathrm{nB}}(Y)&=\sum_{s\in\mathcal S_n} \frac{n(s)}{n}\left[\left\{\hat\mu(Y,1,s)-\hat\mu(Y,0,s)\right\}-\hat\tau\right]^2\\
    &=\sum_{s\in\mathcal S_n}\frac{n(s)}{n}\left[\left\{\hat\mu(Y,1,s)-\hat\mu(Y,0,s)\right\}^2-2\hat\tau\left\{\hat\mu(Y,1,s)-\hat\mu(Y,0,s)\right\}+\hat\tau^2\right]\\
    &=\sum_{s\in\mathcal S_n}\frac{n(s)}{n}\left\{\hat\mu(Y,1,s)-\hat\mu(Y,0,s)\right\}^2-\hat\tau^2\\
    &=\sum_{s\in\mathcal S_n}\frac{n(s)}{n}\left\{\hat\mu^2(Y,1,s)+\hat\mu^2(Y,0,s)-2\hat\mu(Y,1,s)\hat\mu(Y,0,s)\right\}-\hat\tau^2\\
    &=\hat V_\mathrm{nB}(Y)+\sum_{s\in\mathcal S_n} \sum_{a\in\{0,1\}}\frac{n(s)}{n}\left\{-\hat\mu_2(Y,a,s)+\hat\sigma^2(Y,a,s)+\hat\mu^2(Y,a,s)\right\}.
\end{align*}
For $-\hat\mu_2(Y,a,s)+\hat\sigma^2(Y,a,s)+\hat\mu^2(Y,a,s)$, we have the reduction for simplicity. Without loss of generality, we directly assume $n$ i.i.d. samples $y_i$ within strata.
\begin{align*}
    \hat\sigma^2+\bar y^2&=\frac{1}{n-1}\sum_{i=1}^n(y_i-\bar y)^2 +\bar y^2\\
    &=\frac{1}{n-1}\sum_{i=1}^n y_i^2 -\frac{n}{n-1}\bar y^2 +\bar y^2\\
    &=\frac{1}{n}\sum_{i=1}^n y_i^2 + \frac{1}{n(n-1)}\sum_{i=1}^n y_i^2 -\frac{1}{n-1}\bar y^2\\
    &=\frac{1}{n}\sum_{i=1}^n y_i^2 + \frac{\hat\sigma^2}{n}.
\end{align*}
Hence within strata we have
\[
-\hat\mu_2(Y,a,s)+\hat\sigma^2(Y,a,s)+\hat\mu^2(Y,a,s)=\frac{\hat\sigma^2(Y,a,s)}{n_a(s)}.
\] 
Plug in the form of $\breve V_{\mathrm{nB}}(Y)$ and we obtain
\[\hat V_\mathrm{nB}(Y)-\breve V_\mathrm{nB}(Y)=-\sum_{s\in\mathcal S_n} \sum_{a\in\{0,1\}}\frac{n(s)}{n}\frac{\sigma^2(Y,a,s)}{n_a(s)}.\]

Directly, we have
\begin{align*}
    \breve V_{\mathrm{nW}}(Y)&=\sum_{s\in\mathcal S_n}\sum_{a\in\{0,1\}}\frac{n(s)}{n}\left[\frac{\{n_a(s)-1\}}{n_a(s)}\frac{n(s)}{n_a(s)}\hat\sigma^2(Y,a,s)\right]\\
    &=\sum_{s\in\mathcal S_n}\sum_{a\in\{0,1\}}\frac{n(s)}{n}\left[\frac{\{n_a(s)\}}{n_a(s)}-\frac{n(s)}{n^2_a(s)}\right]\hat\sigma^2(Y,a,s)\\
    &=\hat V_\mathrm{nW}(Y)-\sum_{s\in\mathcal S_n}\sum_{a\in\{0,1\}}\frac{n(s)}{n}\frac{n(s)}{n^2_a(s)}\hat\sigma^2(Y,a,s).
\end{align*}
Finally,
\begin{align*}
    \hat V_\mathrm{\hat\tau}-\breve V_{\hat\tau}&=\hat V_\mathrm{nB}(Y)-\breve V_{\mathrm{nB}}(Y)+\hat V_\mathrm{nW}(Y)-\breve V_{\mathrm{nW}}(Y)\\
    &=\sum_{s\in\mathcal S_n} \sum_{a\in\{0,1\}}\frac{n(s)}{n}\left[-\frac{1}{n_a(s)}+\frac{n(s)}{n^2_a(s)}\right]\hat\sigma^2(Y,a,s)\\
    &=\sum_{s\in\mathcal S_n} \sum_{a\in\{0,1\}}\frac{n(s)}{n}\cdot\frac{n_{1-a}(s)}{n^2_a(s)}\hat\sigma^2(Y,a,s).
\end{align*}
\end{proof}

\end{document}